\documentclass[twocolumn,trackchanges]{aastex62}
\usepackage{amsmath}

\newcommand\kms{km s$^{-1}$}
\newcommand\masyr{mas yr$^{-1}$}
\newcommand\teff{$T_{eff}$}
\newcommand\logg{$\log g$}
\newcommand\vsini{$v_{rot}\sin i$}

\newcommand\msun{M$_\odot$}
\newcommand\vmax{$V_{max}$}
\newcommand\smax{$S_{max}$}

\begin{document}
\title{Close companions around young stars}

\author[0000-0002-5365-1267]{Marina Kounkel}
\affil{Department of Physics and Astronomy, Western Washington University, 516 High St, Bellingham, WA 98225}
\author[0000-0001-6914-7797]{Kevin Covey}
\affil{Department of Physics and Astronomy, Western Washington University, 516 High St, Bellingham, WA 98225}
\author{Maxwell Moe}
\affil{Steward Observatory, University of Arizona, 933 N. Cherry Ave., Tucson, AZ 85721, USA}
\author[0000-0001-5253-1338]{Kaitlin M. Kratter}
\affil{Steward Observatory, University of Arizona, 933 N. Cherry Ave., Tucson, AZ 85721, USA}
\author[0000-0002-2011-4924]{Genaro Su\'arez}
\affil{Instituto de Astronom\'{i}a, Universidad Nacional Aut\'{o}noma de M\'{e}xico, Unidad Acad\'{e}mica en Ensenada, Ensenada 22860, Mexico}
\author[0000-0002-3481-9052]{Keivan G. Stassun}
\affil{Department of Physics and Astronomy, Vanderbilt University, VU Station 1807, Nashville, TN 37235, USA}
\author[0000-0001-8600-4798]{Carlos Rom\'{a}n-Z\'{u}\~{n}iga}
\affil{Instituto de Astronom\'{i}a, Universidad Nacional Aut\'{o}noma de M\'{e}xico, Unidad Acad\'{e}mica en Ensenada, Ensenada 22860, Mexico}
\author[0000-0001-9797-5661]{Jesus Hernandez}
\affil{Instituto de Astronom\'{i}a, Universidad Nacional Aut\'{o}noma de M\'{e}xico, Unidad Acad\'{e}mica en Ensenada, Ensenada 22860, Mexico}
\author[0000-0001-6072-9344]{Jinyoung Serena Kim}
\affil{Steward Observatory, University of Arizona, 933 N. Cherry Ave., Tucson, AZ 85721, USA}
\author[0000-0002-5855-401X]{Karla Pe\~{n}a Ram\'{i}rez}
\affil{Centro de Astronom\'{i}a (CITEVA), Universidad de Antofagasta, Av. Angamos 601, Antofagasta, Chile}
\author[0000-0002-1379-4204]{Alexandre Roman-Lopes}
\affil{Department of Physics \& Astronomy, Universidad de La Serena, Av. Juan Cisternas, 1200 North, La Serena, Chile}
\author[0000-0003-1479-3059]{Guy S Stringfellow}
\affil{Center for Astrophysics and Space Astronomy, Department of Astrophysical and Planetary Sciences, University of Colorado, 389 UCB, Boulder, CO 80309-0389, USA}
\author[0000-0002-7916-1493]{Karl O Jaehnig}
\affil{Department of Physics and Astronomy, Vanderbilt University, VU Station 1807, Nashville, TN 37235, USA}
\author[0000-0002-5936-7718]{Jura Borissova}
\affil{Instituto de F\'isica y Astronom\'ia, Universidad de Valpara\'iso, Av. Gran Breta\~na 1111, Playa Ancha, Casilla 5030, Chile}
\affil{Millennium Institute of Astrophysics (MAS), Santiago, Chile}
\author[0000-0003-2053-0749]{Benjamin Tofflemire}
\affil{Department of Astronomy, The University of Texas at Austin, Austin, TX 78712, USA}
\author[0000-0001-9626-0613]{Daniel Krolikowski}
\affil{Department of Astronomy, The University of Texas at Austin, Austin, TX 78712, USA}
\author[0000-0001-9982-1332]{Aaron Rizzuto}
\affil{Department of Astronomy, The University of Texas at Austin, Austin, TX 78712, USA}
\author[0000-0001-9811-568X]{Adam Kraus}
\affil{Department of Astronomy, The University of Texas at Austin, Austin, TX 78712, USA}
\author[0000-0003-3494-343X]{Carles Badenes}
\affil{PITT PACC, Department of Physics and Astronomy, University of Pittsburgh, Pittsburgh, PA 15260, USA}
\author{Pen\'{e}lope Longa-Pe\~{n}a}
\affil{Centro de Astronomía (CITEVA), Universidad de Antofagasta, Avenida Angamos 601 Antofagasta-Chile}
\author{Yilen G\'{o}mez Maqueo Chew}
\affil{Instituto de Astronom\'{i}a, Universidad Nacional Aut\'{o}noma de M\'{e}xico, A.P. 70-264, 04510, Mexico, D.F., Mexico}
\author[0000-0003-1086-1579]{Rodolfo Barba}
\affil{Department of Physics \& Astronomy, Universidad de La Serena, Av. Juan Cisternas, 1200 North, La Serena, Chile}
\author[0000-0002-1793-3689]{David L. Nidever}
\affil{Department of Physics, Montana State University, P.O. Box 173840, Bozeman, MT 59717-3840, USA}
\affil{National Optical Astronomy Observatory, 950 North Cherry Ave, Tucson, AZ 85719}
\author{Cody Brown}
\affil{Department of Physics, Montana State University, P.O. Box 173840, Bozeman, MT 59717-3840, USA}

\author[0000-0002-3657-0705]{Nathan De Lee}
\affil{Department of Physics, Geology, and Engineering Technology, Northern Kentucky University, Highland Heights, KY 41099}
\affil{Department of Physics and Astronomy, Vanderbilt University, VU Station 1807, Nashville, TN 37235, USA}
\author[0000-0002-2835-2556]{Kaike Pan}
\affil{Apache Point Observatory and New Mexico State University, P.O. Box 59, Sunspot, NM, 88349-0059, USA}
\author[0000-0002-3601-133X]{Dmitry Bizyaev}
\affil{Apache Point Observatory and New Mexico State University, P.O. Box 59, Sunspot, NM, 88349-0059, USA}
\affil{Sternberg Astronomical Institute, Moscow State University, Moscow, Russia}
\author{Daniel Oravetz}
\affil{Apache Point Observatory and New Mexico State University, P.O. Box 59, Sunspot, NM, 88349-0059, USA}
\author{Audrey Oravetz}
\affil{Apache Point Observatory and New Mexico State University, P.O. Box 59, Sunspot, NM, 88349-0059, USA}

\email{marina.kounkel@wwu.edu}
\begin{abstract}

Multiplicity is a fundamental property that is set early during stellar lifetimes, and it is a stringent probe of the physics of star formation. The distribution of close companions around young stars is still poorly constrained by observations. We present an analysis of stellar multiplicity derived from APOGEE-2 spectra obtained in targeted observations of nearby star-forming regions. This is the largest homogeneously observed sample of high-resolution spectra of young stars. We developed an autonomous method to identify double lined spectroscopic binaries (SB2s). Out of 5007 sources spanning the mass range of $\sim$0.05--1.5 \msun, we find 399 binaries, including both RV variables and SB2s. The mass ratio distribution of SB2s is consistent with a uniform for $q<0.95$ with an excess of twins with $q>0.95$. The period distribution is consistent with what has been observed in close binaries ($<10$ AU) in the evolved populations. Three systems are found to have $q\sim$0.1, with a companion located within the brown dwarf desert. There are not any strong trends in the multiplicity fraction (MF) as a function of cluster age from 1 to 100 Myr. There is a weak dependence on stellar density, with companions being most numerous at $\Sigma_*\sim30$ stars/pc$^{-2}$, and decreasing in more diffuse regions. Finally, disk-bearing sources are deficient in SB2s (but not RV variables) by a factor of $\sim$2; this deficit is recovered by the systems without disks. This may indicate a quick dispersal of disk material in short-period equal mass systems that is less effective in binaries with lower $q$.

\end{abstract}

\keywords{binaries: spectroscopic, stars: pre-main sequence}

\section{Introduction}

Approximately half of the solar-type stars in the solar neighborhood are found in multiple stellar systems \citep{duquennoy1991,raghavan2010}. The period distribution of companions in these systems is log-normal, and the multiplicity fraction (MF) of close companions with separations $<$10 AU is $\sim$20\% \citep{moe2017}. The total MF is mass dependent, however, decreasing to $\sim$30\% for M dwarfs \citep[e.g,.][]{ward-duong2015}, and increasing to almost 100\% for OB stars \citep{sana2014}. In addition to this dependence on primary mass, the MF has been found to depend on metallicity, in that metal poor solar-type stars are more likely to have a close companion \citep{moe2018,badenes2018}. A recent review of multiplicity has been conducted by \citet{duchene2013}.

Binary systems form early in a star's lifetime; companions are commonly observed around the youngest protostars \citep[e.g.,][]{chen2010,tobin2016a}. The primordial multiplicity distribution may be altered through dynamical interactions between the young stars, such as through Kozai-Lidov interactions \citep[e.g.,][]{moe2018a}, but by the time these stars disperse from their birth environments, their multiplicity properties are expected to resemble, on average, those found in the field population.

A number of studies of multiplicity in nearby star-forming regions have been conducted, typically through high resolution imaging. The current diffraction limited observations, however, struggle to resolve companions closer than 10 AU even in the closer ($<$200 pc) regions. However, a number of differences have been observed between the MF measured in these star-forming regions and the field. Most notably, the Taurus Molecular clouds are known for having a very high MF, in excess of 60--70\% \citep[e.g.,][]{kraus2011}, significantly higher than $\sim$40\% found in the field \citep{raghavan2010}. A similar excess has also been found recently in the Orion Nebular Cluster (ONC), where companions with separations of 10--100 AU are twice as numerous as the rate that is found in the field \citep{duchene2018}. At larger separations, however, the MF in the ONC drops to the levels consistent with the measurements in the field, possibly due to earlier dynamical interactions with other cluster members \citep{reipurth2007}. In the larger Orion Molecular Clouds, however, outside of the ONC, the companions with separations of 100--1000 AU have been found to be $\sim$2 times more common in the densely populated stellar regions than in the more diffuse regions \citep{kounkel2016a}, suggesting a density dependent mechanism for the fragmentation of protostellar envelopes.

In order to analyze the distribution of companions at the separations closer than 10 AU, spectroscopic studies have been conducted to search for radial velocity (RV) variable systems \citep[e.g.,][]{mathieu1992,melo2003,maxted2008,tobin2009,tobin2015,kuruwita2018}. Most of these studies, however, have focused on individual regions, and comparing their results is difficult due to differences in temporal coverage. MF measured by these studies was broadly consistent with what is observed in the field, but small sample sizes make a more robust analysis difficult. A direct comparison has been made between the Chameleon I and Taurus \citep{nguyen2012} regions, and between the ONC and NGC 2264 \citep{kounkel2016}, which have found MFs that are largely consistent. On the other hand, \citet{jaehnig2017} analyzed the APOGEE IN-SYNC observations of Orion, Perseus, and the Pleiades, finding an evidence for a possible evolution of the MF with age.

\citet{fernandez2017} conducted a first large study of the double lined spectroscopic binaries (SB2s) in the young clusters in the IN-SYNC data. They identified 104 potential systems, doubling the number of known young SB2s, although their approach was largely reliant on the visual examination of the data, and thus it was difficult to perform any robust statistics on the sample.

The presence of multiples does affect the early stages of stellar evolution because of the effect binaries have on protoplanetary disks. Companions at separations closer than 200 AU are more likely to form from disks rather than core fragmentation \citep{tobin2018,moe2018}. Disks around closely-separated binaries tend to be less luminous at sub-mm wavelengths \citep[e.g.,][]{jensen1996,harris2012,akeson2019} than around single stars. Disks in multiple systems also have shorter lifetimes than disks around single stars \citep[e.g.,][]{cieza2009,kraus2011}, indicating that companions dynamically disrupt both the circumstellar and circumbinary disks. However, a number of close circumbinary planets have been discovered \citep[e.g.,][]{doyle2011,orosz2012}, indicating some circumbinary disks must survive long enough to form planets.

In this paper, we present a study of the MF as a function of age, mass, separation, environment, and the presence of a protoplanetary disk. We base this study on the analysis of the APOGEE observations of nearby star forming regions and young clusters. We search this dataset for RV variables\footnote{The term SB1 is not used to avoid the confusion, because some of the sources identified as RV variable are also found to be SB2s} and SB2s with separations up to 10 AU, and report here the largest uniformly observed sample of close companions around young stars. In Section \ref{sec:sample} we discuss the APOGEE observations, data processing, and the sample construction. In Section \ref{sec:rvv} we identify RV variables, and in Section \ref{sec:sb2} we discuss the methods for an automated identification of SB2s. In Section \ref{sec:complete} we construct a sample of synthetic binaries  from which we infer the completeness of our search methods as a function of each system's orbital parameters. In Section \ref{sec:result} our measured and synthetic samples to test for differences between the multiple properties of young stars and those in the field. Finally, in Section \ref{sec:concl} we discuss the implications of the observations on the formation and evolution of young close multiples.

\section{Sample definition}\label{sec:sample}

\subsection{APOGEE Observations}

High resolution near-infrared spectra of several nearby star-forming regions and young clusters have been obtained by the Apache Point Observatory Galactic Evolution Experiment (APOGEE) spectrograph, which is mounted on the 2.5 m Sloan Digital Sky Survey (SDSS) telescope \citep{gunn2006,blanton2017}. The APOGEE spectrographs can observe up to 300 sources simultaneously across a 1.5$^\circ$ field of view, and the instrument covers the spectral range of 1.51--1.7 $\mu$m with the typical R$\sim$22,500 \citep[although it may vary between the individual fibers,][]{wilson2010,majewski2017}. The typical limiting magnitude for the observations analyzed here is H$\sim$13 mag.

The analysis in this work focuses on YSOs in the Orion Complex \citep{da-rio2016,da-rio2017,kounkel2018a}, NGC 1333 \citep{foster2015}, IC 348 \citep{cottaar2015}, Taurus Molecular Clouds, NGC 2264, $\alpha$ Per, and Pleiades. The observations originally began as part of the SDSS-III IN-SYNC survey with APOGEE, and later transitioned into the SDSS-IV Young Clusters Survey with APOGEE-2 with an expanded list of regions, increased footprint on the sky, and additional epochs covering a larger baseline for some of the existing targets. Sources were selected for targeting using the presence of infrared excess, optical variability, and when feasible, identification as a YSO in the literature \citep{cottle2018}. The goals of the APOGEE-2 survey was to create a homogeneously selected sample that is representative of the underlying population, although it may not be complete. Sources cannot be observed simultaneously if they are within 72'' of each other, otherwise a fiber collision would occur. In densely concentrated regions, this required a strategy of using multiple plates to cover the same region in the sky but containing different sets of sources. When it was impossible to fill the targets with just the candidate YSOs, the plate included other field sources.

In total, 53,452 spectra were obtained for 14,823 stars across $\sim$150 sq. deg. field of view over 6.5 years, from September 2011 to March 2018 (Table \ref{tab:sample}). A third of the data was made public as part of SDSS DR14 \citep{abolfathi2018}; the remaining spectra will be made released in DR16. 

\subsection{Stellar Parameters}

\begin{figure}
\epsscale{1.1}
\plotone{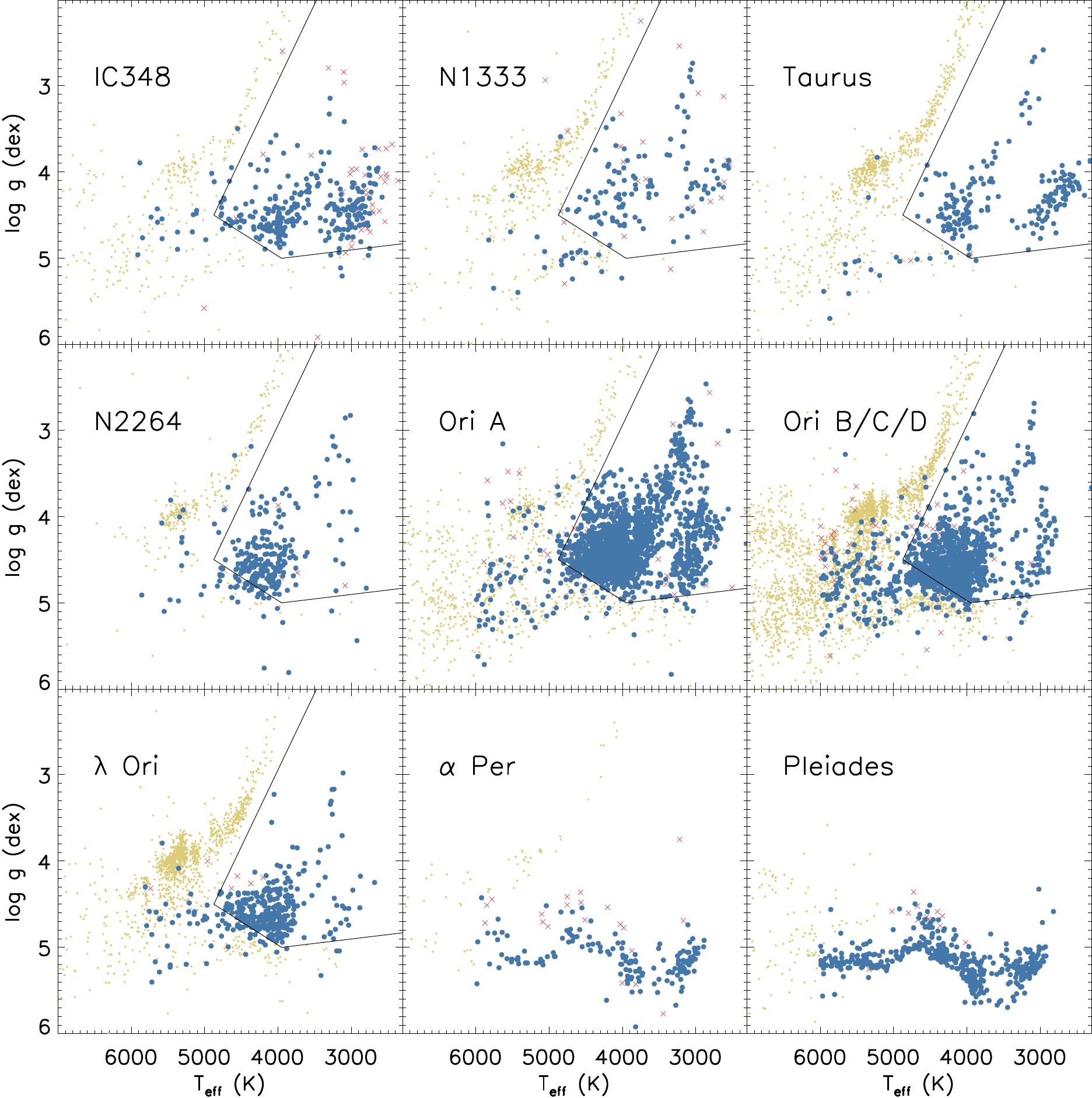}
\caption{Measured stellar parameters for the sources across the regions. Blue circles show all the sources that are part of the curated sample. Red X's are those that have made the kinematical cuts to distinguish them from the field stars, but had too poor SNR to be included in the analysis, either in the spectrum or in the CCF, or their CCF could not be deconvolved into any components. Yellow dots are either field stars or those young stars that were hotter than 6000 K. Black line shows the parameter space in which the sources were considered to be YSOs even if their kinematical parameters did not match that of the corresponding population, corresponding to points (3300,1.7),(4875,4.5),(3950,5.0),(2000,4.8). The gap in the distribution of \teff at $\sim$3500 K is due to the systematics of the IN-SYNC reduction pipeline.
\label{fig:tefflogg}}
\end{figure}

The data were processed and stellar parameters (RV, \teff, \logg, \vsini, and veiling - continuum excess in the spectrum due to accretion) were extracted from all the individual spectra using the pipeline developed as part of the IN-SYNC survey \citep{cottaar2014}, and an uncertainty weighted average was computed for each of the parameters measured for each star over all visits. These parameters are separate from those computed by the APOGEE's primary ASPCAP pipeline \citep{garcia-perez2016}, as ASPCAP is primarily calibrated for the red giants, and is not as successful in extracting accurate stellar parameters for dwarfs and pre-main sequence stars. RVs are largely consistent between both the IN-SYNC and the ASPCAP pipelines, however. There are some systematics in the \teff values measured with the IN-SYNC pipeline, most notable of which occurs at $\sim$3500 K, with this temperature being strongly disfavored, resulting in a gap in the overal distribution, which is apparent in Figure \ref{fig:tefflogg}. A full description of the data processing methods applied to these spectra is presented in \citet{kounkel2018a}. All of the measurements of the stellar properties for all the individual epochs for all sources across the footprint of the survey are included in Table \ref{tab:rv}.

To aid our identification of multiple systems, we supplement the stellar parameters extracted with the IN-SYNC pipeline with an analysis of the a cross correlation analysis. A cross correlation function (CCF) was computed for each spectrum by correlating the normalized spectra with normalized template from the PHOENIX spectral library \citep{husser2013} that best matched in \teff\ and \logg\ to the source. The template was broadened by a 10 \kms\ rotation kernel to achieve a smooth profile as 10 \kms\ is a typical \vsini\ for the pre-main sequence stars in the sample. The sky lines and the known bad pixels were masked. And, as the spectra observed by APOGEE are split into three separate chips with a chip gap spanning $\sim$50 \AA, the CCF was computed for all the chips separately and added together. Correlation was done across 401 velocity channels, separated by 1 \kms, centered at the velocity measured by the IN-SYNC pipeline at the given epoch rounded to the nearest integer \kms. 

\begin{figure}
\epsscale{1.1}
\plotone{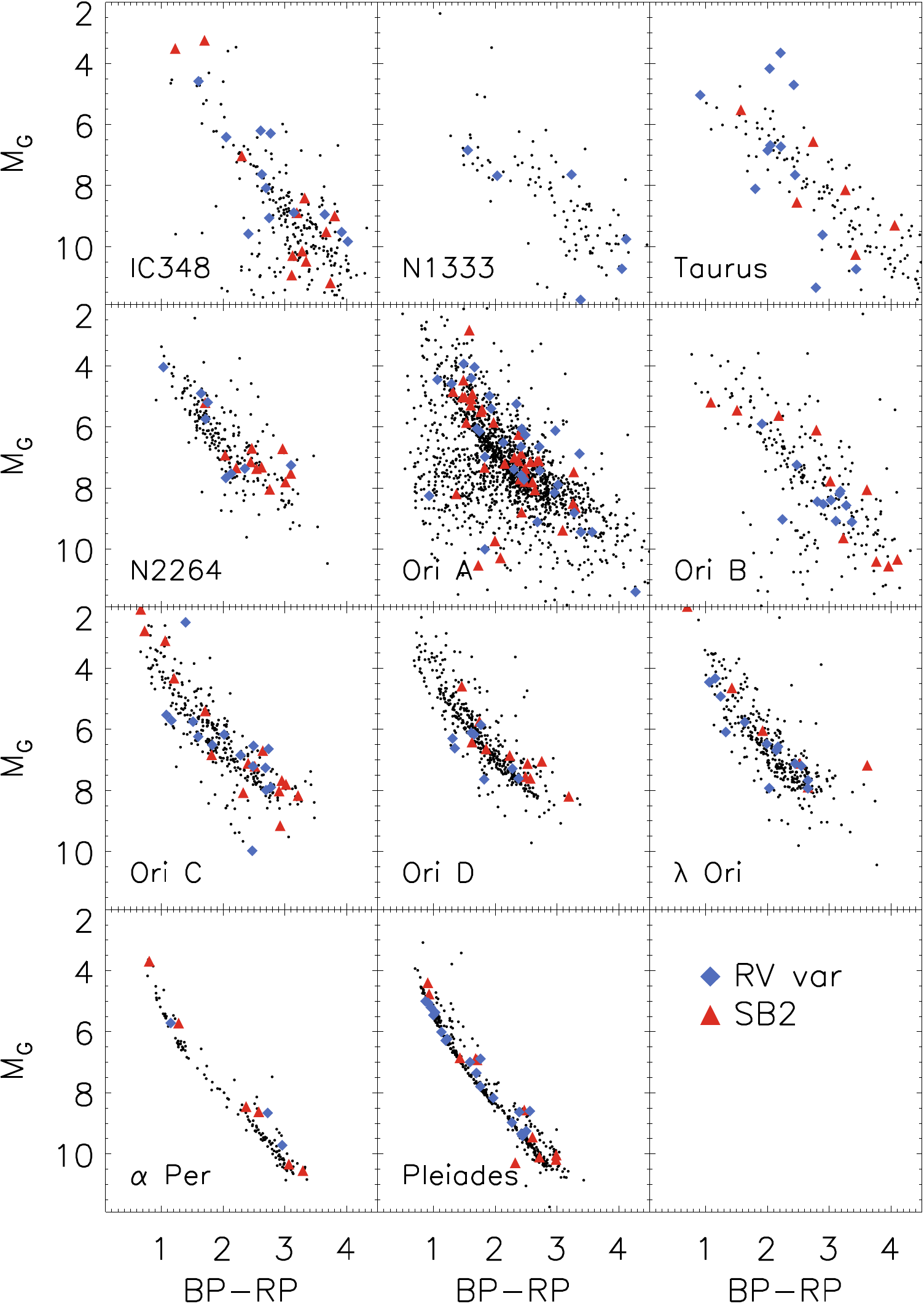}
\caption{HR diagram for each population (not corrected for extinction), with RV variables highlighed with blue diamonds, and SB2s with red triangles.
\label{fig:hr}}
\end{figure}

\begin{figure}
\epsscale{1.1}
\plottwo{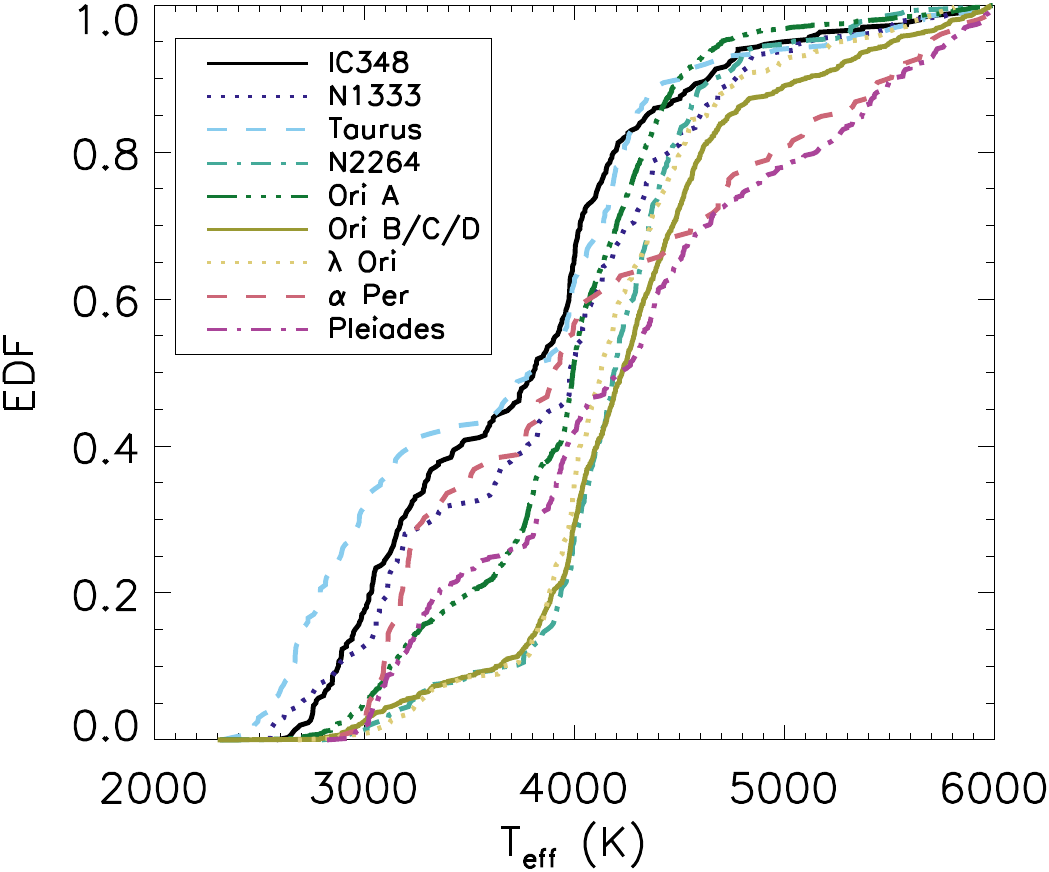}{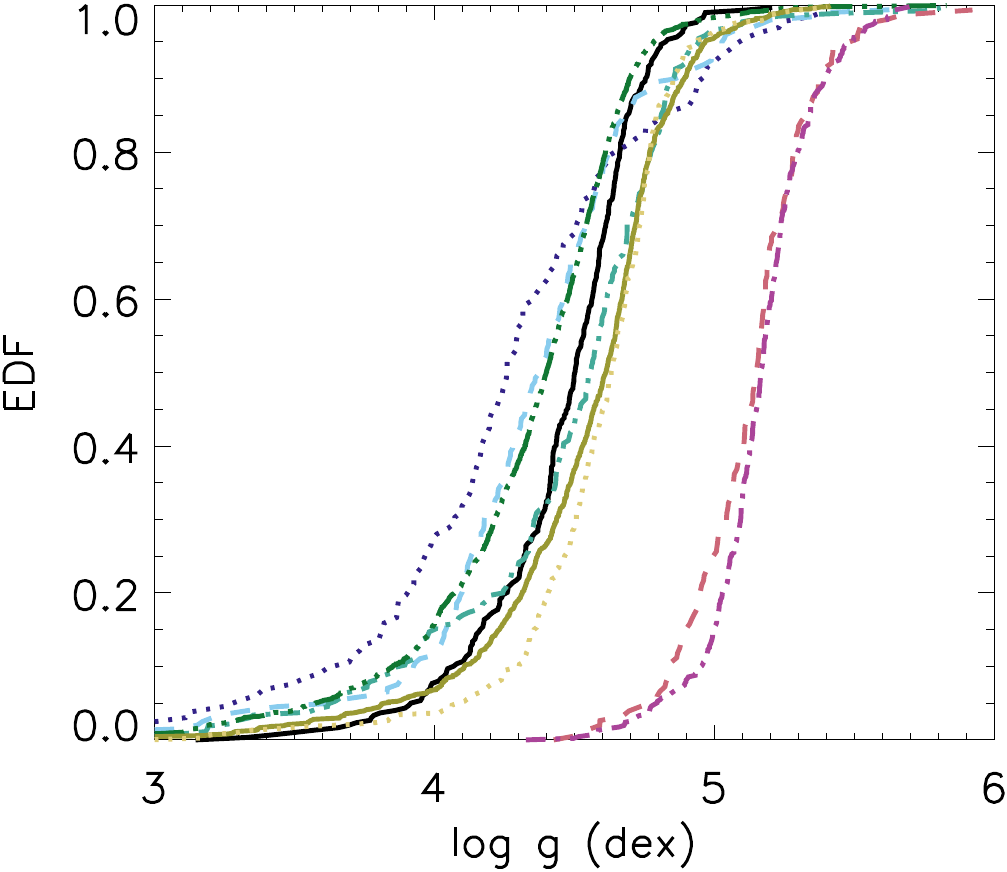}
\plottwo{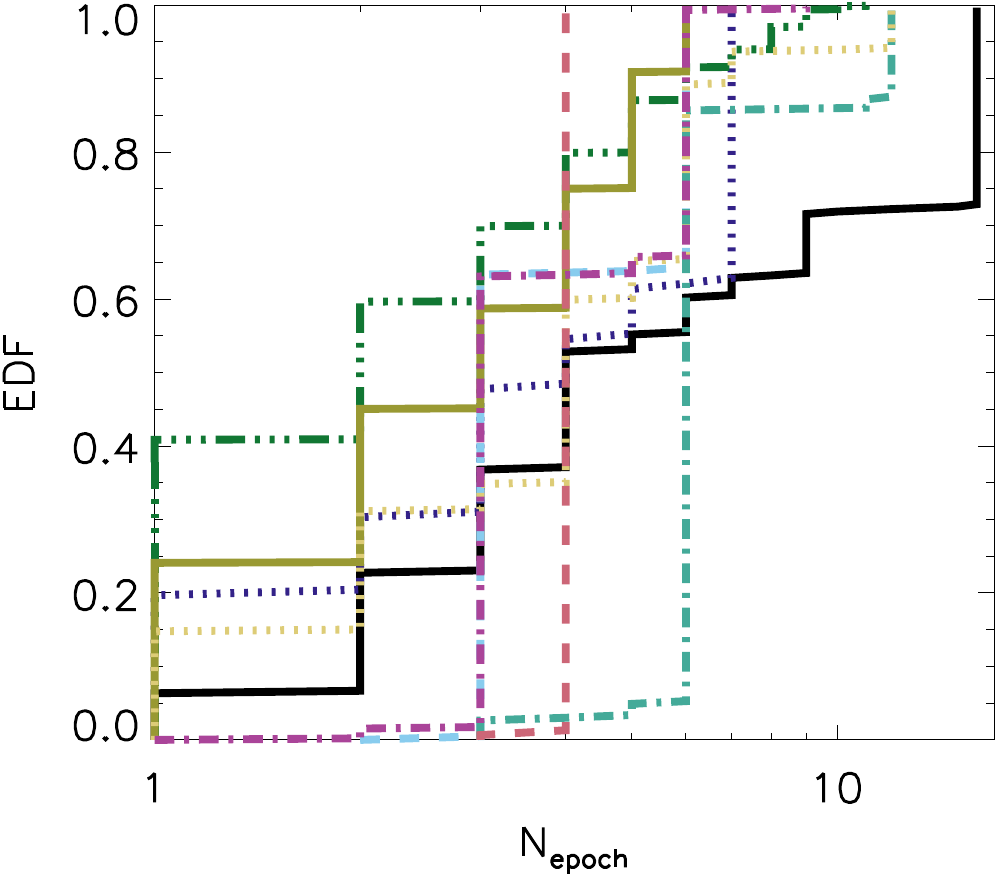}{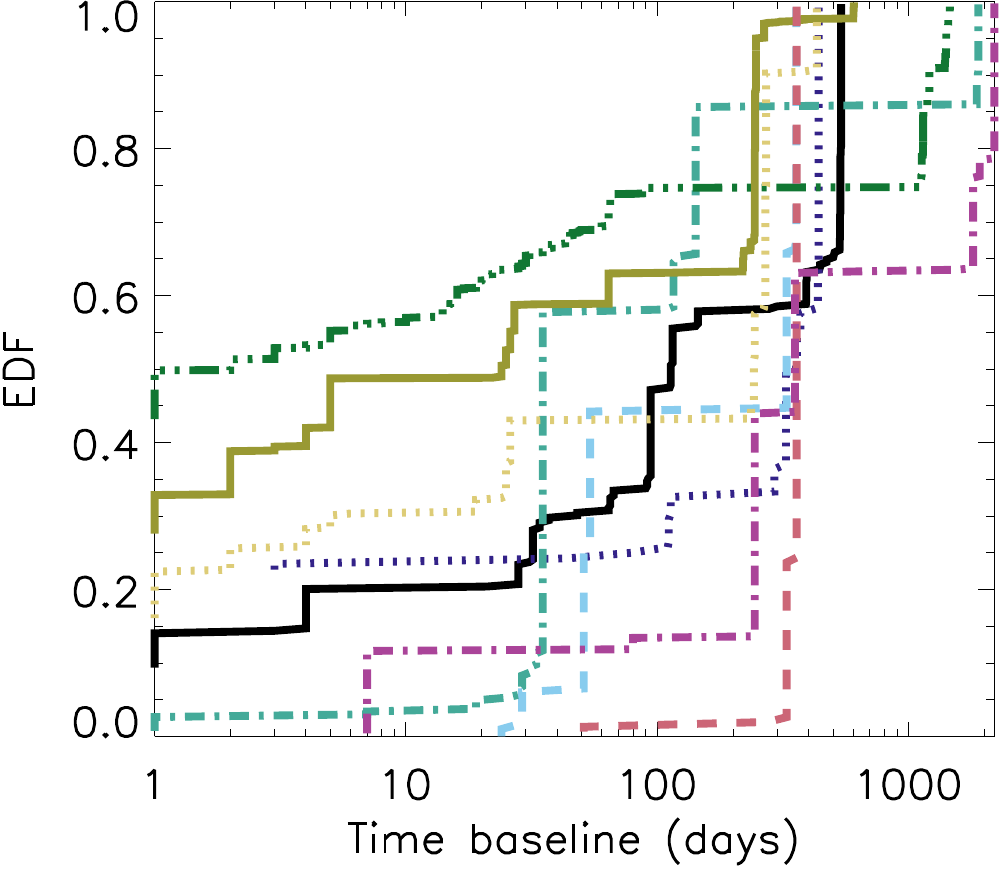}
\caption{Empirical cumulative distribution of the stellar properties for all of the sources in the curated sample, split into the individual regions.
\label{fig:edf}}
\end{figure}

Not all of the spectra obtained in the fields covered by the program were YSOs. If excess fibers were available in a given field, they were usually assigned to field red giants, which comprise the main targets of the APOGEE-2 survey. Additionally, the preliminary analysis of the yield of the young cluster targeting methods quantified a contamination rate of 5--10\% \citep{cottle2018} due to inclusion of nearby dwarfs or distant giants.

\subsection{Membership identification}

To curate the sample limited only to the members of the corresponding star forming regions and young clusters, several cuts were applied. The catalog was cross-matched with \textit{Gaia} DR2 \citep{gaia-collaboration2018,lindegren2018}, requiring agreements in positions of better than 1''. For sources that had \textit{Gaia} astrometric solutions (95\% of the total sample), we required the source to have weighted average RV \textit{or} a combination of parallax and proper motions to be consistent with the typical cluster properties to within 15--25 \kms\, and parallax to within several mas, depending on the cluster, to fully encompass the Gaussian distribution of positions and velocities within a given region (Table \ref{tab:cluster}).

Because \textit{Gaia} DR2 does not have a prescription for the astrometric binaries (which on average would have longer periods than the spectroscopic binaries), the astrometric solutions for some of them may fall outside of these accepted thresholds. High extinction that is frequently associated with YSOs may further degrade the astrometry, even for single stars. To avoid unduly biasing our sample in the presence of these effects we did not impose any quality cut on the Gaia measurements, neither in terms of the uncertainties, nor in terms of the excess astrometric noise, or any other alternative metric. On the other hand, spectroscopic binaries that are not observed with the full phase coverage may appear to have an average RV that is different from the true average by up to several dozen \kms. By requiring a consistency in either (and not necessarily both) of those properties, we attempt to retain all the possible YSO multiples in the sample. Sources that were not part of \textit{Gaia} DR2 (5\% of the sample) are most likely too reddened to have an optical detection. They are included without any RV constraint imposed onto these objects.

While these loose cuts may allow for an inclusion of some field stars as part of the cluster membership, this is statistically less detrimental to the analysis in this work than the exclusion of possible multiples. Any field stars will have a field probability of being a binary with the field orbital parameters, and their number in the sample is highly unlikely to be large enough to significantly alter the measured distribution in each cluster. On the other hand, the sample of spectroscopic binary YSOs is significantly smaller, so that any accidental exclusion of some of them would have a much stronger effect.

To provide the most inclusive sample of cluster members, we also examined the location of the sources in the \teff/\logg\ and $[G_B-G_R]/M_G$ diagrams were examined (Figure \ref{fig:tefflogg},\ref{fig:hr}). Low mass YSOs are very distinct from more evolved stars in these diagrams. Even if sources failed both the distance and the RV cuts, they were included in our analysis if they fall into the appropriate portion of these diagrams. The reason why it was done is because, in some cases, e.g., hierarchical triple systems, it is possible for multiplicity to affect both RV and astrometric measurements.

Another constraint on the curated sample was to impose a \teff$<6000$ K limit. Sources hotter than 6000 K have spectra dominated by hydrogen lines, resulting in a broad CCF and  more uncertain RV measurements, making them poorly suited for a uniform analysis of spectroscopic multiplicity. Moreover, it becomes significantly more difficult to distinguish hotter YSOs from the more evolved field stars from their position in the HR diagram. Finally, we required SNR of the spectra to be $>20$, and that CCF could be decomposed into at least a single Gaussian component (Section \ref{sec:sb2}).

In total, the curated sample contains 5007 stars/19127 individual spectra. Characteristic properties of stars in each cluster are given in Table \ref{tab:cluster}: the number of stars in each individual region, the 5th and 95th percentile of \teff\ and \logg\, baseline spanned by APOGEE, and a typical number of epochs per star. The cumulative distribution of the parameters is shown in the Figure \ref{fig:edf}.

\subsection{Disk classification}

To investigate the relationship between multiplicity and protoplanetary disks, the evolutionary classification on the state of the protoplanetary disk was obtained from the works of \citet{megeath2012} for Orion A and B, \citet{hernandez2007} for $\sigma$ Ori, \citet{suarez2017} for 25 Ori, \citet{hernandez2010} for $\lambda$ Ori, \citet{rebull2010} for Taurus, \citet{rapson2014} for NGC 2264, and \citet{luhman2016} for NGC 1333 and IC348. These catalogs were further supplemented by the WISE classification from \citet{marton2016}, and the remaining sources were assumed to be Class IIIs.

In the sample, there are 1882 disk bearing Class II sources, and 3125 are diskless Class III sources. There does not appear to be any biases in targeting sources of either evolutionary class; the Class II to Class III ratio among the sources in the curated sample is representative of what is typical in the corresponding clusters. It should be noted that there are a few Class I YSOs in the sample, but not enough to do any robust statistics on them separately. As they also have protoplanetary disks, they are grouped together with Class II systems.

\begin{figure}
\epsscale{1.1}
\plotone{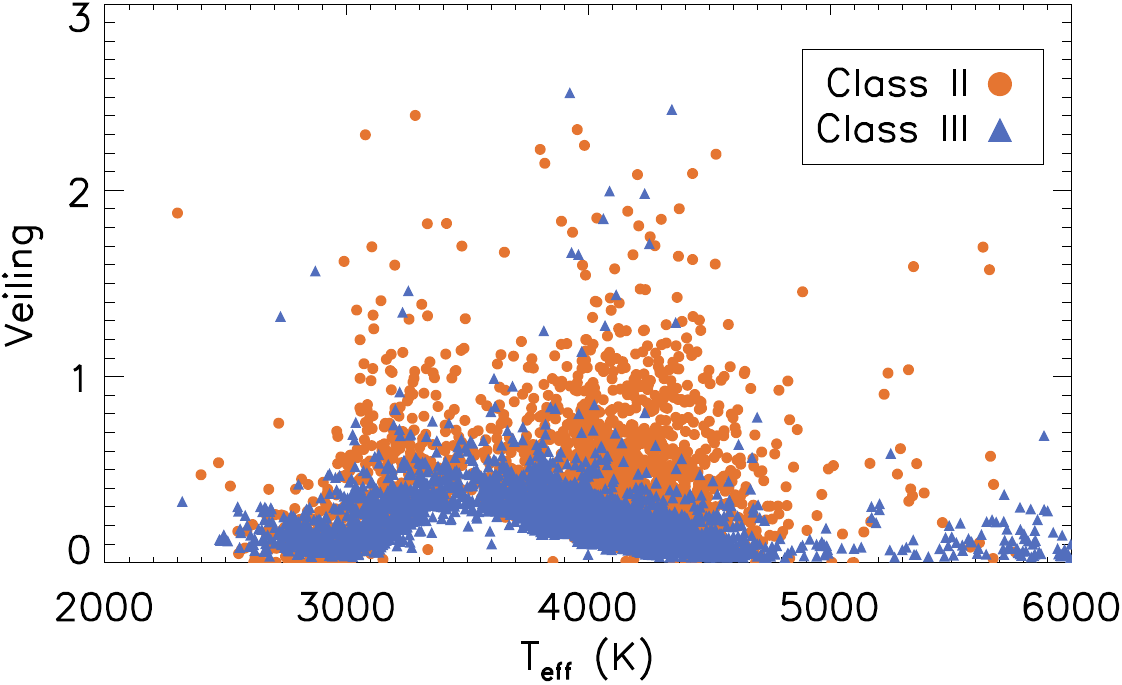}
\caption{Measured veiling of stars as a function of \teff, separated by the evolutionary classes.
\label{fig:veil}}
\end{figure}

In general, spectra of Class II and Class III systems have similar signal to noise ratio, similar fluxes, and similar shape of their CCF profiles. The only quantitative difference between them is the veiling, which has been measured by the pipeline. As mentioned in \citet{kounkel2018a}, the measurement of veiling in the spectrum has a number of systematic effects as a function of \teff, because the pipeline is subject to degeneracy between veiling and metallicity (Figure \ref{fig:veil}). In general, measurements of veiling $<$0.6 may not necessarily be reliable. However, with only a few exceptions, Class II systems do have systematically higher veiling than Class III systems.

\begin{splitdeluxetable*}{ccccccccBccccccccc}
\tabletypesize{\scriptsize}
\tablewidth{0pt}
\tablecaption{Catalog of the sources observed by APOGEE-2 survey towards the young clusters and star-forming regions\label{tab:sample}}
\tablehead{
\colhead{2MASS} &\colhead{$\alpha$} & \colhead{$\delta$} &\colhead{Region} & \colhead{YSO} &\colhead{N} &\colhead{RV}&\colhead{\vsini}&\colhead{\teff}&\colhead{\logg}&\colhead{Veiling}&\multicolumn{5}{c}{Binary}\\
\cline{12-17}
\colhead{ID} &\colhead{deg.} & \colhead{deg.}& \colhead{} & \colhead{class} &\colhead{epoch} &\colhead{\kms}&\colhead{\kms}&\colhead{K}&\colhead{dex}&\colhead{}&\colhead{SB2?}&\colhead{$\chi^2$}&\colhead{Slope}&\colhead{Lit. $\chi^2$}&\colhead{Lit. Slope}&\colhead{q}
}
\colnumbers
\startdata
\\
2M03192494+4859402 & 49.85391998 & 48.99450302 & alphaper & 3 & 4 & -0.489 $\pm$ 0.536 & 20.38 $\pm$ 0.97 & 3207 $\pm$ 22 & 5.223 $\pm$ 0.067 & 0.289 $\pm$ 0.038 & 1 & 0.3 & 0.4 & 0.3 & 0.4 &  \\
2M03193068+4903021 & 49.87783813 & 49.05060196 & alphaper & 3 & 4 & -9.223 $\pm$ 0.176 & 0.076 $\pm$ 0.56 & 3996 $\pm$ 13 & 5.090 $\pm$ 0.054 & 0.290 $\pm$ 0.025 & 1 & 0.4 & 0.4 & 0.4 & 0.4 &  \\
2M03193117+4941171 & 49.87989044 & 49.68811035 & alphaper & 3 & 3 &  0.588 $\pm$ 1.881 & 37.76 $\pm$ 3.00 & 3053 $\pm$ 40 & 5.161 $\pm$ 0.132 & 0.195 $\pm$ 0.068 & 1 &     & 	  &     &  	  &  \\
2M03220975+4834024 & 50.54065704 & 48.56735229 & alphaper & 3 & 4 & 15.332 $\pm$ 0.214 & 14.82 $\pm$ 0.48 & 3873 $\pm$ 15 & 5.346 $\pm$ 0.057 & 0.403 $\pm$ 0.024 & 2 & 1641.8 & 64.5 & 1641.8 & 64.5 & 0.89 $\pm$ 0.02\\
\enddata
\tablenotetext{}{Only a portion shown here. Full table with all deconvolved parameters is available in an electronic form.}
\end{splitdeluxetable*}

\section{RV variables}\label{sec:rvv}

In this section we analyze the results of the IN-SYNC pipeline, identify and correct the systematic offsets in the measured RVs, and identify RV variables.

\begin{deluxetable*}{cccccccc}
\tabletypesize{\scriptsize}
\tablewidth{0pt}
\tablecaption{Stellar properties measured from the IN-SYNC pipeline across the individual epochs\label{tab:rv}}
\tablehead{
\colhead{2MASS} &\colhead{MJD} & \colhead{RV} & \colhead{\vsini} &\colhead{\teff} &\colhead{\logg}&\colhead{Veiling} &\colhead{SNR}\\
\colhead{ID} &\colhead{} & \colhead{\kms} & \colhead{\kms} &\colhead{K} &\colhead{dex}&\colhead{}&\colhead{}
}
\startdata
2M03173879+4850489 & 57821 & -0.77 $\pm$ 0.35 & 0.086 $\pm$ 0.896 & 3617 $\pm$ 72 & 4.97 $\pm$ 0.14 & 0.52 $\pm$ 0.06 & 41.27\\
2M03173879+4850489 & 58097 & -0.00 $\pm$ 0.39 & 0.056 $\pm$ 0.621 & 3464 $\pm$ 63 & 4.92 $\pm$ 0.15 & 0.49 $\pm$ 0.07 & 37.86\\
2M03173879+4850489 & 58148 & -1.11 $\pm$ 0.38 & 0.148 $\pm$ 1.802 & 3507 $\pm$ 50 & 5.09 $\pm$ 0.13 & 0.41 $\pm$ 0.06 & 35.81\\
2M03173879+4850489 & 58179 & -1.29 $\pm$ 0.61 & 0.208 $\pm$ 2.319 & 3448 $\pm$ 75 & 5.03 $\pm$ 0.22 & 0.42 $\pm$ 0.10 & 24.25\\
\enddata
\tablenotetext{}{Only a portion shown here. Full table with all deconvolved parameters is available in an electronic form.}
\end{deluxetable*}

\begin{deluxetable*}{cccccccccccc}
\tabletypesize{\scriptsize}
\tablewidth{0pt}
\tablecaption{Observed clusters and star-forming regions\label{tab:cluster}}
\tablehead{
\colhead{Cluster} & \colhead{Age\tablenotemark{a}} & \colhead{N$_*$} & \colhead{N$_{bin}$} & \colhead{\teff$_{,5-50-95\%}$} &\colhead{\logg$_{,5-50-95\%}$} &\colhead{$\Delta$t} &\colhead{N$_{visits}$}&\colhead{RV} &\colhead{$\mu_\alpha$}&\colhead{$\mu_\delta$}&\colhead{$\pi$}\\
\colhead{} & \colhead{Myr} &  \colhead{} &  \colhead{} & \colhead{(K)} &\colhead{(dex)} &\colhead{(days)}&\colhead{}&\colhead{\kms} &\colhead{\masyr}&\colhead{\masyr}&\colhead{mas}
}
\startdata
IC348 & 3 (a) & 299 &24& 2750--3800--4990 &3.9--4.5--4.8 &538&4&10--25&0--10&-15--0&1.5--5\\
NGC 1333 & 1 (a) & 132 &7& 2680--3980--5060 &3.2--4.3--5.1&438&4 & 5--25 & 5--20 & -20--0& 2--6 \\
Taurus & 2 (b) &199&37&2550--3770--5220 &3.4--4.4--5.0&357&3 & 10--30 & -5--20 & -35-- -10 & 4.8--10 \\
NGC 2264 & 3 (c) &265&28&3210--4190--5010&3.6--4.6--4.9&1892&6 & 10--35 & -8--2 & -10--2 & 0.8--2.5\\
ONC & 2 (d) &1274&137&2990--3980--4690&3.5--4.4--4.8&1457&2 & 14--36 & -5--5 & -5--5 & 1.5--3.5 \\
L1641 & 2 (d) &711&40&2990--4000--4780&3.8--4.5--4.8&1476&3 & 14--36 & -5--5 & -5--5 & 1.5--3.5 \\
Orion B & 1 (d) &308&23&3010--4170--5310& 3.5--4.3--4.7&605&3 & 14--36 & -5--5 & -5--5 & 1.5--3.5 \\
Orion C\tablenotemark{b} & 4 (d)&360&34&3060--4210--5540 &3.9--4.6--5.0&628&3 & 14--25 & -5--5 & -5--5 & 1.5--3.5 \\
Orion D\tablenotemark{b} & 6 (d)&473&19&3324--4280--5710&4.4--4.7--5.0&628&2 & 25--36 & -5--5 & -5--5 & 1.5--3.5 \\
$\lambda$ Ori& 4 (d) &379&33&3240--4130--5250&4.1--4.6--4.9&710&4 & 14--36 & -5--5 & -5--5 & 1.5--3.5 \\
$\alpha$ Per& 50 (e) &152&9&3020--3900--5700&4.7--5.1--5.5&358&4 & -15--15& 15--35 & -35-- -20 & 4--8 \\
Pleiades& 125 (f) &455&36&3060--4210--5770&4.8--5.2--5.5&2190&3 & 0--20 & 10--30 & -55-- -35 & 6--9 \\
\enddata
\tablenotetext{a}{a - \citet{luhman2016}; b - \citet{luhman2018}; c - \citet{venuti2018}; d - \citet{kounkel2018a}; e - \citet{balachandran2011}; f - \citet{bouvier2018}}
\tablenotetext{b}{Following nomenclature from \citep{kounkel2018a}; sources without membership assignment were split on the basis of their RVs.}
\end{deluxetable*}

\subsection{Identification of RV variables}

Identification of RV variables was performed by applying two criteria to the IN-SYNC pipeline extracted RVs. The first criterion is a reduced $\chi^2>16$, which identifies sources that have a significant scatter in the typical velocity of a star. This is the same criterion that was used in \citet{kounkel2016}, and it translates to a 4$\sigma$ detection of RV variability. The second criterion is derived from the shape of the linear fit to all the available RV measurements for each system as a function of time. We flag sources as RV variable if the slope of their fitted RVs is inconsistent with 0 by $>4\sigma$. This second criterion helps to identify systems with long periods and that have a larger number of observations; these systems may be missed just with a $\chi^2$ cut. For both of these criteria, we required a minimum of 3 epochs, as the determination from just two epochs is both incomplete and more likely to result in false positives (Section \ref{sec:complete}). However, even with these criteria, it is still possible that some of the systems identified may be false positives, both due to the random counting statistics as well as some effects outside of multiplicity that may affect RVs (such as starspots, discussed more in Section \ref{sec:spots}

As was noted in \citet{kounkel2018a}, some of the RVs measured from lower SNR spectra by the IN-SYNC pipeline are not necessarily representative of the true velocity of the star. The spurious RVs are usually shifted by $>50$ \kms, making the source appear to be strongly RV variable despite the stability of the remaining measurements. While SNR$>$20 cut eliminates most of these spurious RVs, a handful of these poor measurements do occur in spectra with SNR$>$20 and thus persist in our sample. To confirm the accuracy of all the measurements in the curated sample, we visually examined the CCFs of all the sources identified as RV variable, as well as all the sources that had pipeline extracted RVs inconsistent with the position of the primary Gaussian component by more than 5 \kms. In total, 20 out of $>$19K observations in our sample ($\sim$0.1\%) were affected by this issue, and their RVs were corrected to the RV determined from the CCF that we compute independently of the IN-SYNC pipeline. This check was not performed for the sources that have failed the membership criteria.

\subsection{Systematics}

Afterwards, a consistency check was performed to find an epoch specific systematic offset across the entire observing field. We calculated a difference between all the individual RV measurements (at a given epoch for a given field) and the weighted averages of the sources that have at least 3 epochs and have not been identified as RV variables. If an epoch had more than 50 such sources, the median offset was calculated for the entire epoch. There are 144 (out of 171) epochs for which this could be performed; 75\% of them have a systematic offset $<$0.1 \kms, and for 92\% the offset is $<$0.2 \kms, which is smaller than the typical $\sigma_{RV}$ as measured by the IN-SYNC pipeline (which is typically $\sim$0.2--0.3 \kms). For only a few epochs the systematic offset is larger than the typical uncertainties of the observations. These offsets are removed from the RVs.

\begin{figure}
\epsscale{1.1}
\plotone{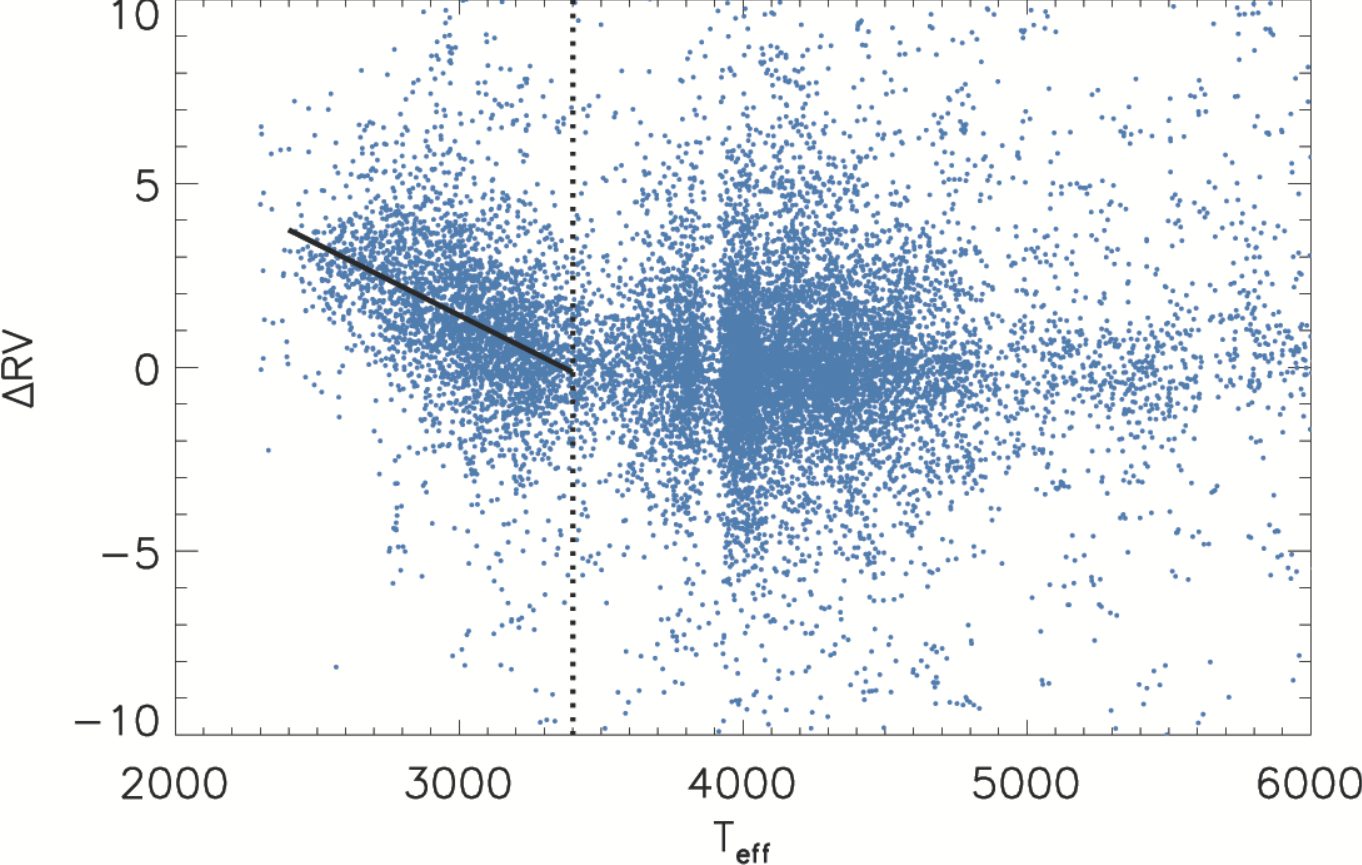}
\caption{Stellar RV with the average cluster velocity subtracted, as a function of \teff. The dotted line shows the \teff=3400 K; the sources cooler than this \teff have a systematic RV offset that can be fitted by $\Delta$RV=12.84-0.0038$\times$\teff (solid line). All of the clusters that are close enough to have a sources cooler than 3400 K observed show the same trend. The gap at $\sim$3500 K and 3900 K are some of the artifacts of the pipeline \teff measurements.
\label{fig:rvdiff}}
\end{figure}

As was noted by \citet{cottaar2014}, RVs of the sources with low \teff\ appear to have a strong systematic offset. For each star forming region, we average the RVs of all the sources that are hotter than 3500 K (excluding those that are identified as multiples) to get the region typical RV. In Figure \ref{fig:rvdiff}, we compare the RV offset of all sources relative to their cluster mean as a function of \teff. Sources cooler than 3400 K appear to be systematically redshifted by $\Delta$RV=12.84-0.0038$\times$\teff, which results in an offset as large as 4 \kms\ at \teff$\sim$2400 K. This effect is present regardless of the RV extraction method, as the RVs extracted by the IN-SYNC pipeline are generally consistent with RVs inferred from the CCFs, as well as with those reported in the APOGEE ASPCAP catalog. Most likely this is due to the synthetic spectra not carrying a sufficient precision in the energy levels for the lower \teff\ spectra.

\subsection{Literature RVs}

To improve the temporal baseline coverage of the data, we cross-match the sources in the curated sample with optical RV surveys, namely \citet{dolan2001}, \citet{furesz2006,furesz2008}, \citet{briceno2007}, \citet{flaherty2008}, \citet{gonzalez-hernandez2008}, \citet{maxted2008}, \citet{sacco2008}, \citet{nguyen2012}, \citet{hernandez2014}, \citet{kounkel2016,kounkel2017b}, and \citet{jackson2018}. However, this creates a very inhomogeneous dataset, as some of these surveys do not report individual RV measurements and instead average several measurements together. Many of them also do not list the dates of the individual observations, or even the related uncertainties. We use these data only for the purpose of identifying candidate RV variables in Lit $\chi^2$ and Lit Slope columns of Table \ref{tab:sample}, and do not rely on them for our further analysis of the MF or sample completeness in each region.

The RVs that are listed in Table \ref{tab:rv} are corrected to remove all of the aforementioned systematics (i.e, effect of temperature, and epoch dependent offsets). In total, 123 sources can be identified as RV variables in the APOGEE curated sample on the basis of the $\chi^2$, 161 from the slope, and 174 from either method, out of 2774 sources in the curated sample that have at least 3 epochs. Including the archival data, there are 205 sources that can be identified as RV variables from the $\chi^2$ cut, 255 from the slope (only including those observations with a well-defined date), and 297 in total, out of 3352 sources with at least 3 epochs (Table \ref{tab:sample}). 

\section{SB2}\label{sec:sb2}

\begin{deluxetable*}{cccccc}
\tabletypesize{\scriptsize}
\tablewidth{0pt}
\tablecaption{Deconvolved CCFs\label{tab:deconvolved}}
\tablehead{
\colhead{2MASS} & \colhead{MJD} & \colhead{FWHM$_1$} &\colhead{AMP$_1$} &\colhead{$v_1$} &\colhead{Flag$_1$}\\
\colhead{ID} & \colhead{} & \colhead{\kms} &\colhead{} &\colhead{\kms} &\colhead{}
}
\startdata
2M03173879+4850489 & 57821 & 20.1$\pm$ 0.9 & 0.77$\pm$0.03 &   -2.11$\pm$   0.36 & 4 \\
2M03173879+4850489 & 58097 & 17.9$\pm$ 1.1 & 0.63$\pm$0.03 &   -0.16$\pm$   0.46 & 4 \\
2M03173879+4850489 & 58148 & 18.2$\pm$ 1.4 & 0.69$\pm$0.05 &   -2.24$\pm$   0.60 & 4 \\
2M03173879+4850489 & 58179 & 25.9$\pm$ 0.7 & 0.50$\pm$0.01 &   -4.35$\pm$   0.31 & 4 \\
\enddata
\tablenotetext{}{Only a portion shown here. Full table with all deconvolved parameters, including those of the secondary peaks, is available in an electronic form.}
\end{deluxetable*}

In this section we develop and test a metric for an autonomous identification of SB2s and discuss some of the limiting cases that may confuse this metric.

\subsection{Initial decomposition}

Identification of SB2s was done on the shape of the CCF. To characterize the coherent signal in the CCF, and to relate it to the physical stellar properties, we also computed CCFs of the synthetic spectra of synthetic binary systems. The preliminary sample was constructed only for the purposes of calibrating and testing the pipeline, as well as exploring the initial effects on the parameter space, and it is not astrophysically significant beyond this; a more detailed synthetic sample with the full analysis of completeness is later discussed in Section \ref{sec:complete}.

The synthetic binaries were constructed using the PHOENIX spectral library \citep{husser2013}, combining two templates with \teff\ and \logg\ set by interpolating two randomly selected stellar masses onto a 5 Myr PARSEC-COLIBRI isochrone \citep{marigo2017}. Both spectra were convolved with a velocity kernel corresponding to a random \vsini (same for both stars), offset from each other by a velocity that corresponded to random separations, modulated by a random factor that accounts for the orientation of the system and eccentricity. Then the spectra were added together, flux scaled to represent a random distance, and the combined spectrum was interpolated over the wavelength range of a typical APOGEE spectrum. Random flux noise that is typical of the empirical APOGEE spectra was applied. In total, we computed 90,000 synthetic spectra of binary stars, as well as 10,000 spectra of the control single stars.

CCFs were deconvolved into Gaussians using a Python package GaussPy \citep{lindner2015}. The package relies on the Autonomous Gaussian Deconvolution which fits a profile described by a sum of $k$ Gaussians:
\begin{equation}
\sum_{k=1}^{N}\mathrm{AMP}_k\exp\left[-4\ln 2(x-v_k)^2/\mathrm{FWHM}_k^2 \right]
\end{equation} where $\mathrm{AMP}_k$, $v_k$, and $\mathrm{FWHM}_k$ describe the amplitude, mean position, and width of each of the Gaussian components.

The initial estimates of all the parameters, as well as the number of components ($N$) that are being fitted are chosen by the procedure through derivative spectroscopy. The only parameter that is supplied to the procedure is $\alpha$ which filters the derivative of the function and controls the relative balance between real variance and noise. GaussPy may struggle to find a solution if the fitted function does not level out to the baseline of zero. Therefore, to improve GaussPy's performance and to limit the number of peaks that would be fitted, from each CCF we subtracted the median, or 20\% of the peak, whichever was largest. The CCF was also extended by 100 additional points set to zero along each end. GaussPy was trained on the CCFs produced by the synthetic spectra, optimizing the recovery of Gaussians with $v$ and FWHMs to accurately reflect the injected radial and rotational velocities. Based on this training exercise, we adopt $\log\alpha=$1.5 for the analysis, as it provided the optimal balance of filtering out the noise structure from the CCF and fitting the primary peaks.

There may be other techniques that could be used to decompose CCF into Gaussians. For example, Finite Mixture Models such as `mclust' rely on well-established statistical methods based on maximum likelihood estimation \citep{mclachlan2000,fraley2002}. Performance of the GaussPy has not yet been tested against such techniques. Nonetheless, it was chosen due to the ease of implementation.

We report all of the GaussPy fitted parameters in the Table \ref{tab:deconvolved}. After the deconvolution, three cuts (described below) were applied to filter the sample. The reliability of the identified components is encoded in numerical quality flags on a scale from 1 to 4, depending on the success of passing these filters, with the quality flag of 4 being the most reliable \ref{tab:flag}.

\begin{deluxetable*}{cccc}
\tabletypesize{\scriptsize}
\tablewidth{0pt}
\tablecaption{Flag quality\label{tab:flag}}
\tablehead{
\colhead{Flag} & \colhead{Fraction of} & \colhead{Summary}\\
\colhead{} & \colhead{multiply deconvolved spectra} & \colhead{}
}
\startdata
1 & 42.4\% & Failed FWHM/amplitude test; could be noise \\
2 & 34.8\% & Fail the symmetry test; could be falsely multiply deconvolved \\
3 & 11.4\% & Inconclusive SB2/Spotted star pair \\
4 & 11.4\% & The primary peaks in all stars, and the secondary peaks of bona fide SB2s \\
\enddata
\end{deluxetable*}

First, Gaussians with an amplitude $<0.15$ or $>3$, as well as a FWHM$<1$ or $>500$ \kms\ were rejected. This cut removes the spurious CCF fits that do not correspond to realistic stellar properties. The components that failed these cuts were assigned a quality flag of 1. In the total sample, out of the initial sample of 3804 spectra that had been identified to have multiple components, only 2192 satisfied this cut (1525 of 2298 spectra in the curated sample).

\begin{figure*}
\epsscale{1.1}
\plottwo{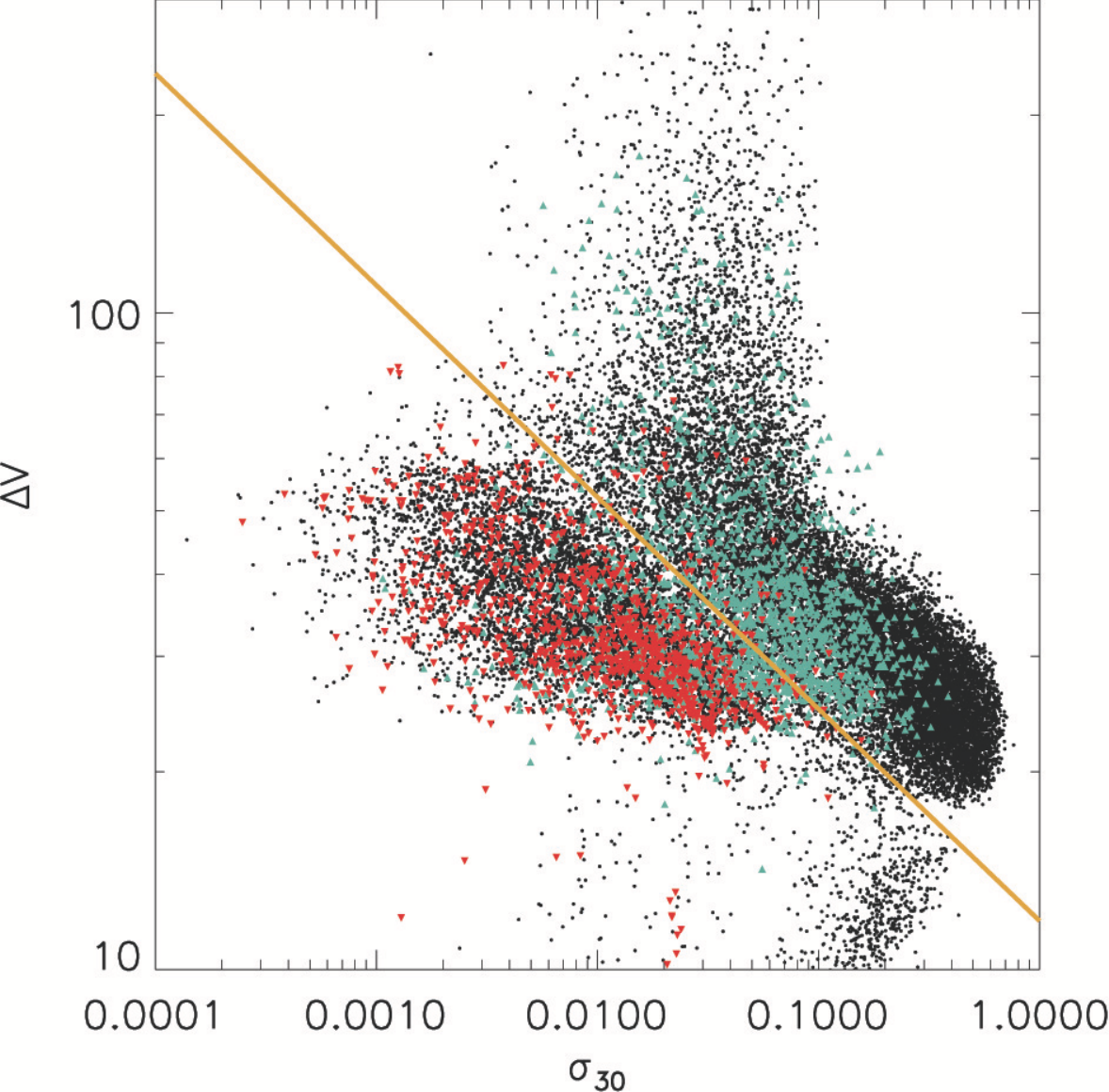}{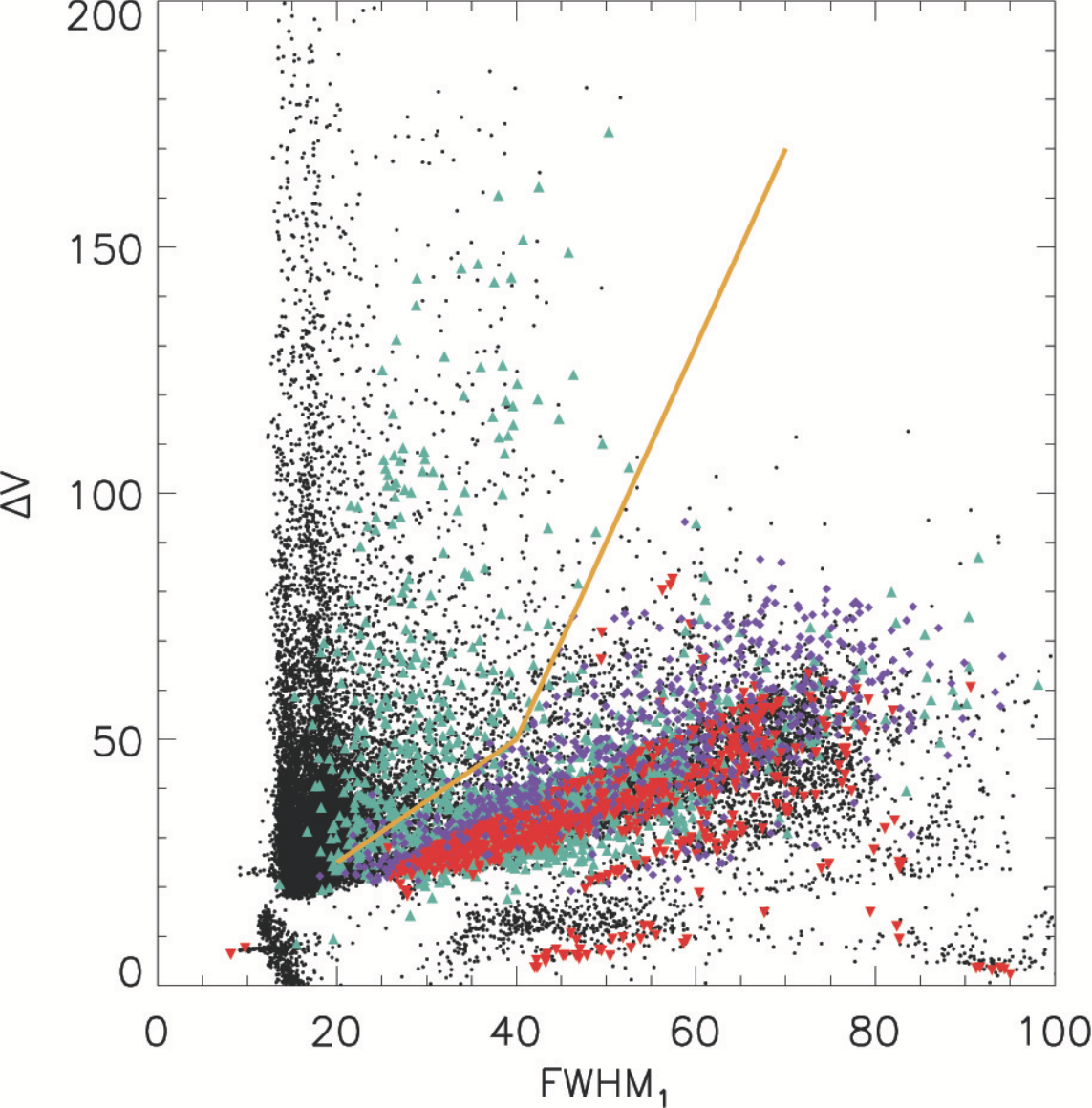}
\caption{Cuts imposed onto the deconvolved Gaussians in the CCF. Left panel shows the antisymmetric noise versus the velocity separation of the peaks. Black dots correspond to the components originating from the synthetically produced spectra, red inverted triangles highlight the multiply decomposed peaks within the control single sources. Teal triangles correspond to the components derived from the APOGEE spectra. Right panel shows the FWHM of the largest amplitude component versus the RV separation between multiply deconvolved peaks. Purple diamonds show the deconvolved parameters of the synthetic single spotted stars. The components that lie below the lines are rejected.
\label{fig:ccf_cut}}
\end{figure*}

A second cut removed sources whose Gaussian fits erroneously separated a single smooth peak of the CCF into multiple Gaussians. These spurious fits were identified by computing the CCFs's antisymmetric noise \citep[$\sigma_a$ in $R=h/\sqrt{2}\sigma_a$ from][]{tonry1979} within a 30 \kms\ radius ($\sigma_{30}$) centered on the Gaussian with the largest amplitude. Figure \ref{fig:ccf_cut}a shows $\sigma_{30}$ as a function of the velocity separation $\Delta v$ between the multiple Gaussians recovered from a single spectrum, including both synthetic binaries and a sample of synthetic control single stars. We use the location of multiple components extracted from CCFs for control stars to identify areas of parameter space in Figure \ref{fig:ccf_cut}a where spurious extraction reside. Eliminating spurious extractions from empirical CCFs in this region as well, we discarded secondary components if the difference in the velocity of the multiple peaks $\Delta v$ is smaller than the FWHM of the primary peak and $\log_{10}(\Delta$RV$)<-0.32258\log_{10}(\sigma_{30})+1.07419$. While there are some CCFs that do appear asymmetric upon visual examination that would fail this cut (e.g. bottom right panel in Figure \ref{fig:ccfs}), this limited the false positive identification of multiple significant components extracted from synthetic single star CCFs to 0.65$\pm$0.08\%. The secondary components that failed this cut were assigned the quality flag of 2. Within the footprint of young clusters, 728 sources/1325 visits were deconvolved into multiple components that pass both cuts, of which 573 sources/1022 visits were part of the curated sample.

\subsection{Spotted star analysis}\label{sec:spots}

An asymmetric CCF that can be characterized by a multi-Gaussian fit may not necessarily indicate the presence of a binary. Young stars with convective envelopes do have a strong degree of magnetic activity, and significant portions of their surface are often covered by spots \citep[e.g.,][]{bouvier1989,scholz2005}. A spot's apparent velocity will depend both on its location, and the star's rotation period. In many cases, the spot will have a radial velocity that differs significantly from the stellar systemic velocity, producing a line/CCF profile that resembles an SB2 (e.g. similar to the right panel of Figure \ref{fig:ccfs}).

In order to investigate the effect that spots would have on the CCF, we generated synthetic spectra of artificially spotted stars in a manner similar to the spectra of artificial binaries. Instead of broadening a synthetic spectrum by a kernel that would correspond to a given \vsini\ as was done previously, a stellar disk was spatially split into 500$\times\log$ \vsini\ evenly spaced regions. The spectrum from each region was then shifted to the apparent velocity due to the projected rotation velocity, and all spectra were coadded. For non-spotted stars, this simply approximates convolution with a broadening kernel. Then, a number of spots (up to 10) were generated, covering a total area up to 50\% of the stellar disk. These spots are centered at randomly chosen regions, and all of the regions covered by a spot are set to the temperature of the spot ($T_{spot}$). $T_{spot}$ is chosen for each \teff\ according to the $T_{spot}$/\teff\ relation given by \citet{berdyugina2005}. A total of 10,000 spotted spectra of single stars were generated, and their CCFs were analyzed using the same procedure as described above.

In this synthetic sample, 8.2$\pm$0.3\% of the CCFs were deconvolved to have multiple components. While many of the CCFs produced for the synthetic spotted stars had a high degree of high frequency structure that can only be indicative of spots, there were also a significant number of clean CCFs with a slight asymmetry that, upon blind examination, cannot be distinguished from the CCFs of SB2s with low $\Delta v$. The recovery rate and CCF shape did not depend strongly on the number of spots injected, the total area covered by spots, or on the spot's latitudinal or longitudinal positions (which was tracked only in a single spot systems). However, spots cannot produce multiple peaks in the CCF whose $\Delta v$ is significantly larger than the star's \vsini. Therefore, a third cut was introduced, flagging secondary components with FWHM$_1>0.8\Delta$RV \kms\ or FWHM$_1>0.25\Delta$RV+27.5 \kms\ as likely spot signatures, this removed 97.8$\pm$0.5\% of the synthetic spotted sample that remained from the previous cut. Potential components that trigger this likely spot cut were assigned a quality flag of 3, and the remaining binary candidates that pass all the previous tests were assigned a flag of 4. It should be noted that a number of bona fide SB2s likely fail this cut: in the spotless synthetic sample, 22\% of the previously identified binaries are rejected. Further monitoring would be required to separate spotted stars and true binaries more conclusively (by detecting the orbital period, strong variability in the shape of the CCF, or through identifying spots/eclipses in the stellar light curve). This cut may not be necessary for the analysis of the stars on the main sequence. Nonetheless, we restricted our sample of SB2s to only those sources for which we can conclusively discard the possibility of the second peak originating due to star spots, i.e., the sources that had multiple components with the quality flag of 4 in at least one epoch.

\subsection{Final sample and stellar properties of SB2s}

\begin{figure}
\epsscale{1.1}
\plottwo{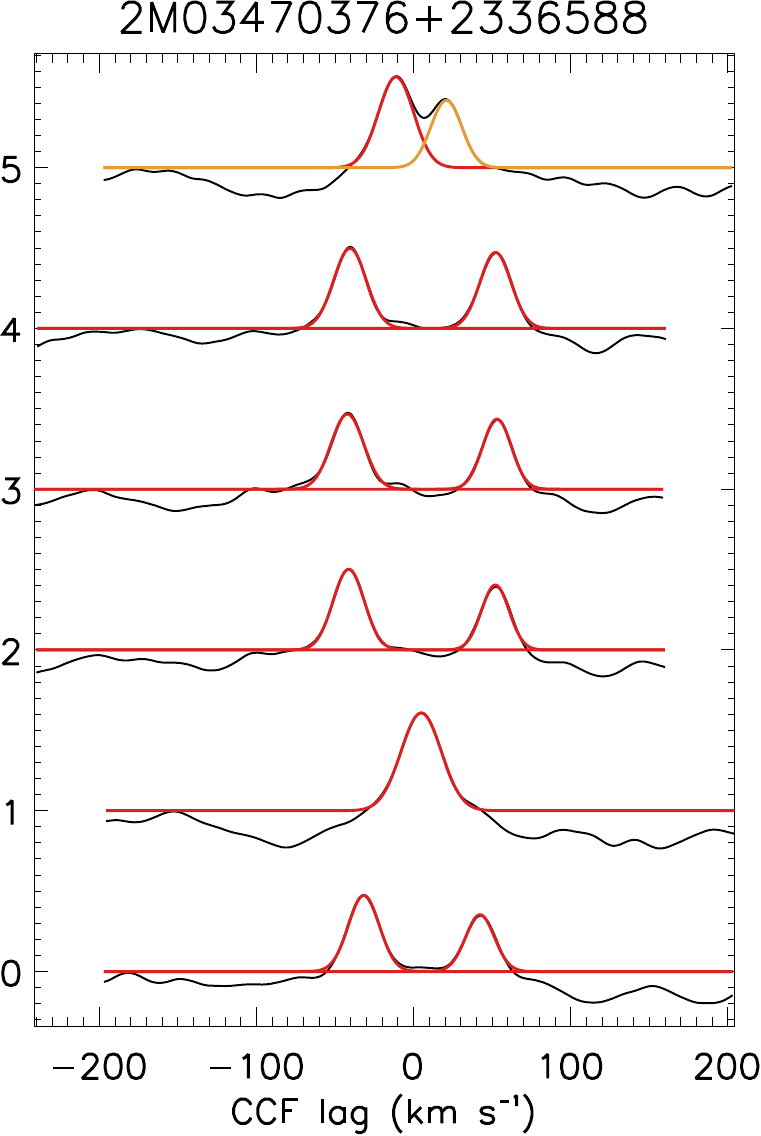}{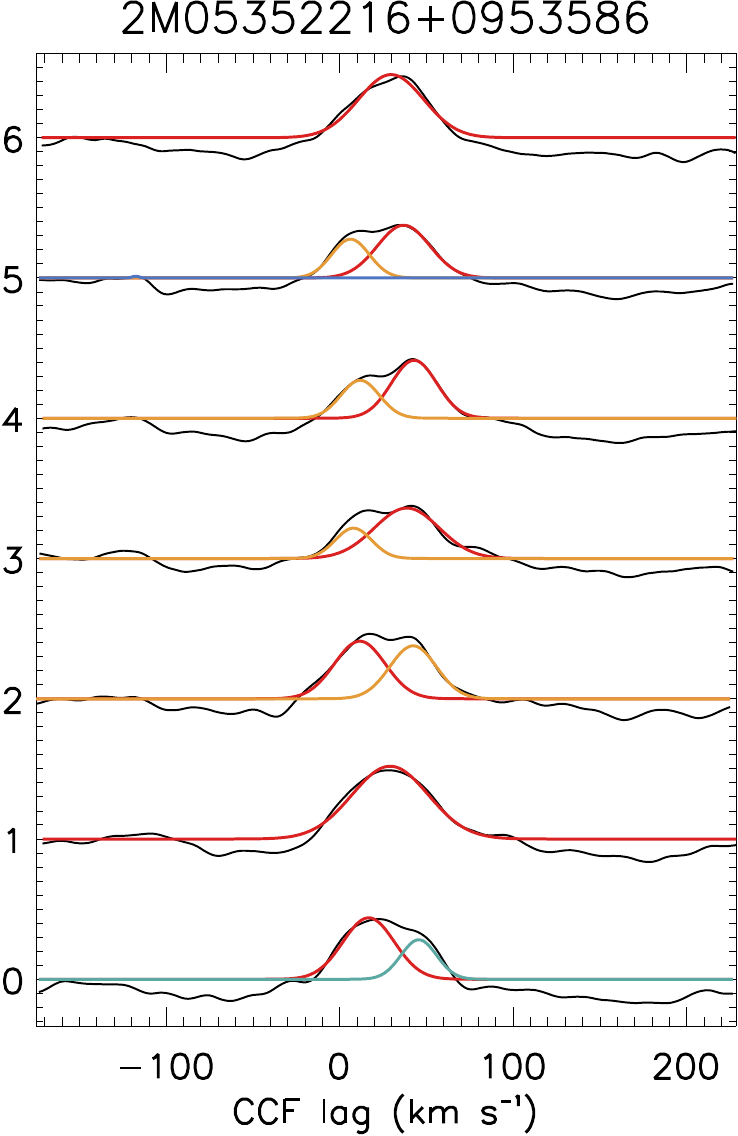}
\caption{Deconvolved CCFs of the sources in the primary sample for sources with at least one epoch that contains multiple profiles with Flag 3 or 4. Red shows the profiles with F=4, yellow are those with F=3, cyan are F=2, and blue are F=1. Only two sources shown. The full figure set with 573 figures is available in the online version.
\label{fig:ccfs}}
\end{figure}

We identify 229 sources/432 visits as SB2s in the full sample (14,823 sources), of which 141 sources/268 visits are in the curated sample of 5007 sources (Figure \ref{fig:ccfs}). Only 20\% of SB2s can also be identified as RV variables (32\% of the sources with at least 3 epochs). Comparing these results to the sample of 104 SB2s from \citet{fernandez2017} who used the earlier APOGEE/IN-SYNC sample in their analysis, we identify 65 of them as SB2s with a flag of 4, 24 with a flag of 3, 1 with a flag of 2, and 6 with a flag of 1. Only 8 systems are not recovered at all, usually due to the secondary component having too low of an amplitude for GaussPy to recover. 

\begin{figure}
\epsscale{1.1}
 \centering
		\gridline{\fig{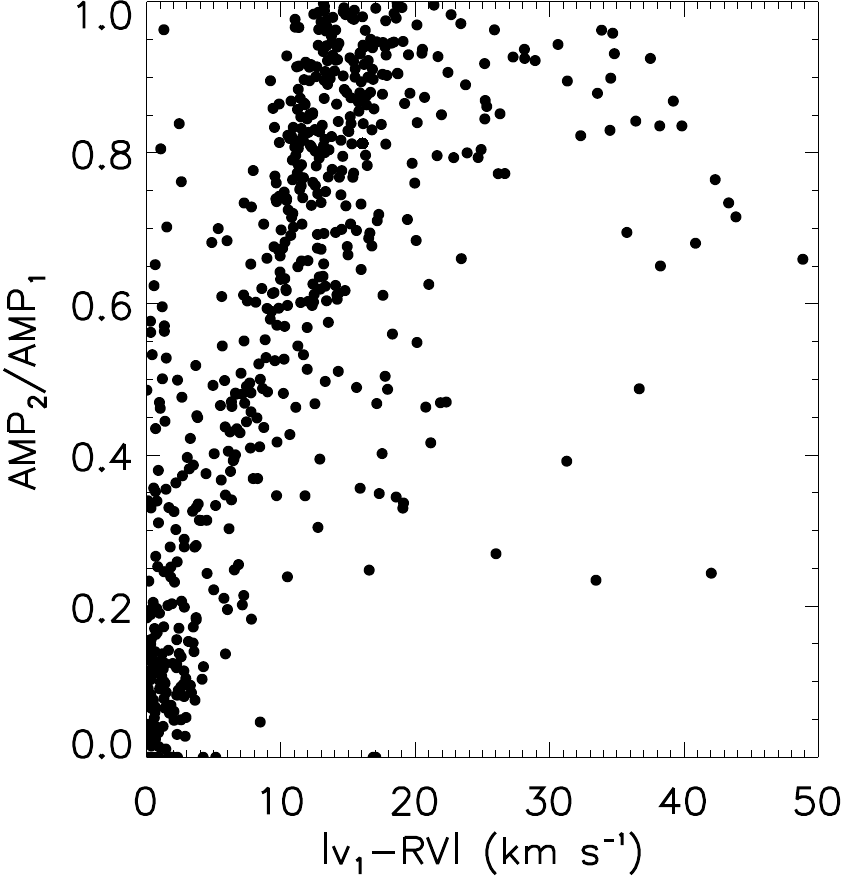}{0.22\textwidth}{}
             \fig{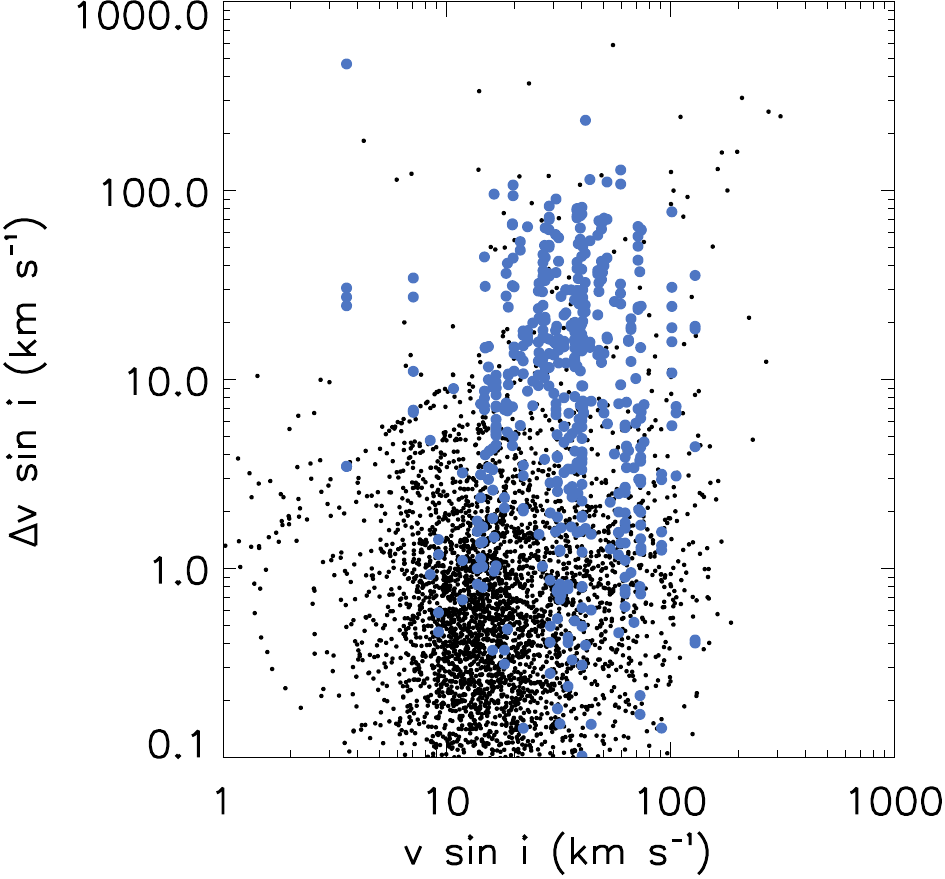}{0.25\textwidth}{}
        }
\caption{Left: The velocity of the primary component in an SB2 relative to the RV value determined from the pipeline as a function of the amplitude ratio between two peaks in the CCF. Right: Stability of \vsini\ of SB2s (blue points) in comparison to the rest of the sample (black points): sources with \vsini\ variability of more than 10 \kms\ relative to the average between epochs are more likely to be SB2s.
\label{fig:sb2pr}}
\end{figure}

We compare the RV extracted using GaussPy to those RVs that have been extracted by the IN-SYNC pipeline. If the ratio of the amplitudes of the two peaks is AMP$_2$/AMP$_1>0.6$ then typically the pipeline RV was positioned in between two components (Figure \ref{fig:sb2pr}, left panel). This occurs regardless of the velocity separation between the peaks. Conversely, in the high flux ratio regime (AMP$_2$/AMP$_1<0.6$) the pipeline RV agrees with the strongest component.

The \vsini\ values determined by the IN-SYNC pipeline can also be used to identify SB2s. The IN-SYNC \vsini\ for SB2s are often inflated and unstable from one epoch to the next (Figure \ref{fig:sb2pr}). By contrast, the FWHMs extracted from SB2 CCFs accurately reflect the actual \vsini\ used to construct the synthetic sample (Section \ref{sec:complete}). In the data, there is also a good agreement between the measured \vsini\ and FWHM in the sources not identified as SB2s. In the empirical spectra previously flagged as SB2s, however, the IN-SYNC pipeline \vsini\ is typically larger by a factor of $\sim 2$ than that implied by the FWHM of the CCF peak, because the IN-SYNC pipeline is unable to resolve two components and conflates both of them into one. The IN-SYNC pipeline measures a \vsini\ value for each epoch individually, and a weighted average is then computed for each system. For stars that have not been flagged as SB2s, individual measurements are usually consistent with each other, within the errors, but for the SB2s the scatter is typically $>$10 \kms\ (Figure \ref{fig:sb2pr}, right panel). This is the case for almost 80\% of the identified SB2s; conversely, only 25\% of the systems that do have $\Delta$\vsini$>10$ \kms\ are not identified as SB2s by previous cuts. The reason for this strong variability is because SB2 systems tend to have short periods. Thus the components corresponding to each star in the binary would have different RV in different observations, and the width Gaussian envelope that would conflate both components would vary between the epochs.

It is difficult to determine the reliability of the IN-SYNC pipeline estimates for other stellar parameters such as \logg, \teff, or veiling for the sources identified as SB2s, but the ranges of these parameters appear to be representative of what is typically found in the individual clusters. The sources identified as binaries (both RV variables, but especially SB2s) do appear to be brighter, however. The location of the binary sequence is ill-defined for pre-main sequence stars because stars of similar masses have a large range of stellar luminosities due to rapid evolution along pre-main sequence tracks. But, spectroscopic binaries do appear to be systematically `younger' on average on the HR diagram compared to the sources not flagged for multiplicity because of the excess flux from the companions (Figure \ref{fig:hr}). Individual binaries, however, are not guaranteed to have higher fluxes compared to the rest of the population, as this is a function of the flux ratio between the secondary and the primary. Similarly, not all of the sources that lie on the binary sequence will be identified as spectroscopic binaries.

A few of the systems we identify as SB2s and RV variable have been previously resolve through high resolution imaging. Most notable of these systems is 2M04132722+2816247 (=V1096 Tau), for which both stars have been detected with radio interferometry \citep{galli2018}. Other systems include 2M04184061+2819155 \citep[=V892 Tau,][]{monnier2008} and
2M04352089+2254242 \citep[=FF Tau,][]{kraus2008}.

\section{Completeness}\label{sec:complete}

\begin{figure}
\epsscale{1.1}
 \centering
		\gridline{\fig{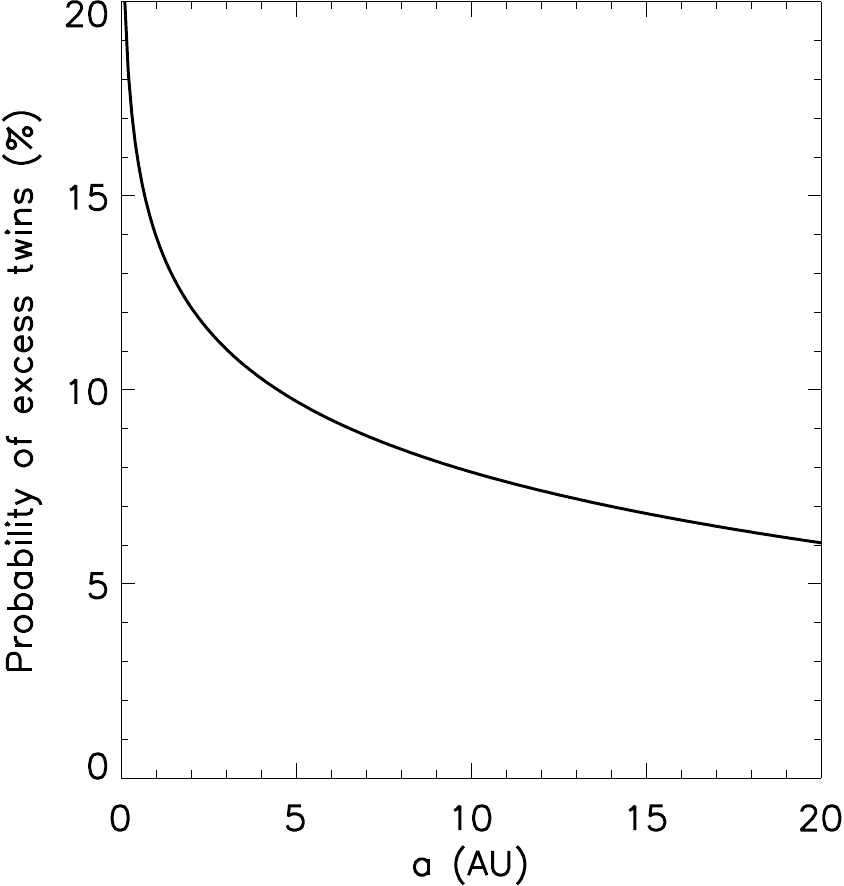}{0.24\textwidth}{}
             \fig{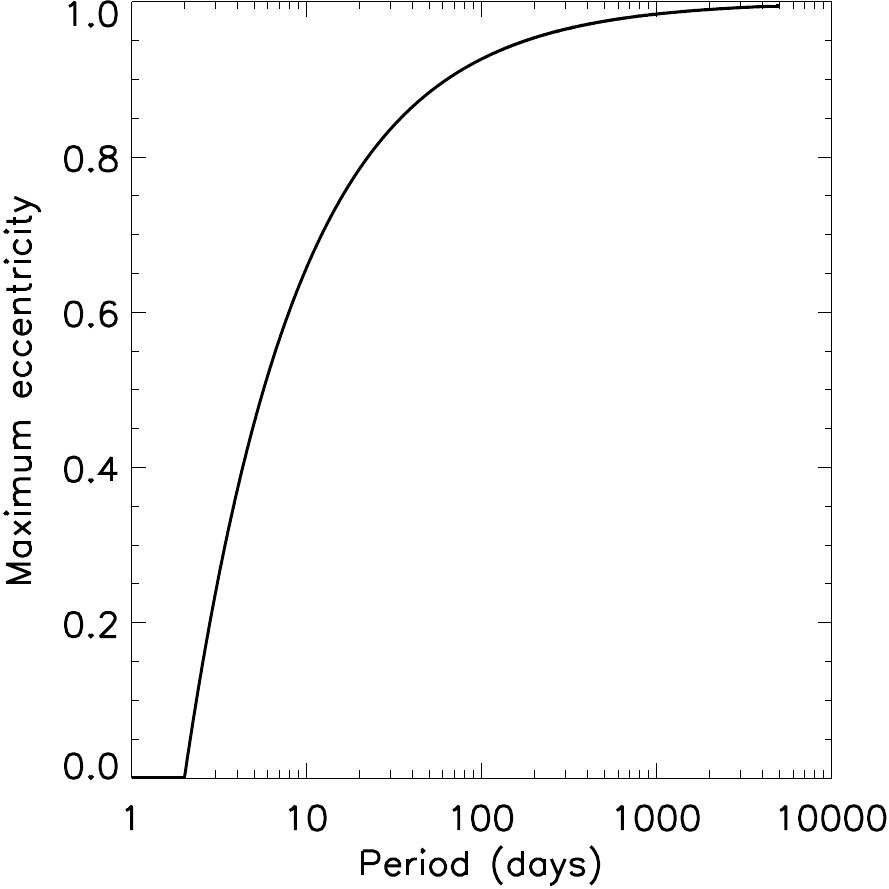}{0.25\textwidth}{}
        }
\caption{Left: Probability of having an excess mass ratio of 0.95$<q<$1.0 (on top of the underlying uniform distribution of $q$) as a function of separation $a$. Right: Maximum eccentricity of a system as a function of orbital period, with the circulization period of 2 days.
\label{fig:orbit}}
\end{figure}

\begin{figure*}
 \centering
		\gridline{\fig{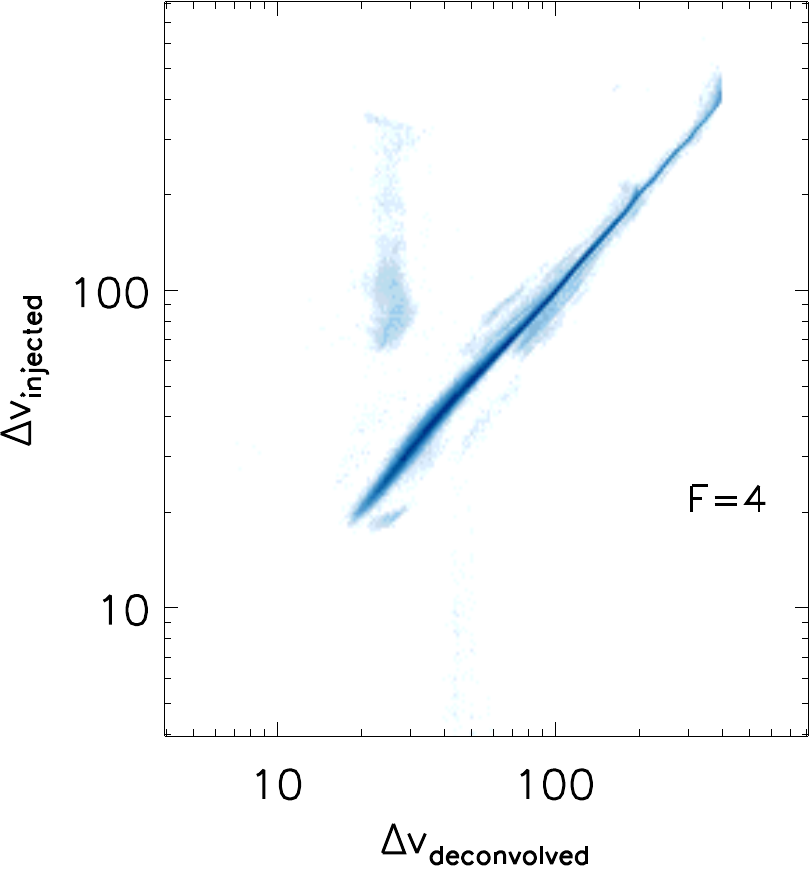}{0.25\textwidth}{}
             \fig{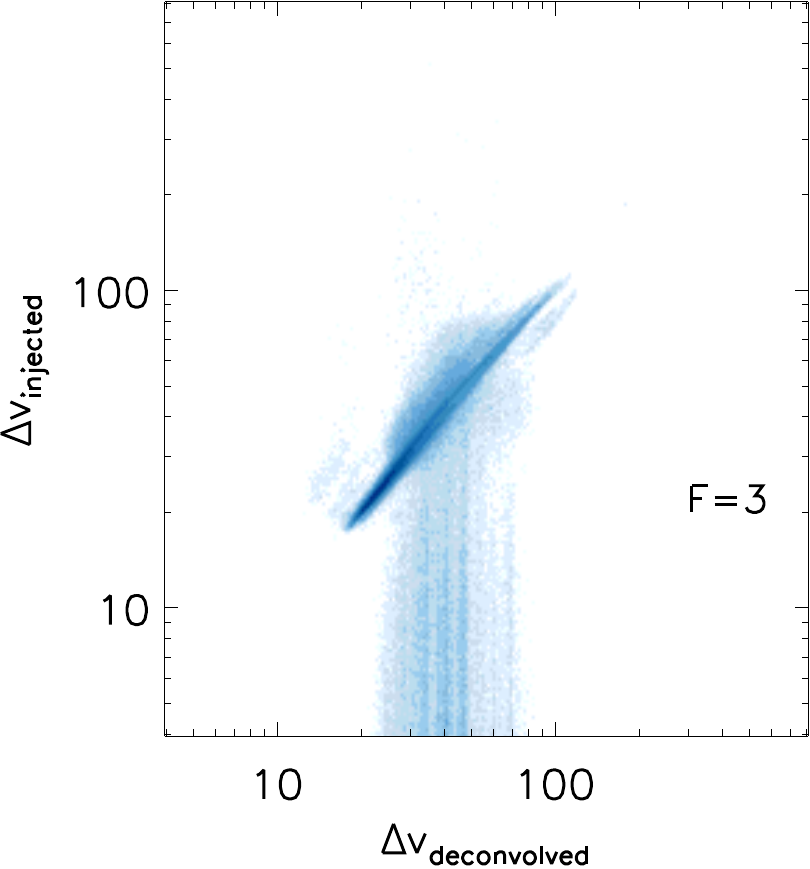}{0.25\textwidth}{}
             \fig{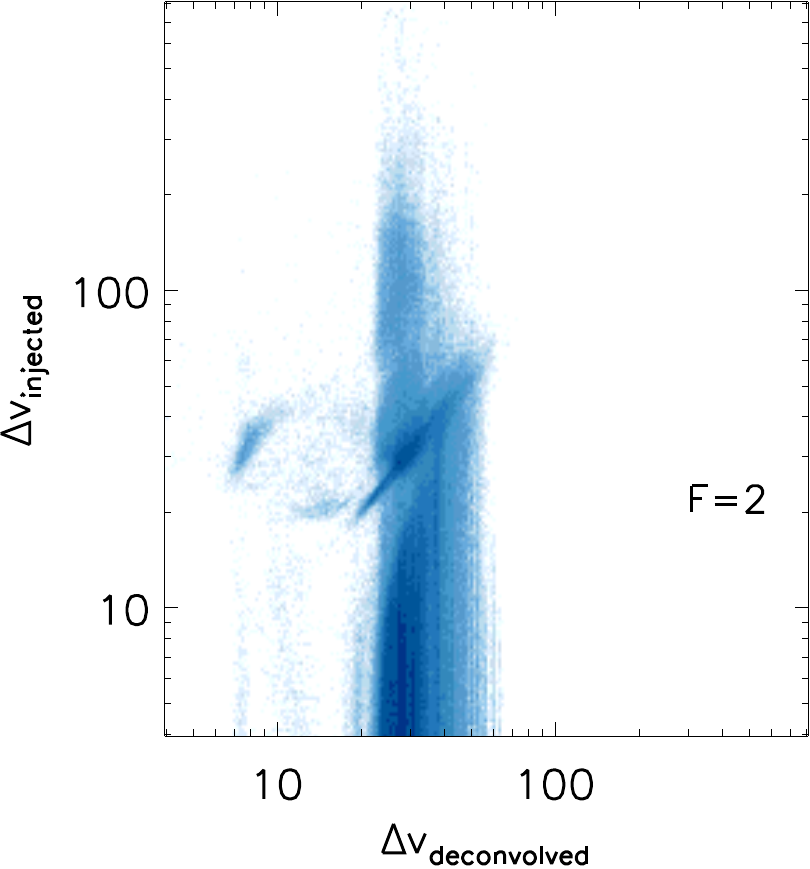}{0.25\textwidth}{}
             \fig{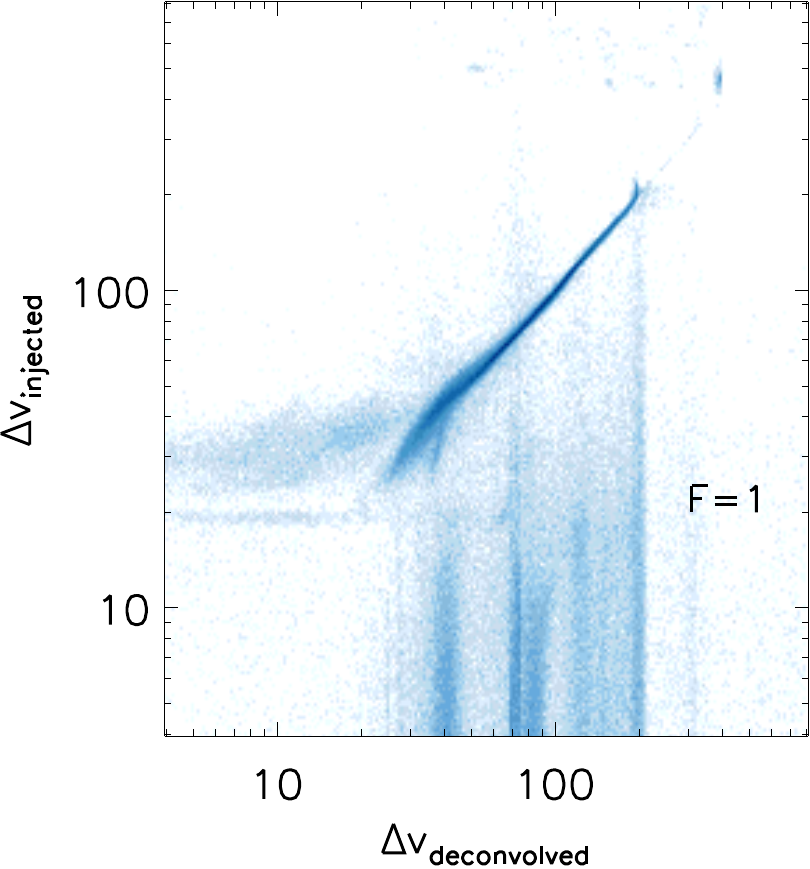}{0.25\textwidth}{}
        }
                \vspace{-0.8cm}
		\gridline{\fig{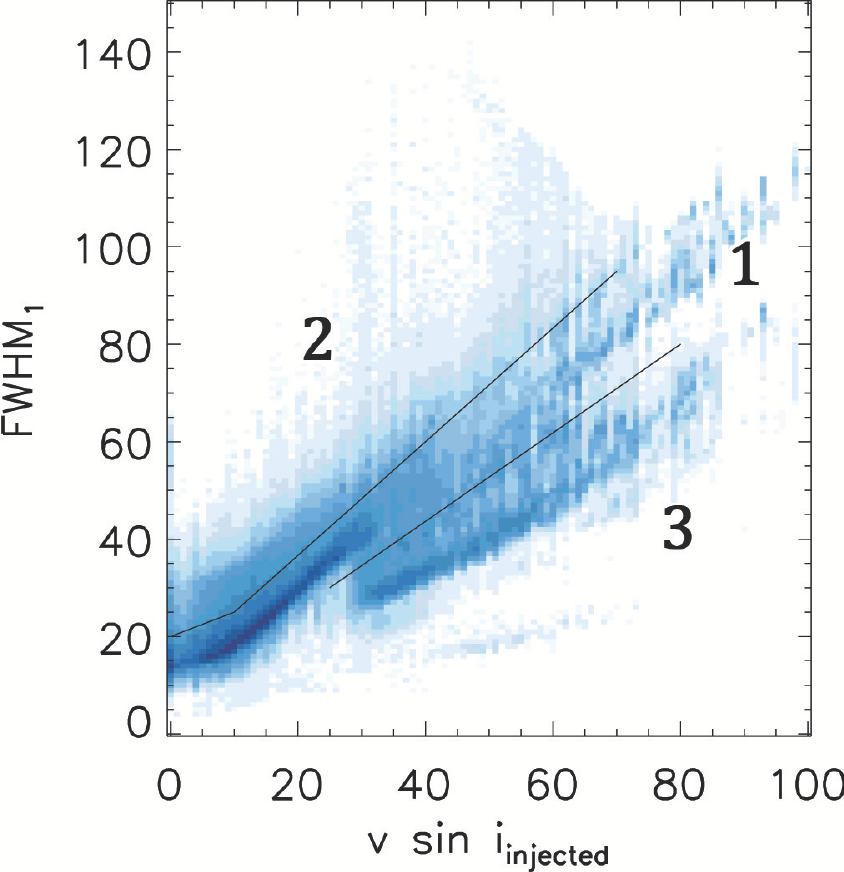}{0.25\textwidth}{}
             \fig{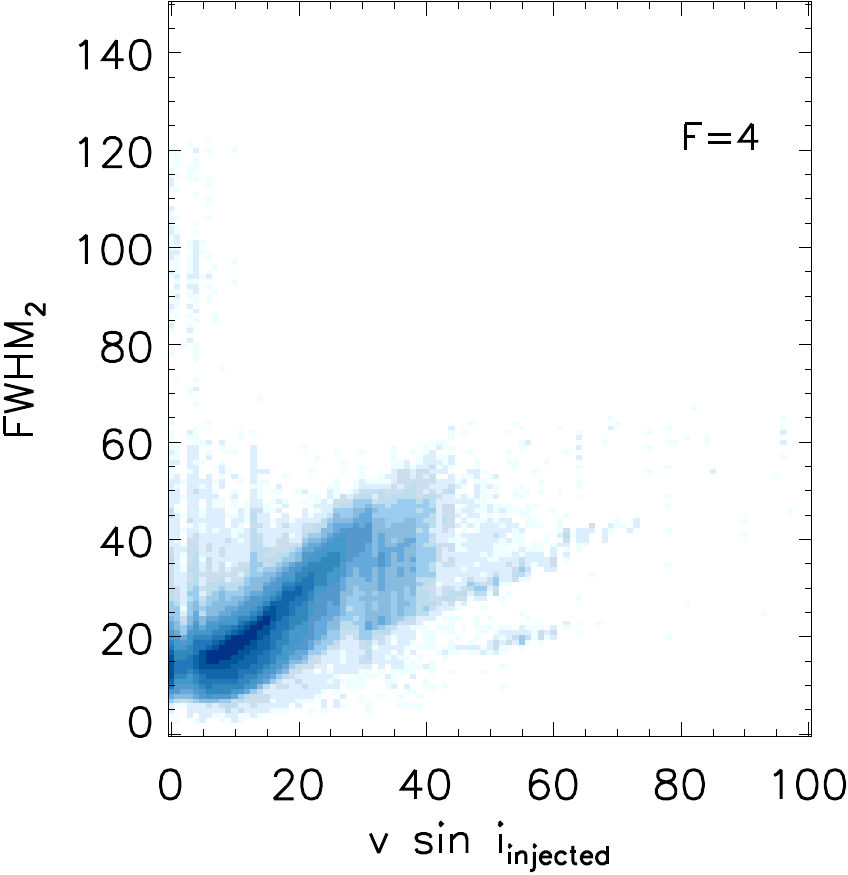}{0.25\textwidth}{}
             \fig{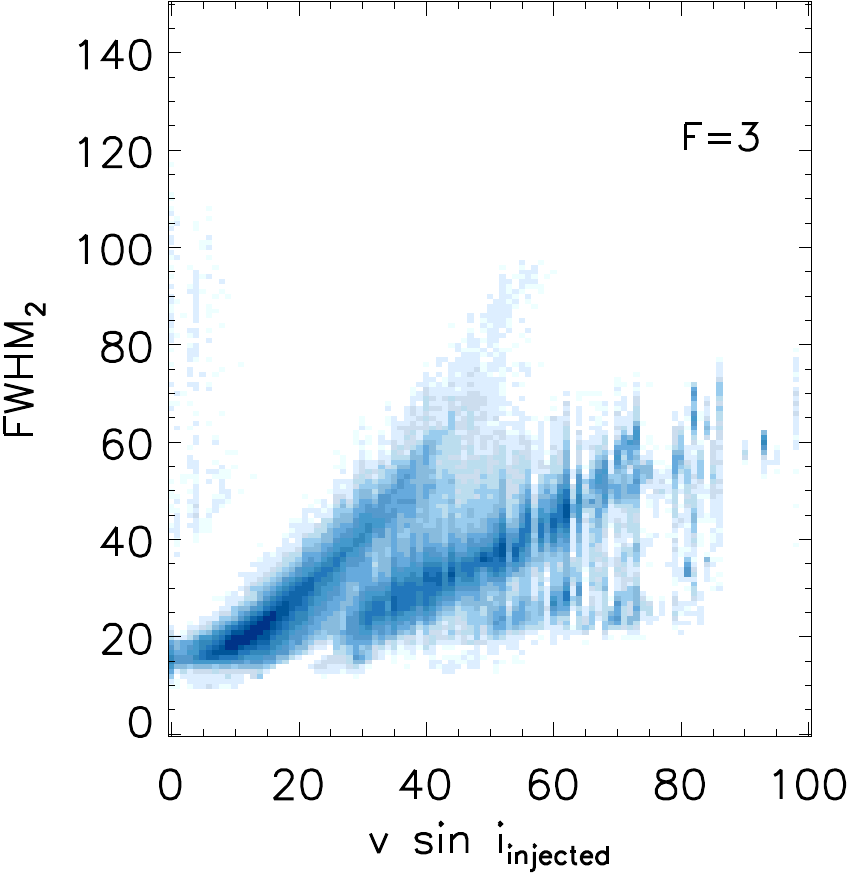}{0.25\textwidth}{}
        }
        \vspace{-0.8cm}
      \caption{Comparison of the injected properties of the synthetic binaries vs. the recovered parameters. Color corresponds to the logarithmic density of the sources. Top: velocity difference between the primary and the secondary; each panel corresponds to the flag assigned to the secondary recovered in the deconvolution. Bottom: \vsini\ vs. FWHM for the primary (regardless of whether companion was recovered or not), companions recovered with flag 4, companions recovered with flag 3. The black line shows the separation between regions where the recovered FWHM corresponds to the original \vsini (1), where unresolved binaries (typically with periods within a few years) inflate FWHM over \vsini (2), and where a symmetric peak is falsely deconvolved into two (3). \label{fig:synt}}
\end{figure*}

\begin{figure*}
 \centering
		\gridline{\fig{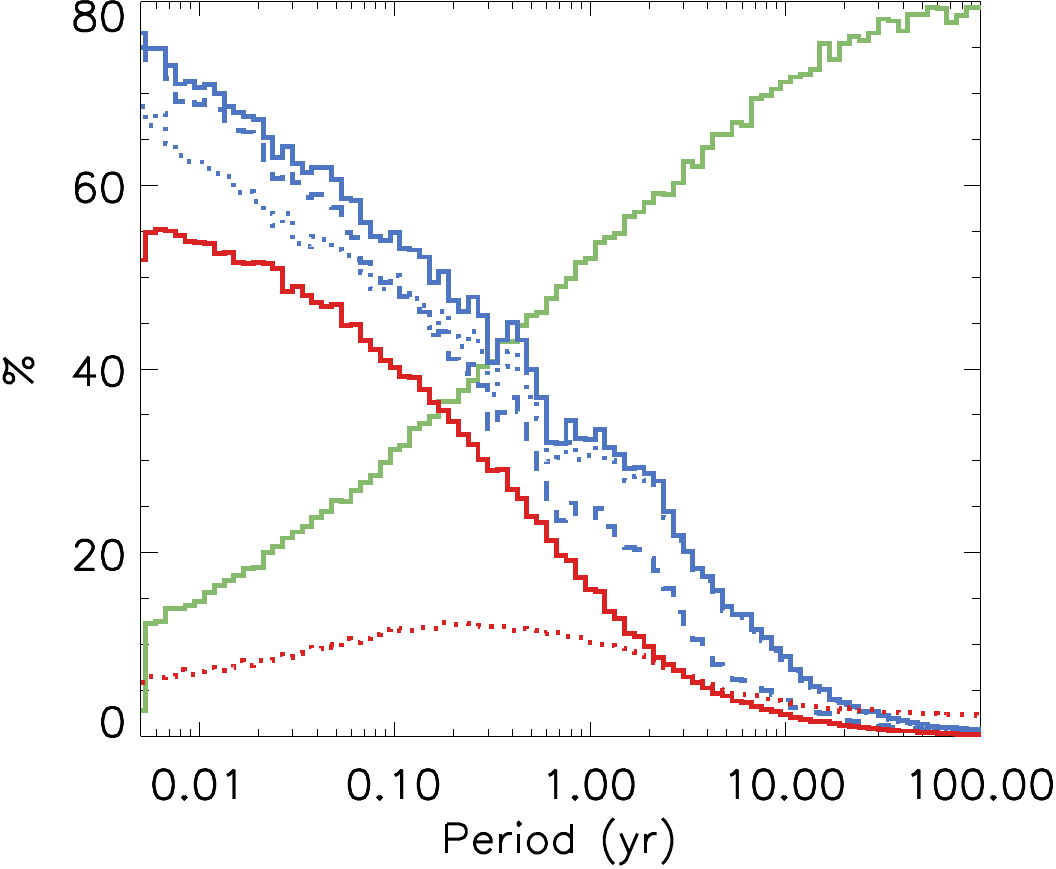}{0.25\textwidth}{}
             \fig{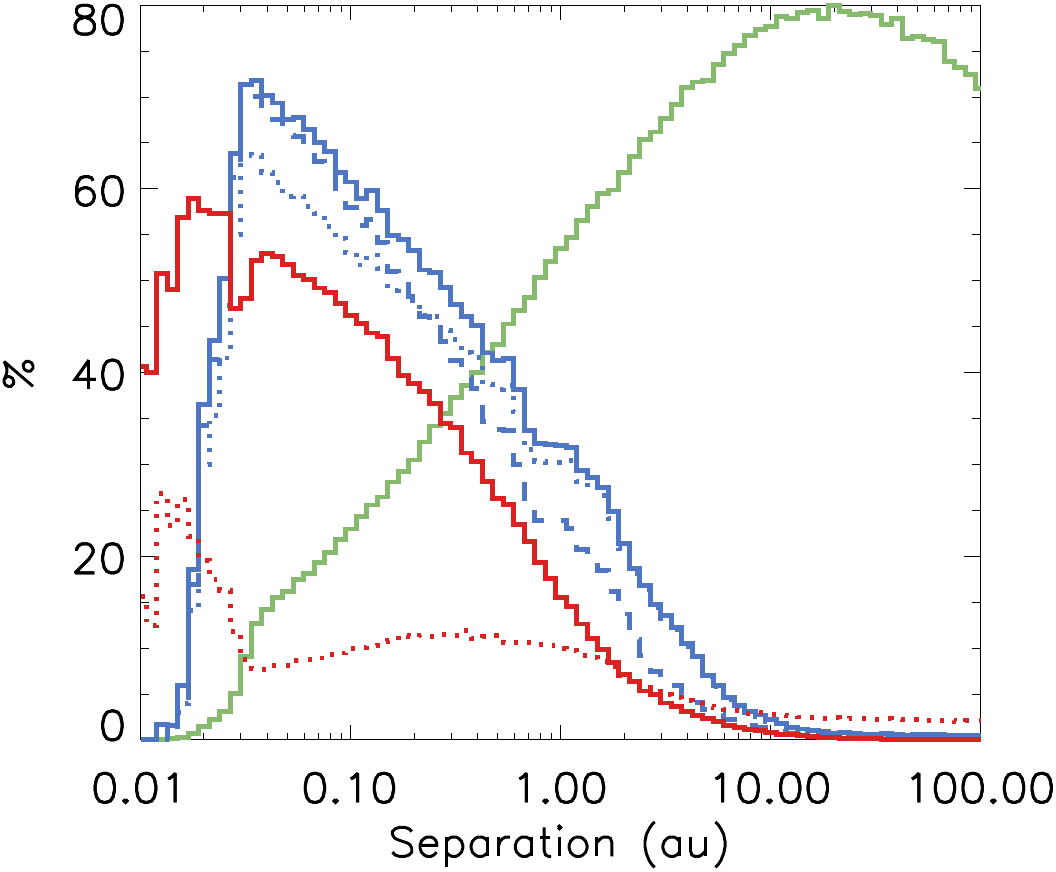}{0.25\textwidth}{}
             \fig{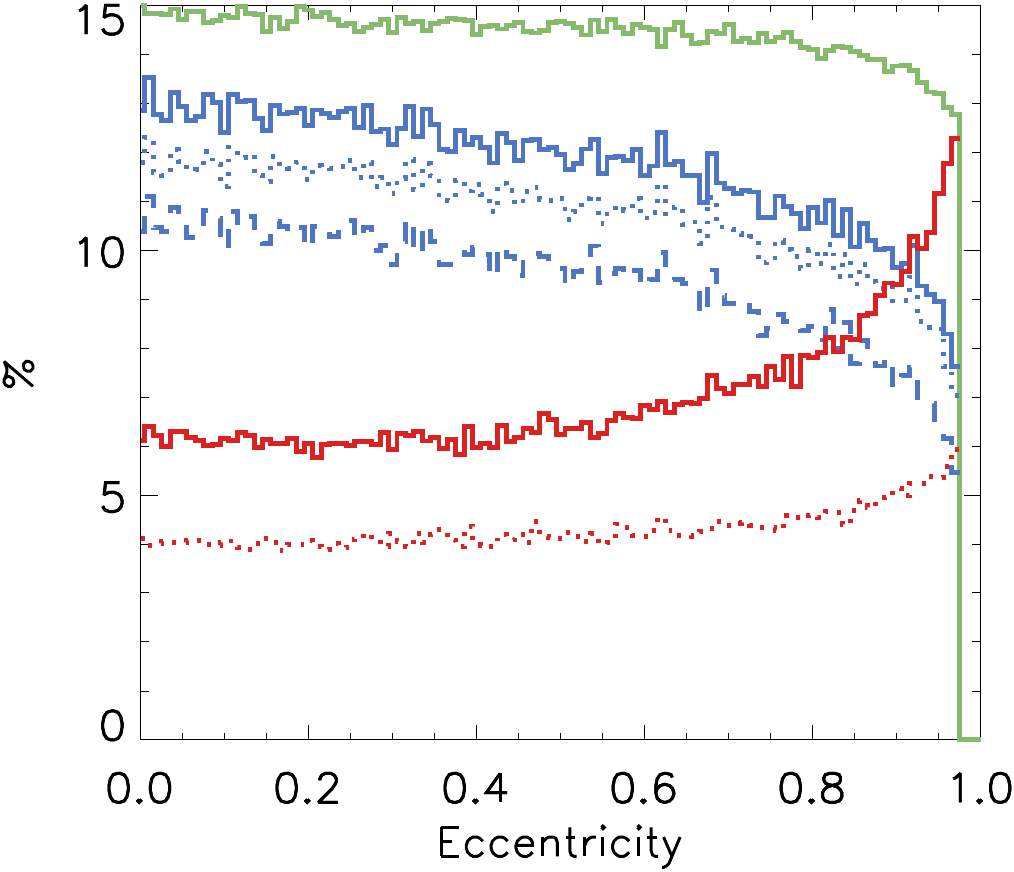}{0.25\textwidth}{}
             \fig{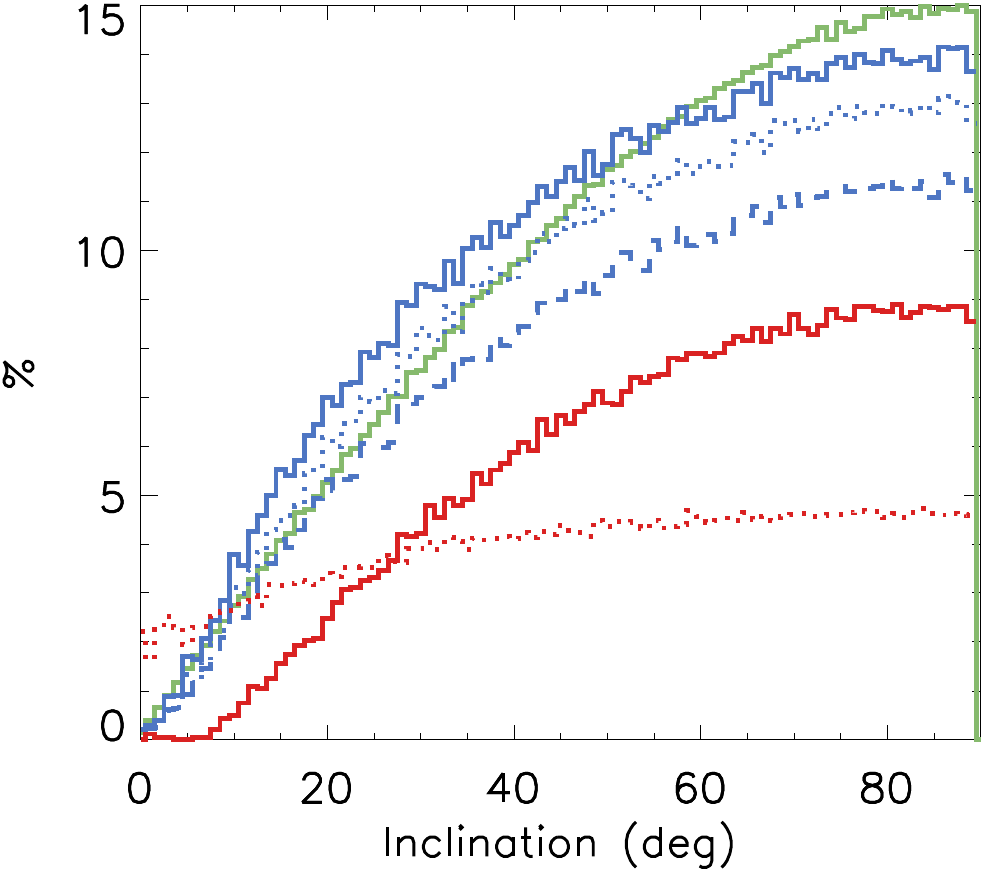}{0.25\textwidth}{}
        }
                \vspace{-0.8cm}
		\gridline{\fig{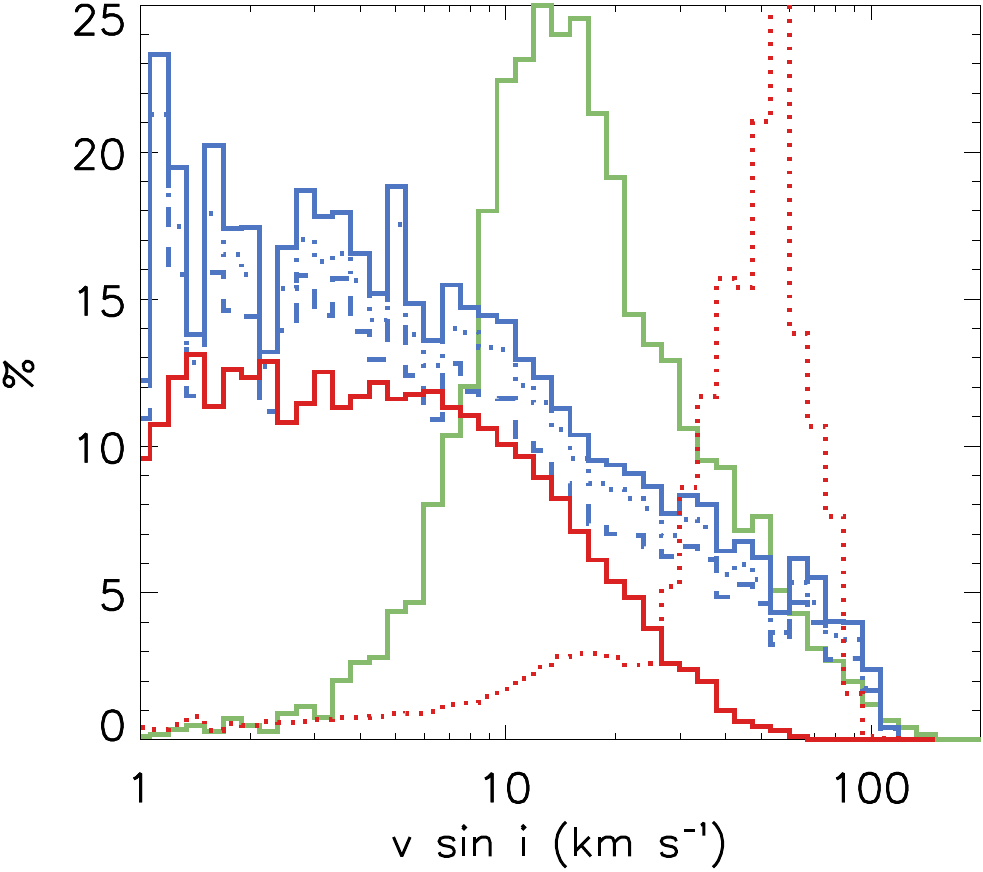}{0.25\textwidth}{}
             \fig{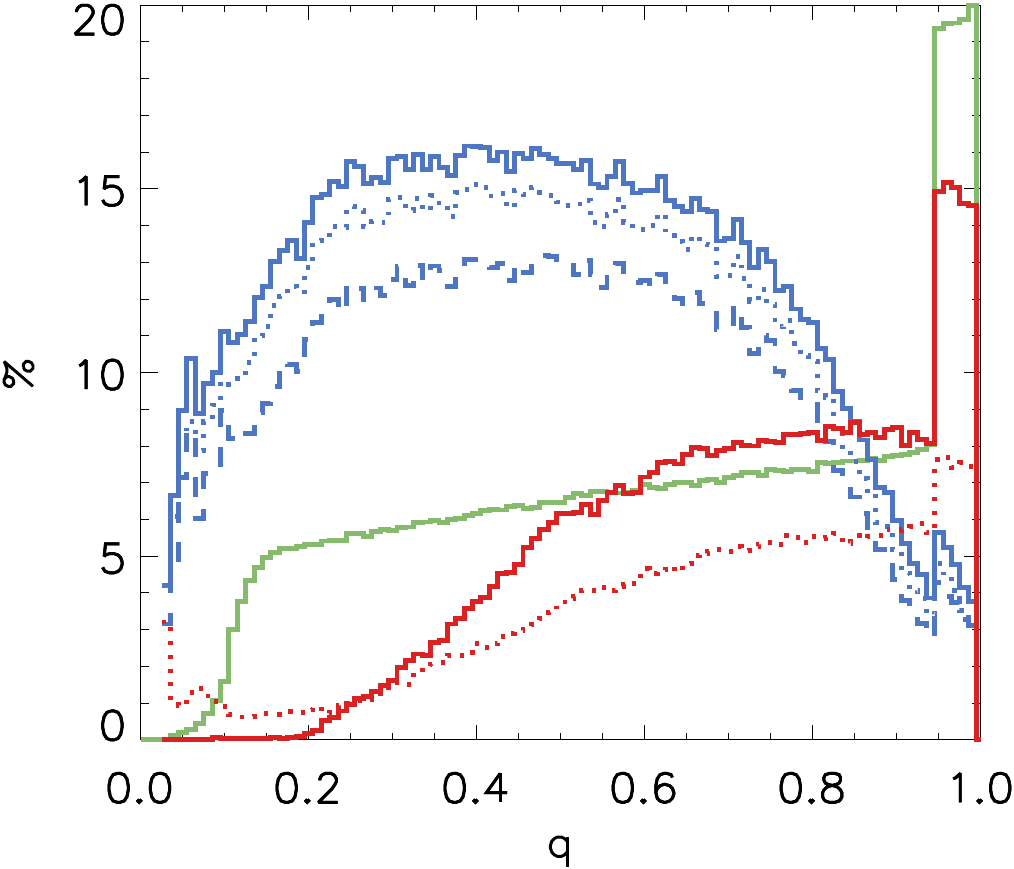}{0.25\textwidth}{}
             \fig{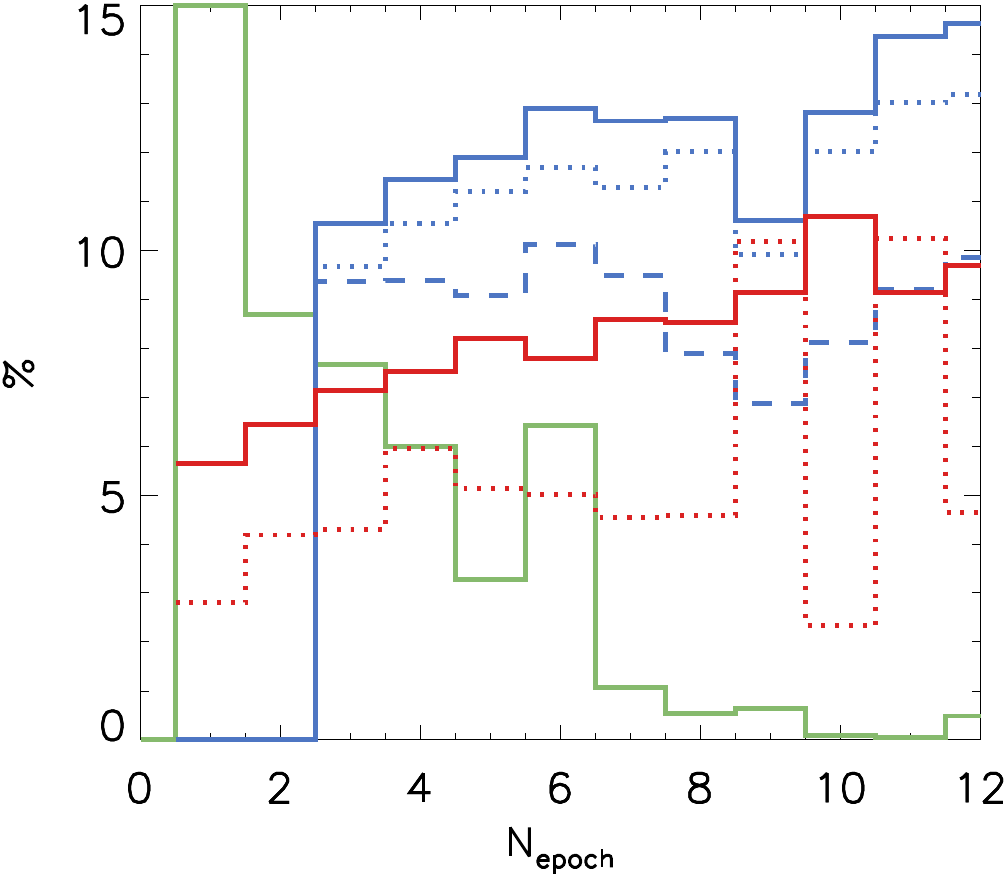}{0.25\textwidth}{}
             \fig{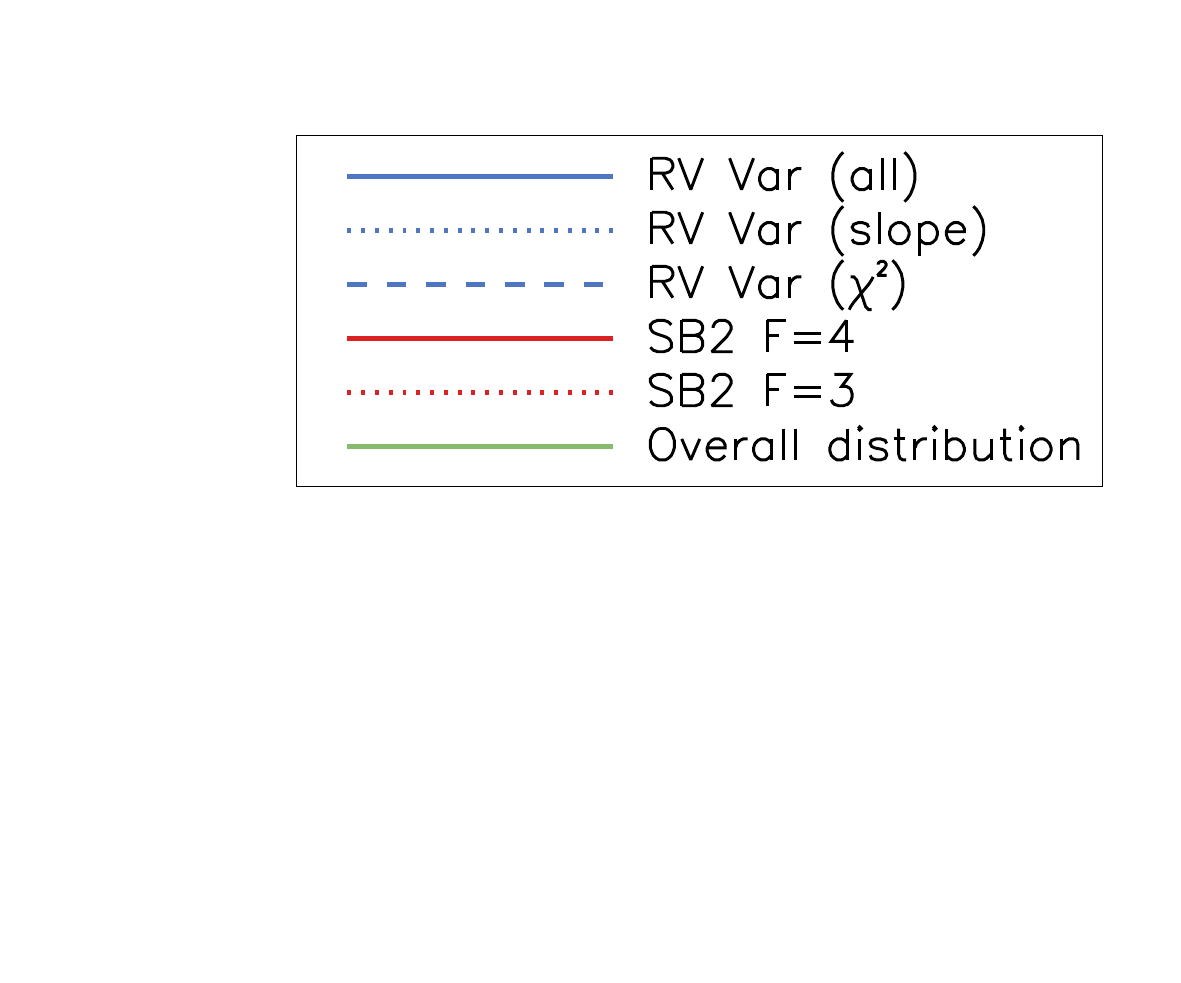}{0.25\textwidth}{}
        }
        \vspace{-0.8cm}
      \caption{Fraction of the identified binaries out of all multiple systems in the synthetic sample marginalized over a single parameter (completeness). Blue line shows the completeness of RV variables, restricted only to the sources with 3 or more epochs. Red line shows the completeness of SB2s. Green line is the histogram (not a fraction) of the underlying distribution of the parameter in the synthetic sample, arbitrarily scaled to fit in each panel. The total distribution of the observed parameters in the identified systems is the completeness multiplied by the green line. \label{fig:detection}}
\end{figure*}

\begin{figure}
 \centering
		\gridline{\fig{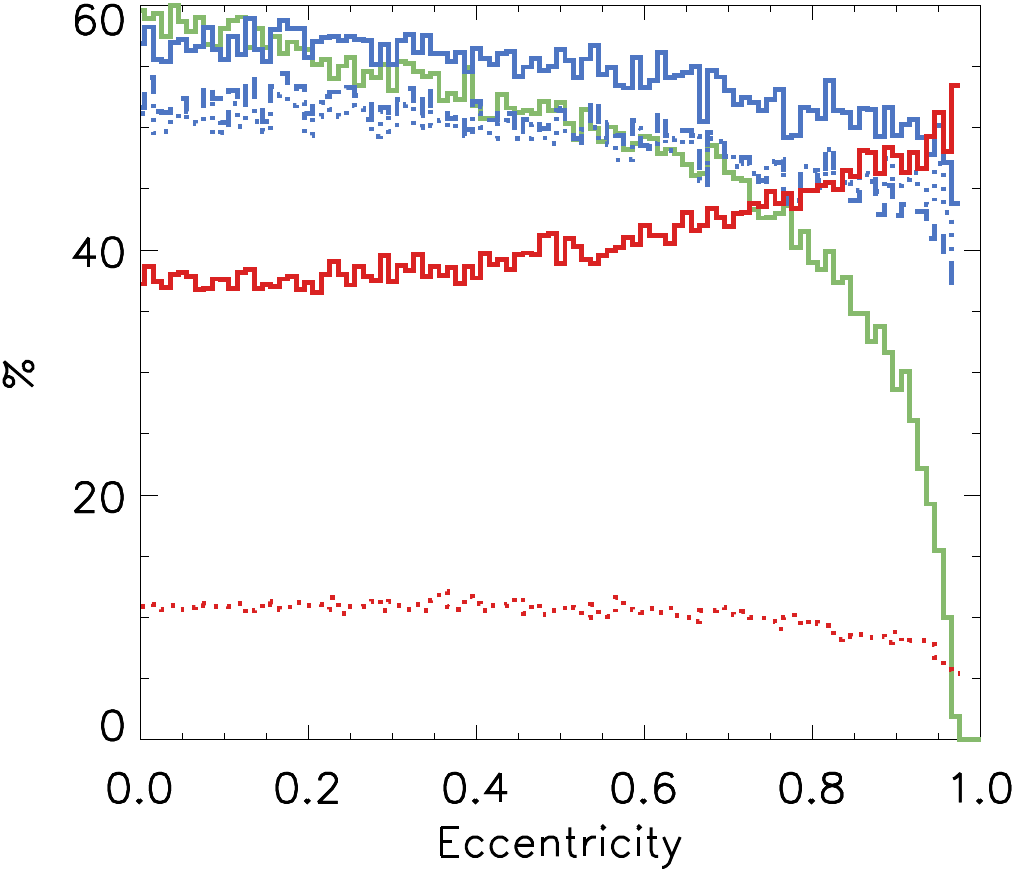}{0.25\textwidth}{}
             \fig{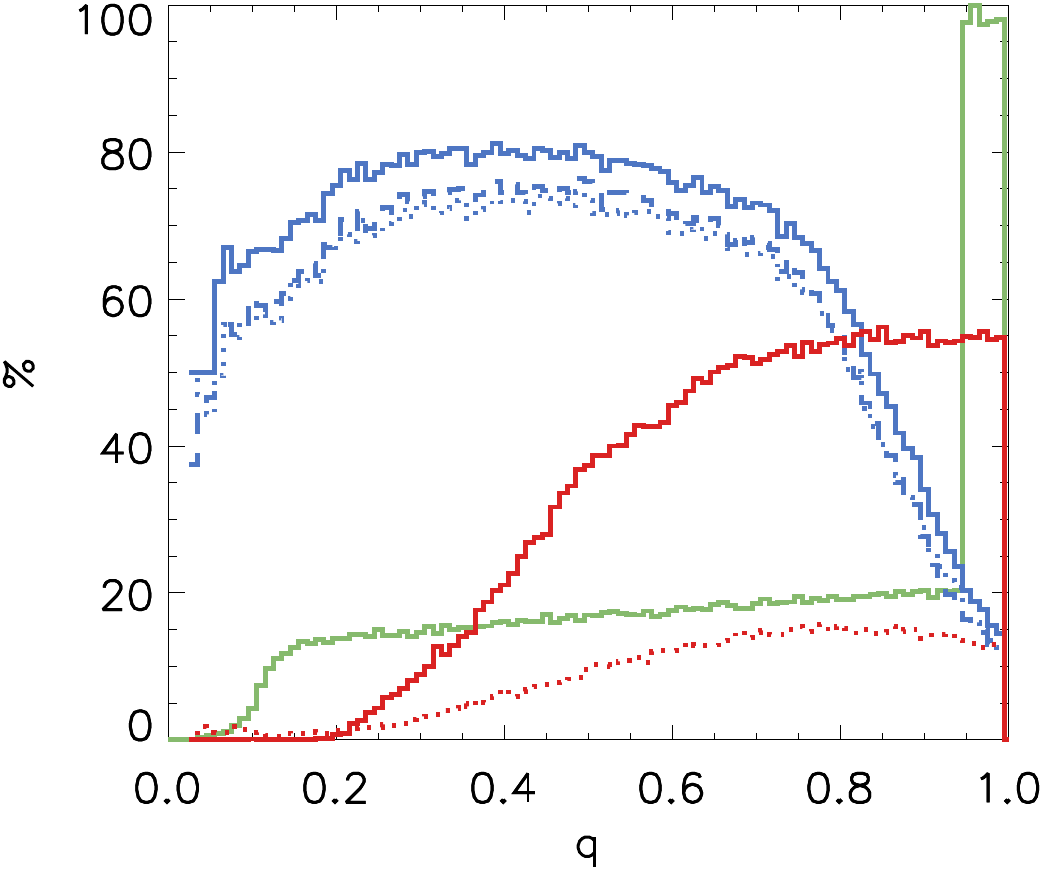}{0.25\textwidth}{}
        }
        \vspace{-0.8cm}
      \caption{Completeness of the recovered binaries (Same as Figure \ref{fig:detection}), but restricted to the systems with orbital periods $<$200 days. Blue line shows the fraction of the recovered of RV variables, restricted to only the sources with 3 or more epochs. Red line shows the completeness of SB2s. \label{fig:detection2}}
\end{figure}

\begin{figure*}
 \centering
		\gridline{\fig{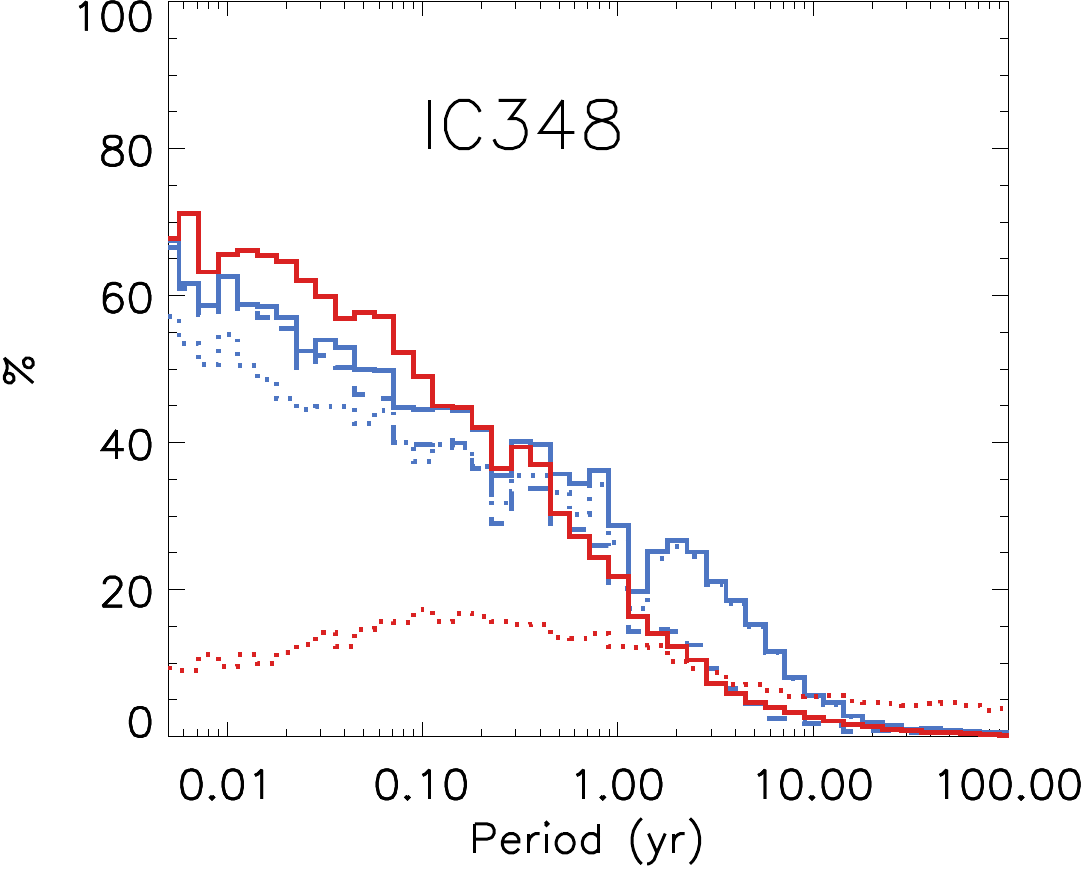}{0.25\textwidth}{}
                  \fig{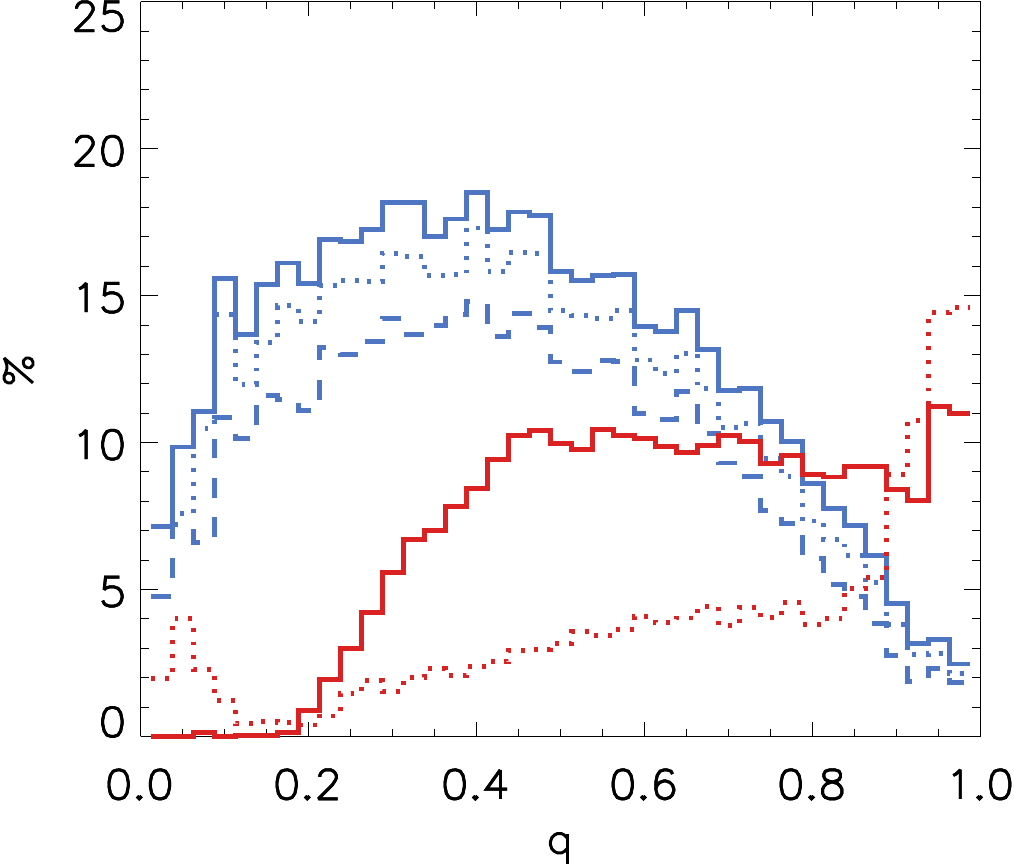}{0.235\textwidth}{}
                  \fig{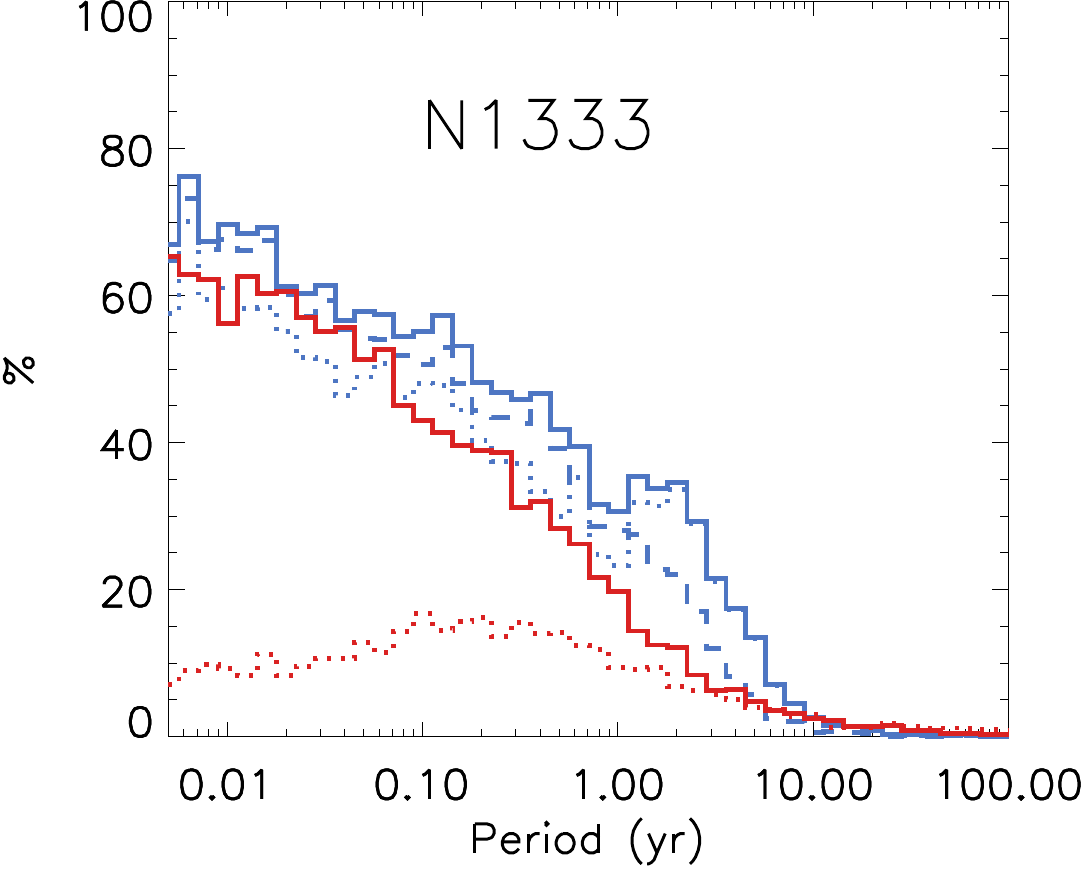}{0.25\textwidth}{}
                  \fig{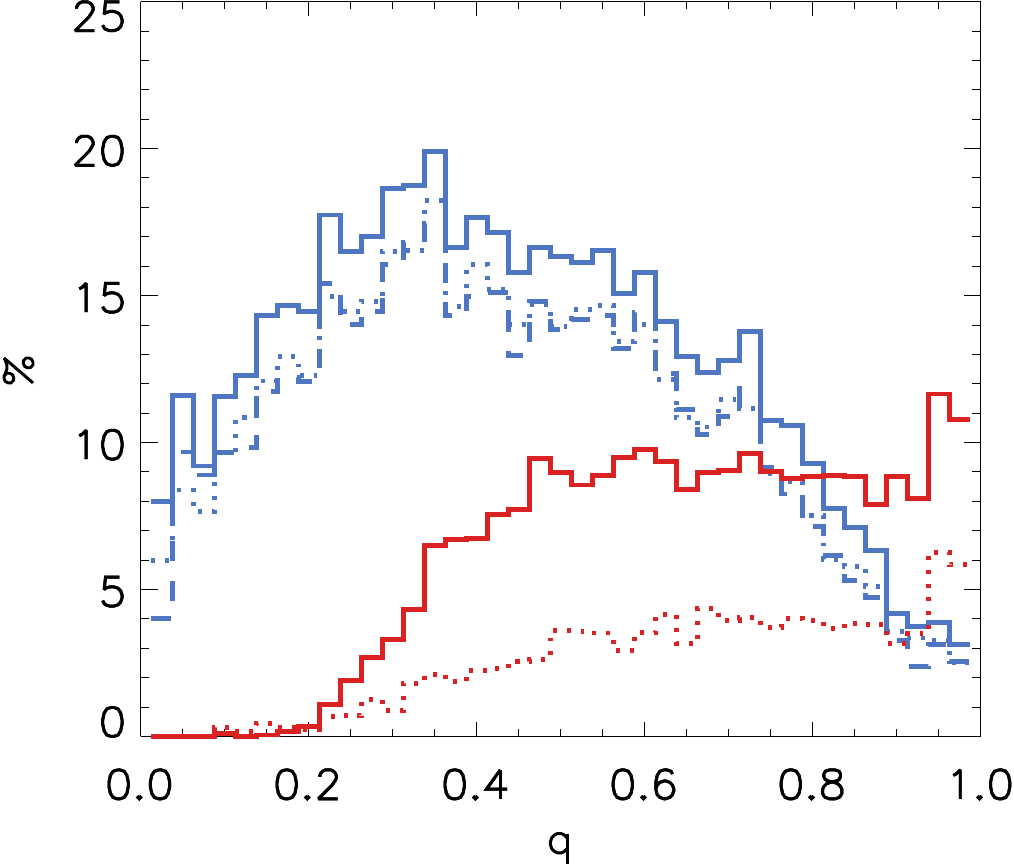}{0.235\textwidth}{}
        }
        \vspace{-0.9cm} 
		\gridline{\fig{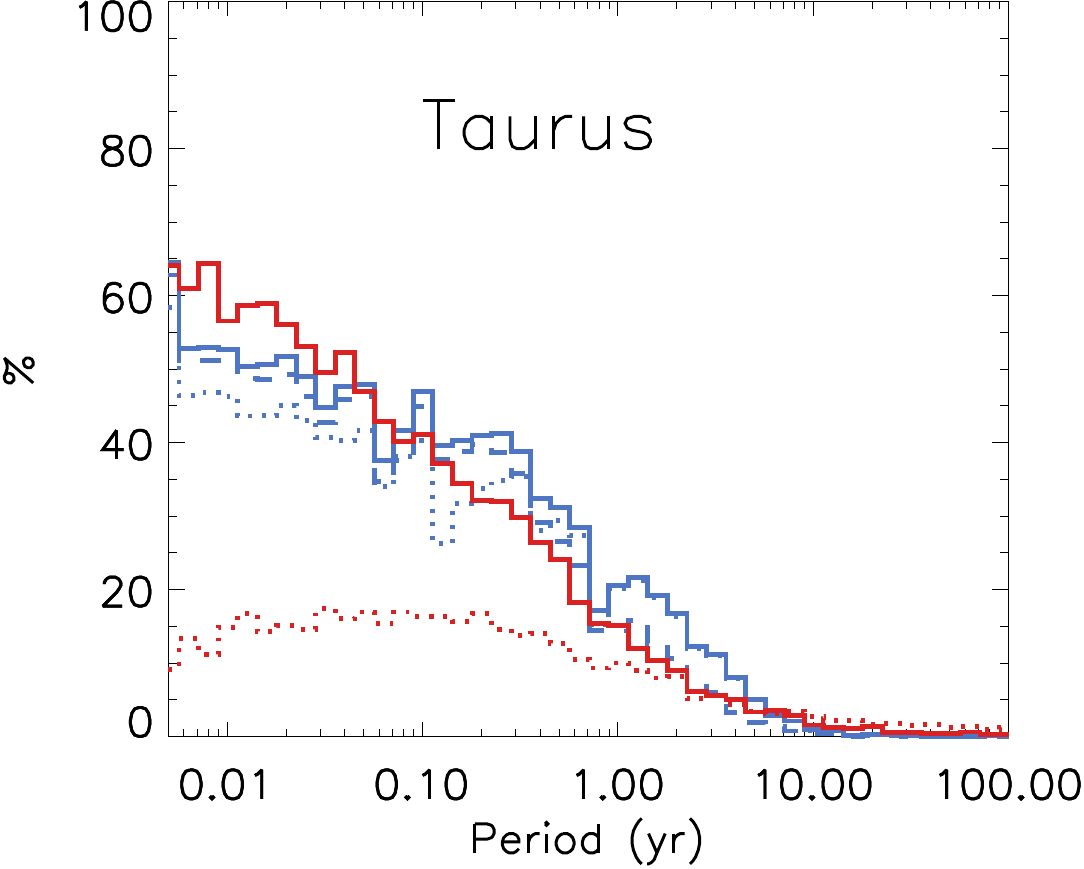}{0.25\textwidth}{}
                  \fig{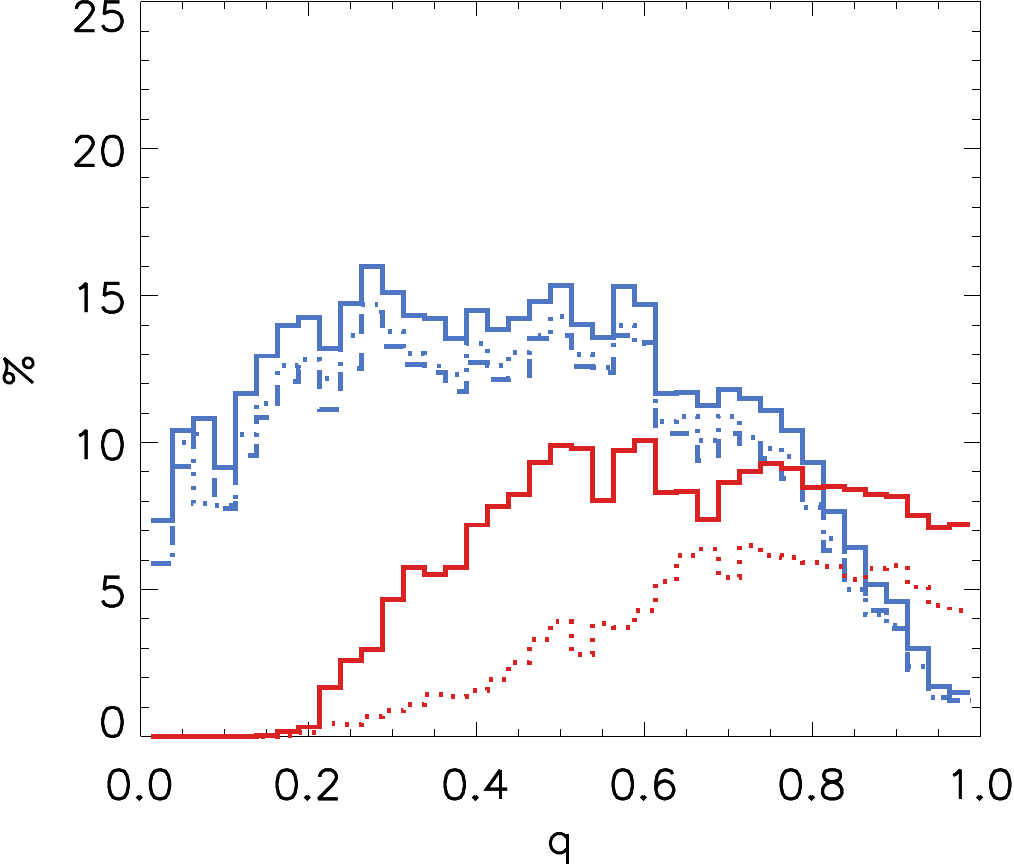}{0.235\textwidth}{}
                  \fig{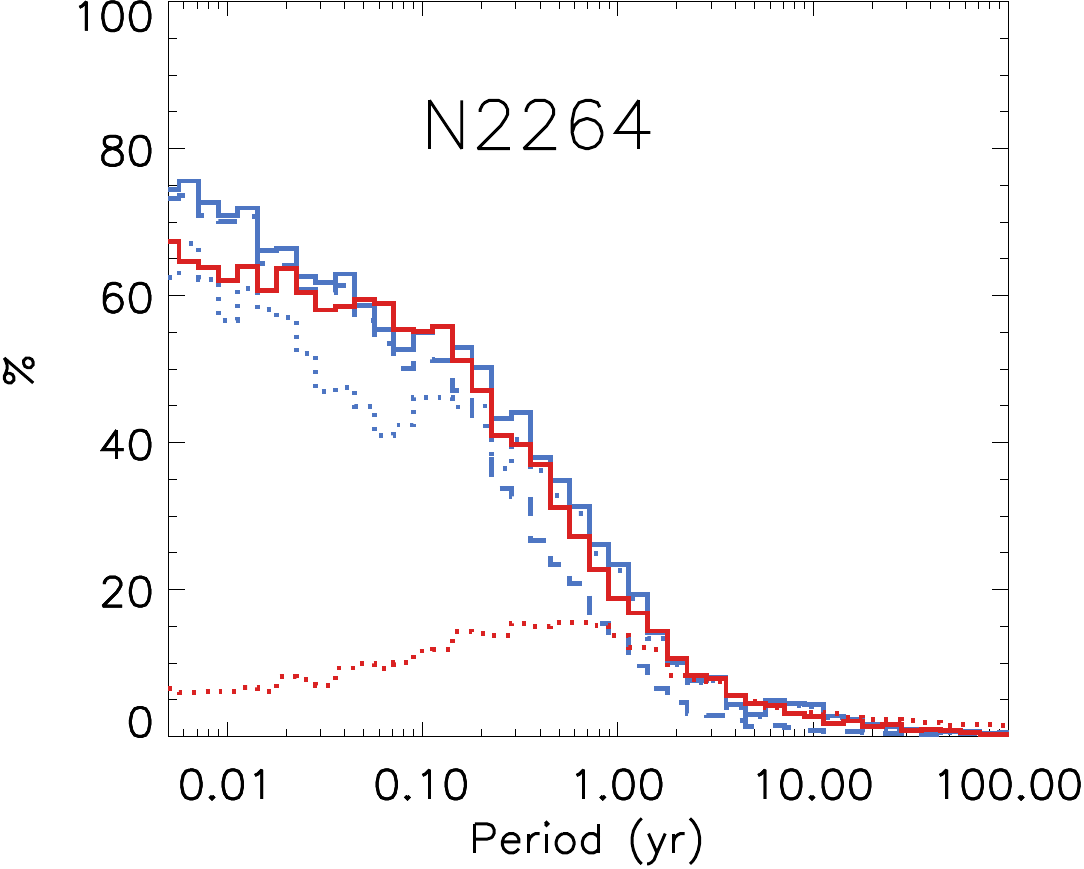}{0.25\textwidth}{}
                  \fig{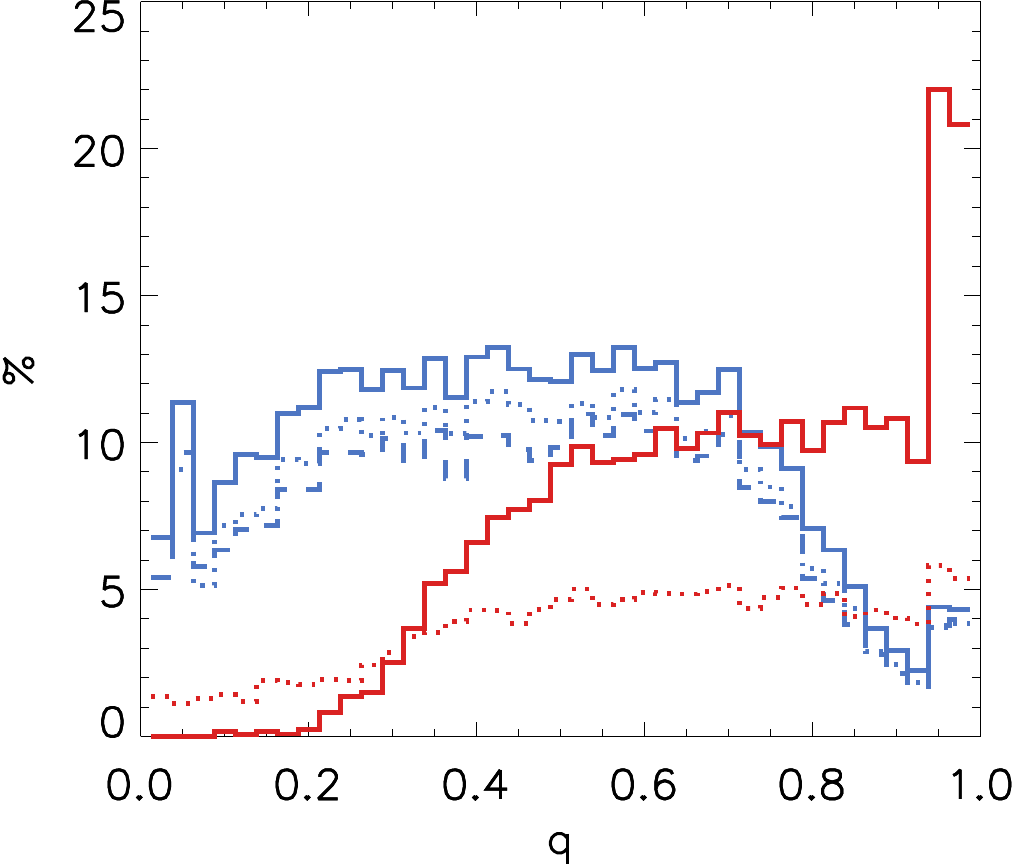}{0.235\textwidth}{}
        }
        \vspace{-0.9cm} 
		\gridline{\fig{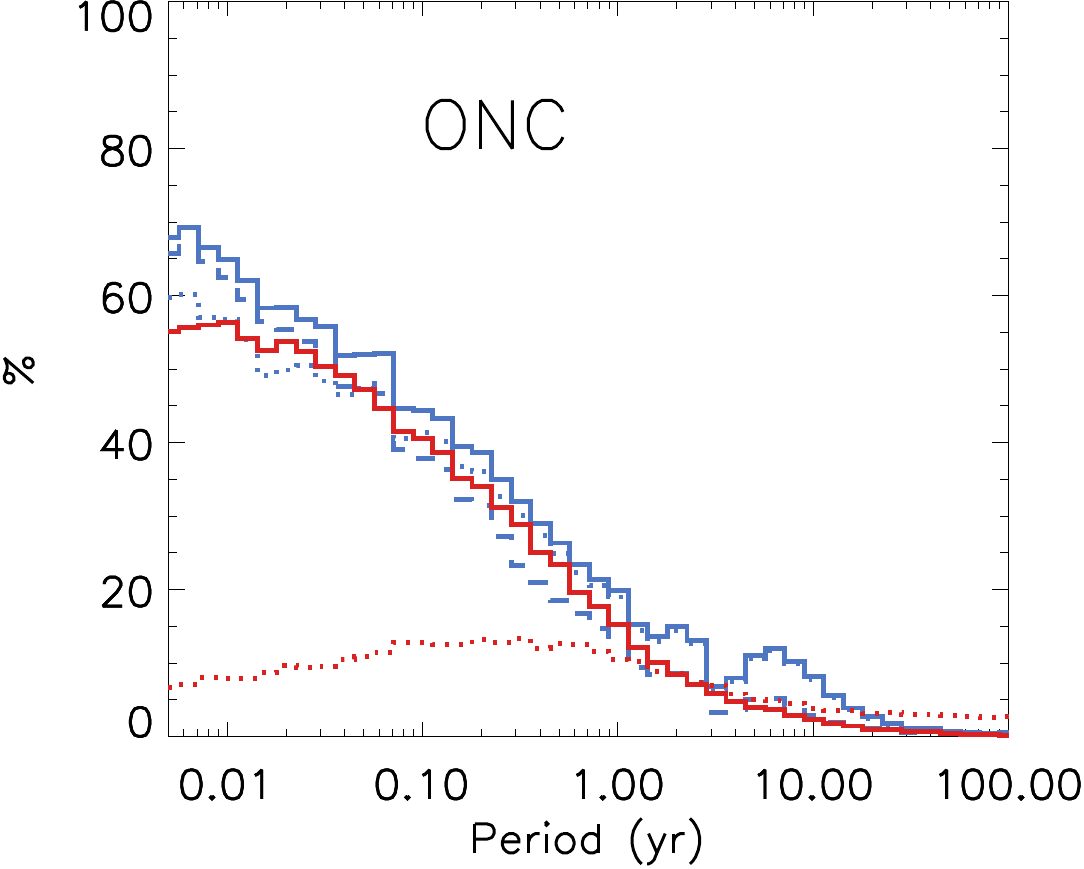}{0.25\textwidth}{}
                  \fig{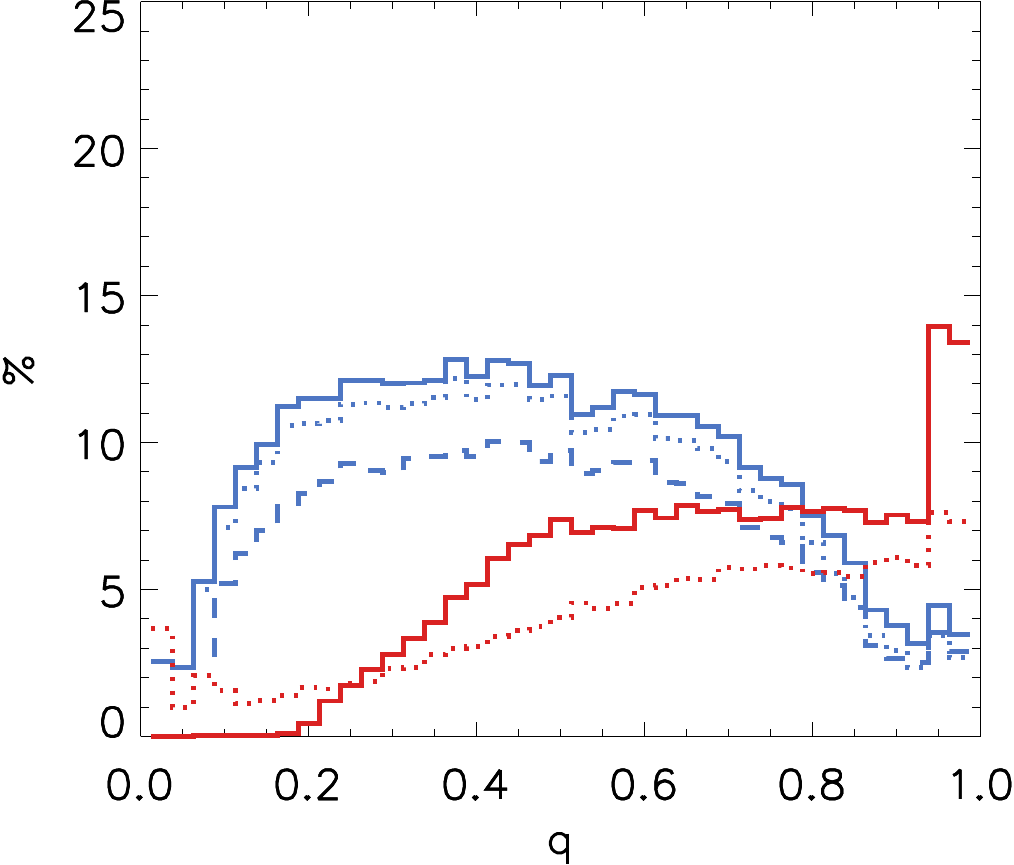}{0.235\textwidth}{}
                  \fig{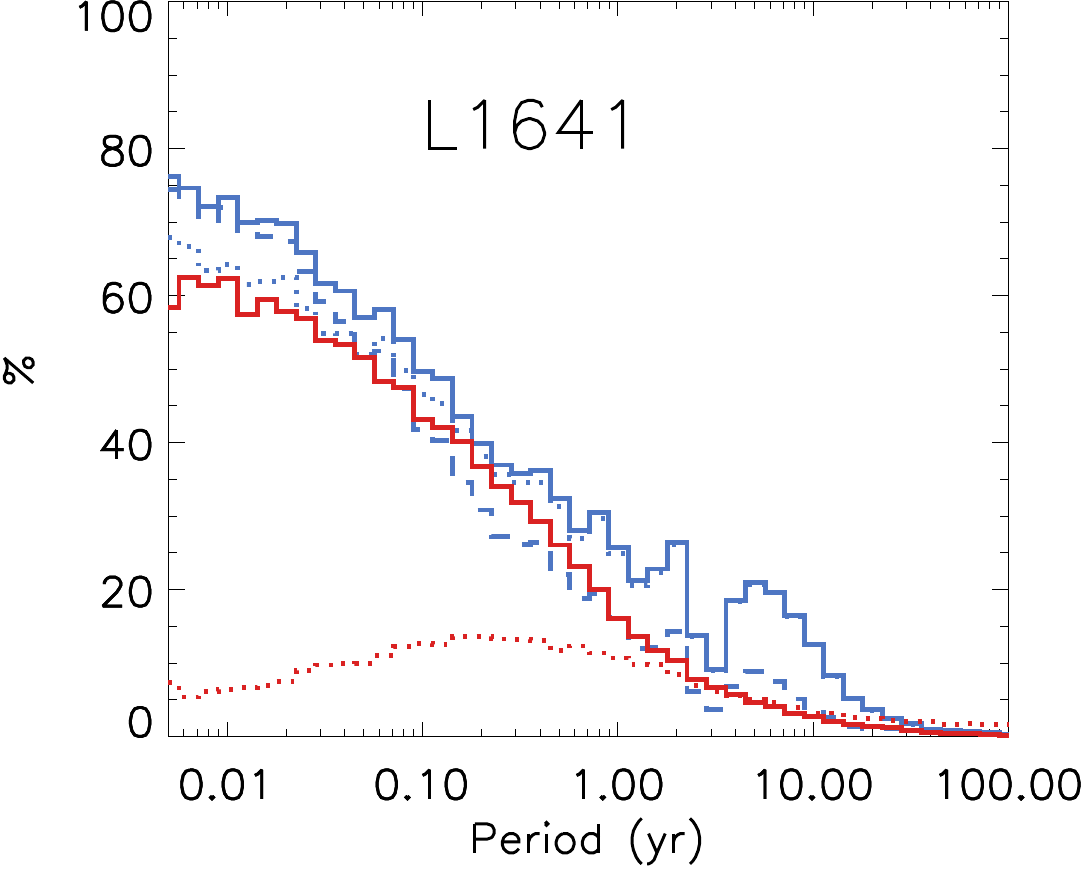}{0.25\textwidth}{}
                  \fig{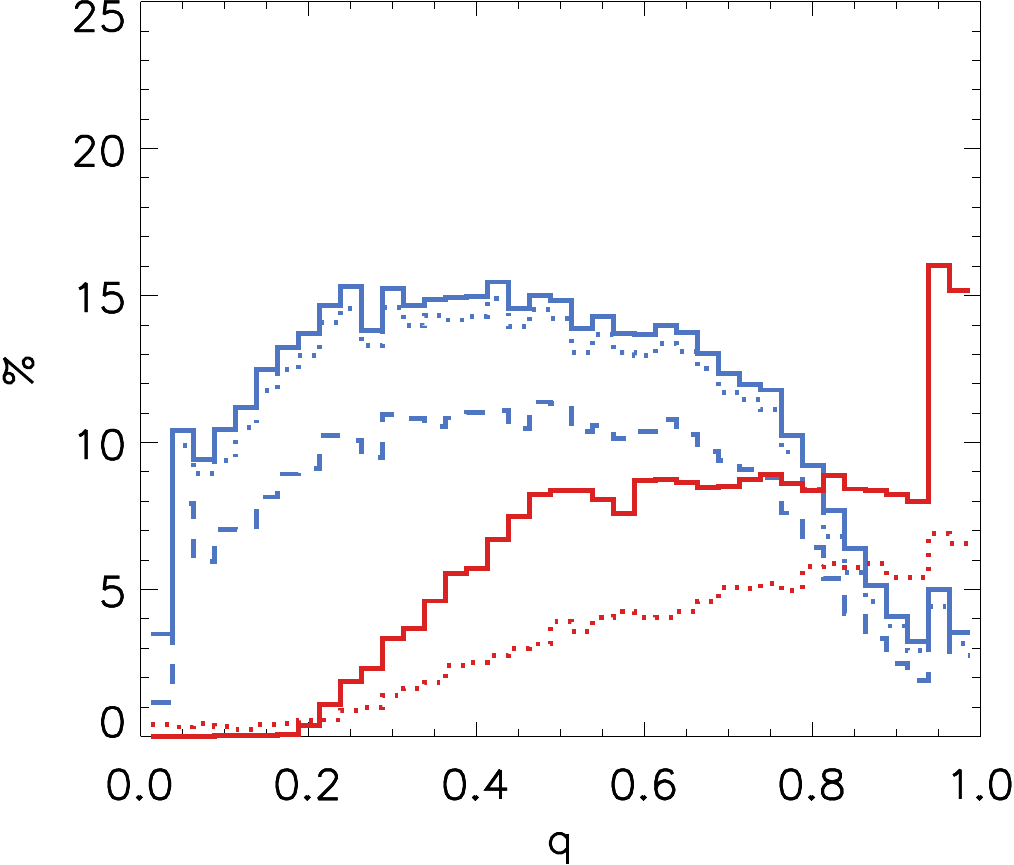}{0.235\textwidth}{}
        }
        \vspace{-0.9cm} 
		\gridline{\fig{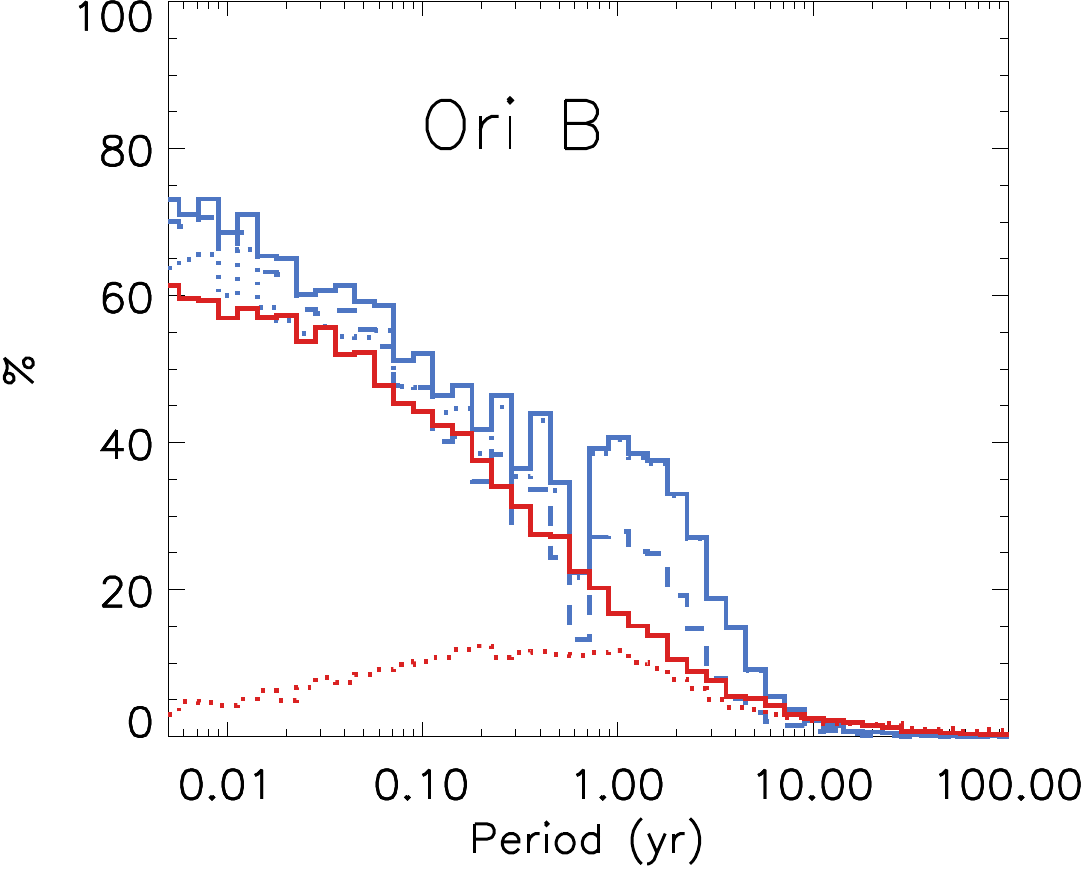}{0.25\textwidth}{}
                  \fig{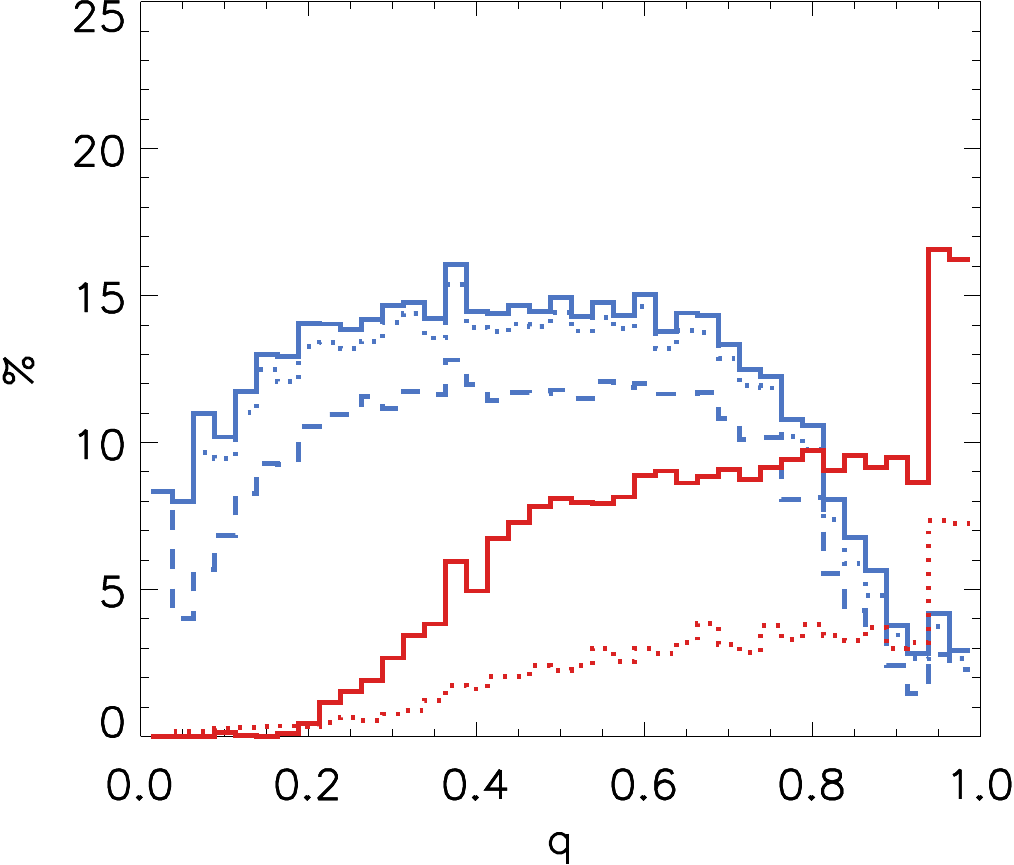}{0.235\textwidth}{}
                  \fig{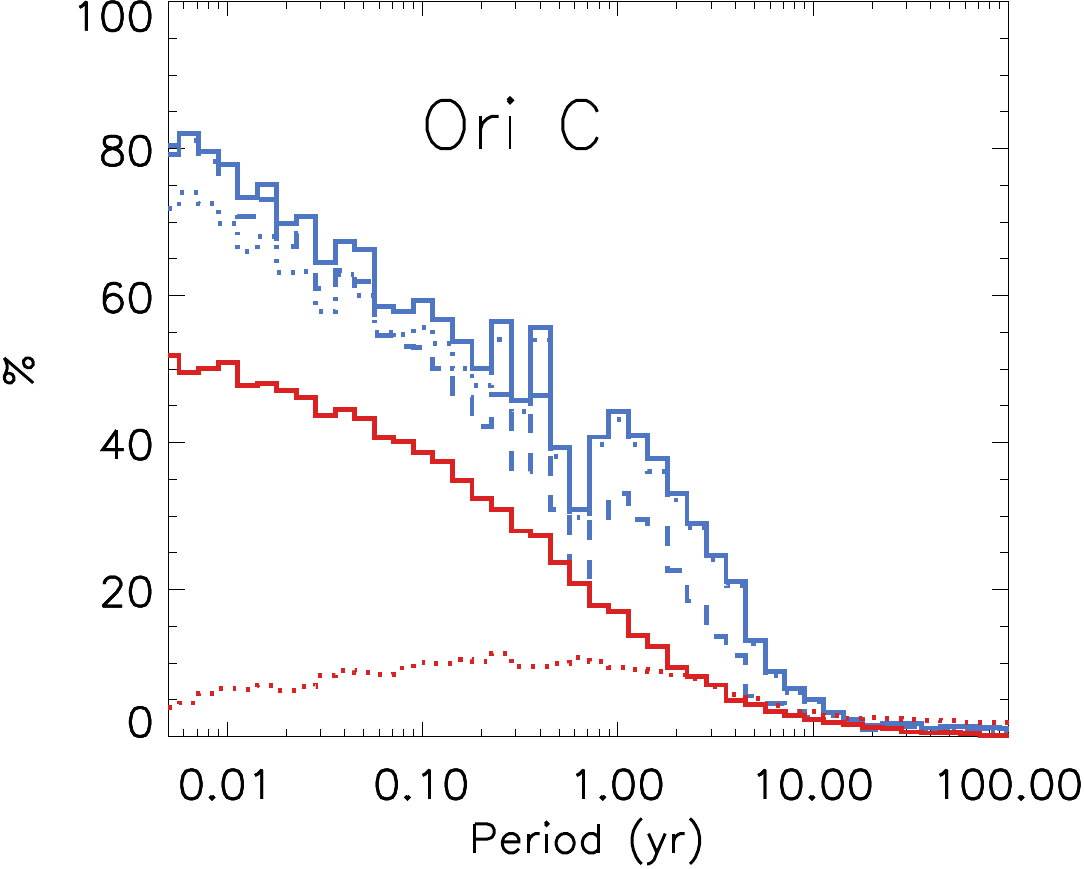}{0.25\textwidth}{}
                  \fig{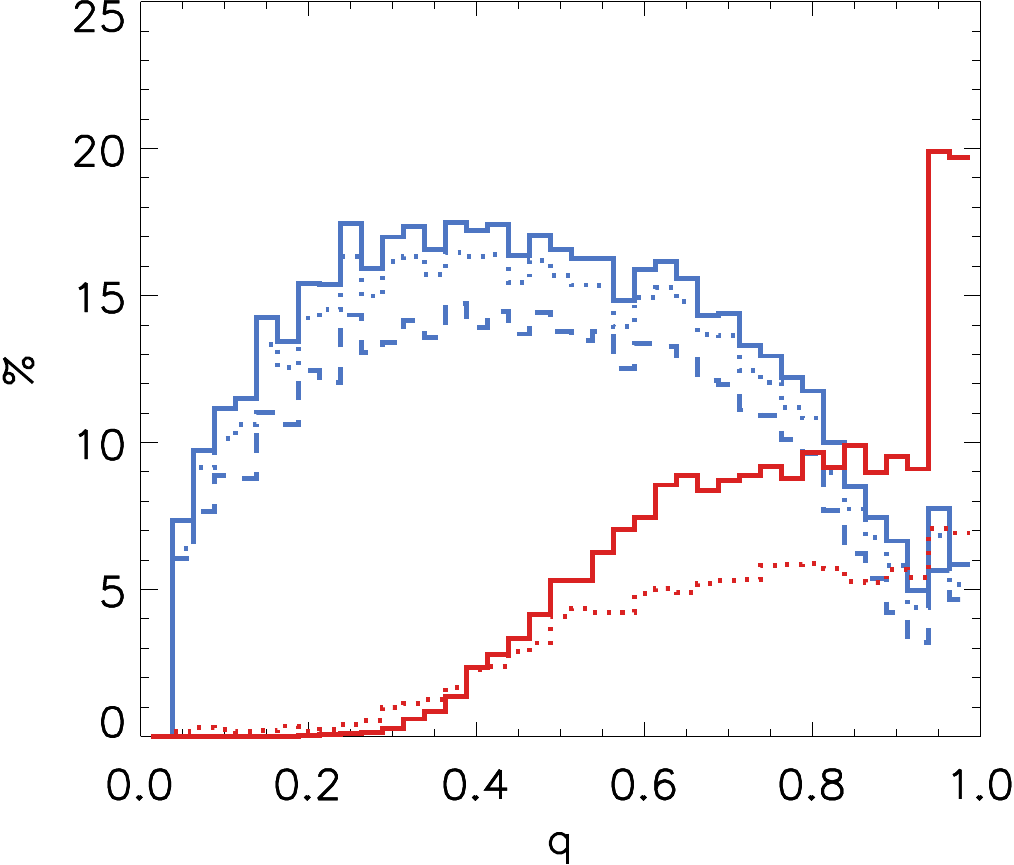}{0.235\textwidth}{}
        }
        \vspace{-0.9cm} 
        \gridline{\fig{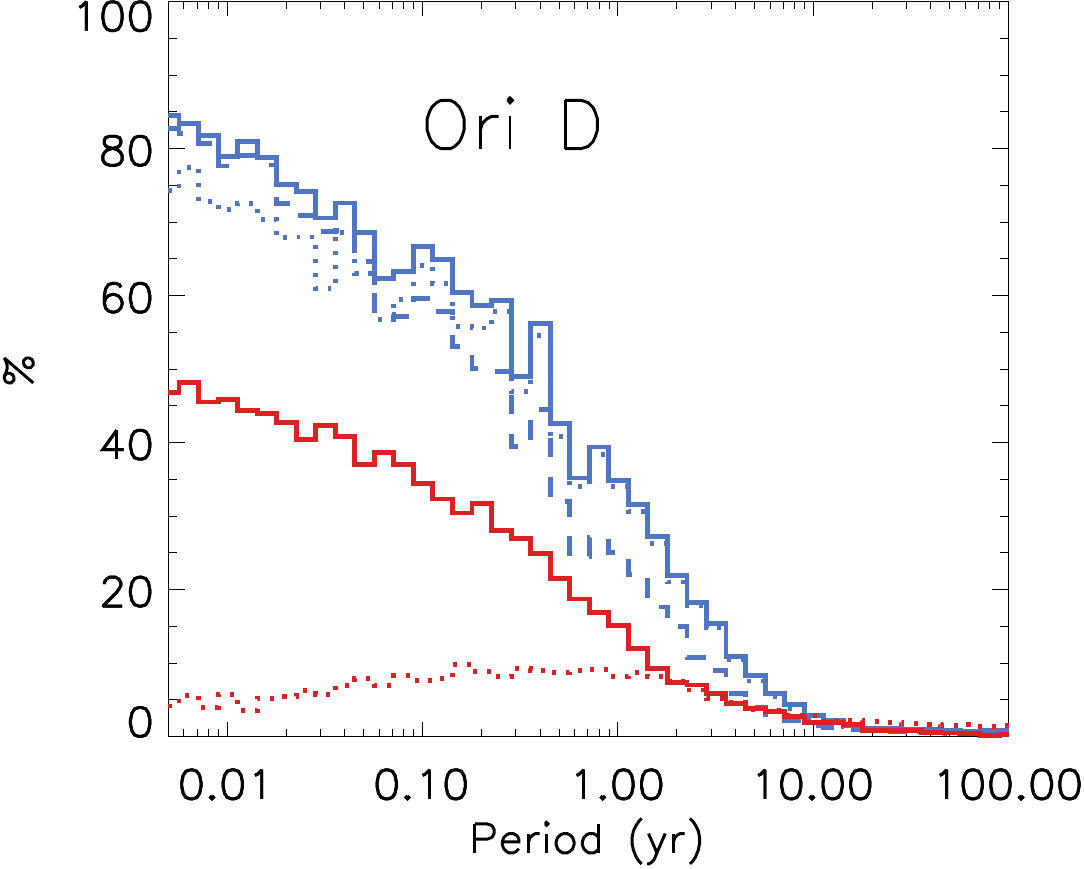}{0.25\textwidth}{}
                  \fig{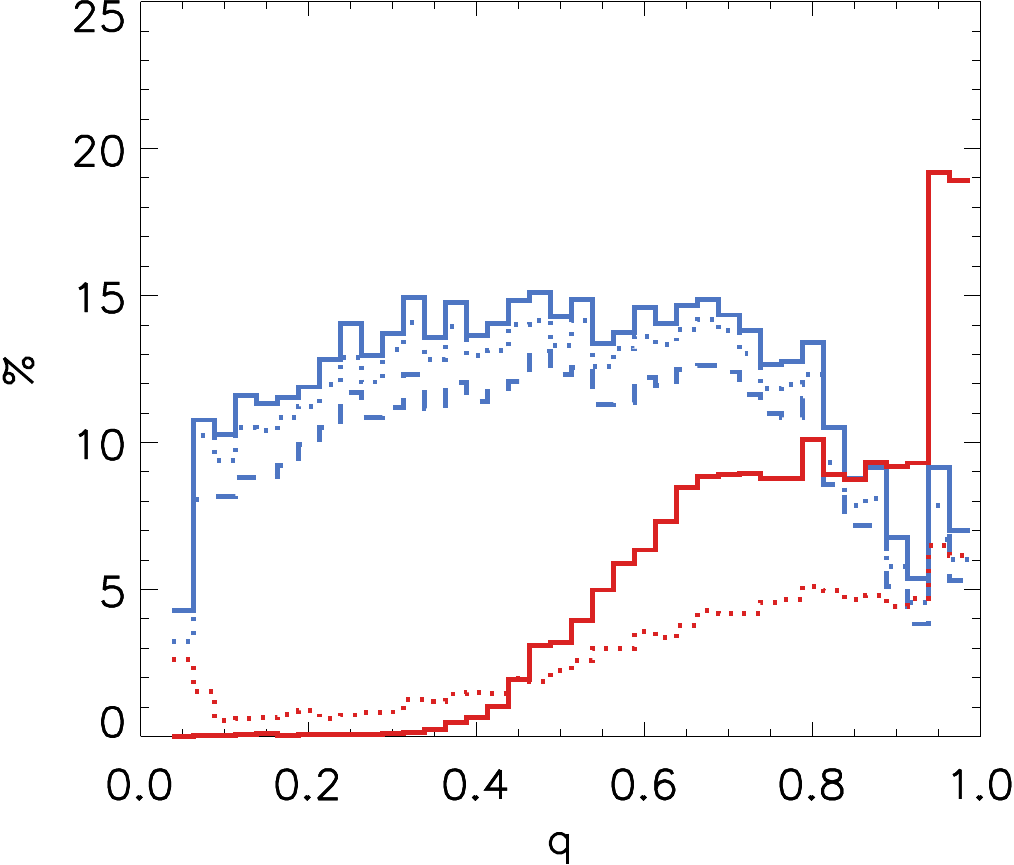}{0.235\textwidth}{}
                  \fig{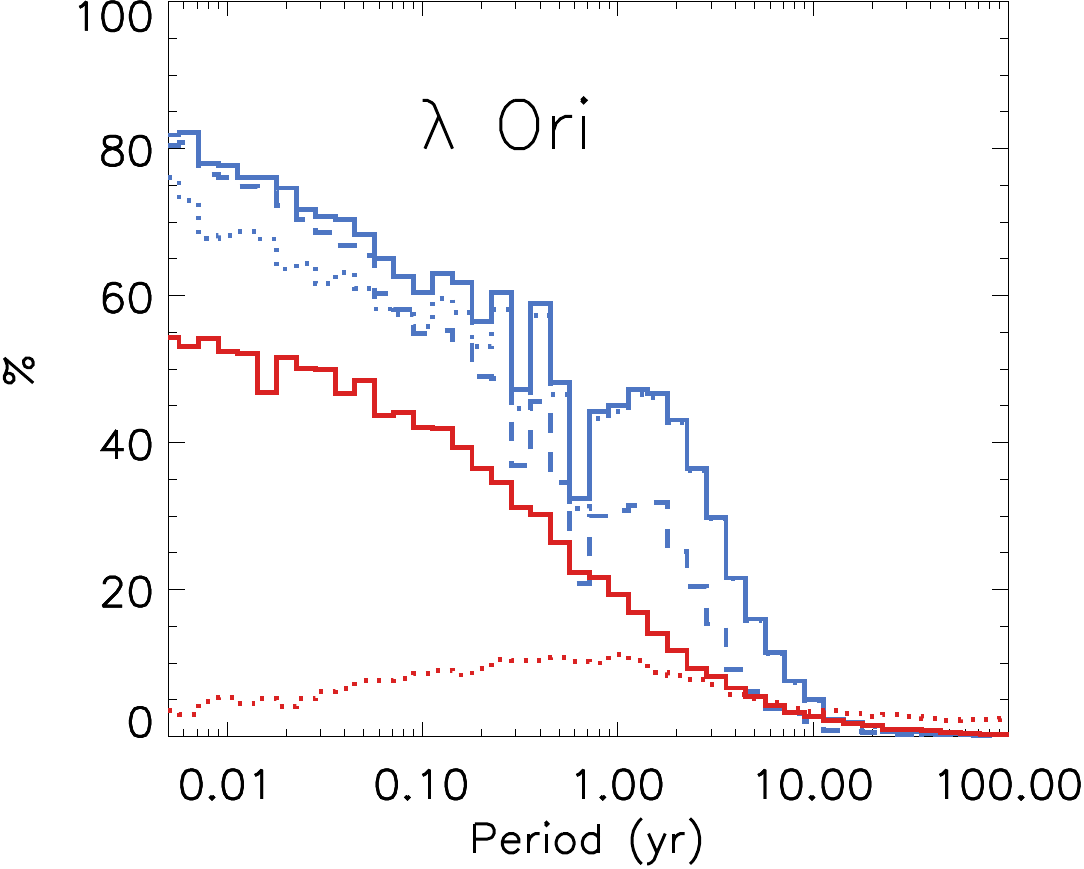}{0.25\textwidth}{}
                  \fig{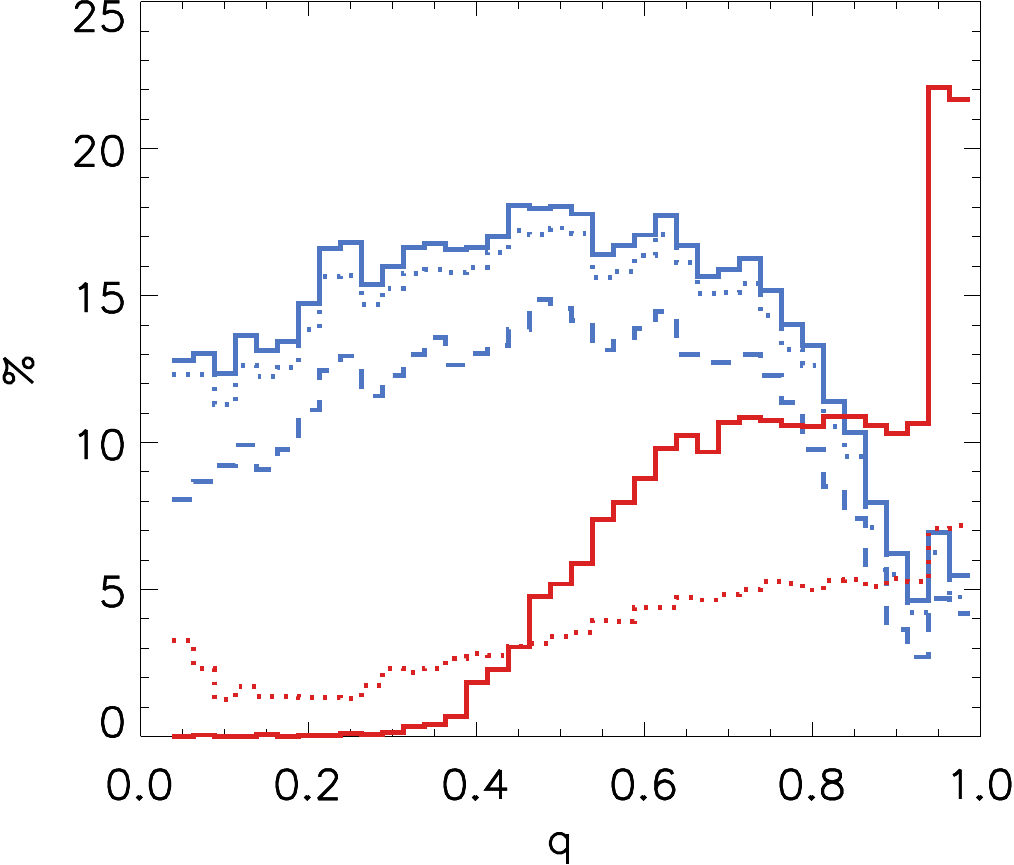}{0.235\textwidth}{}
        }
        \vspace{-0.9cm} 
        \gridline{\fig{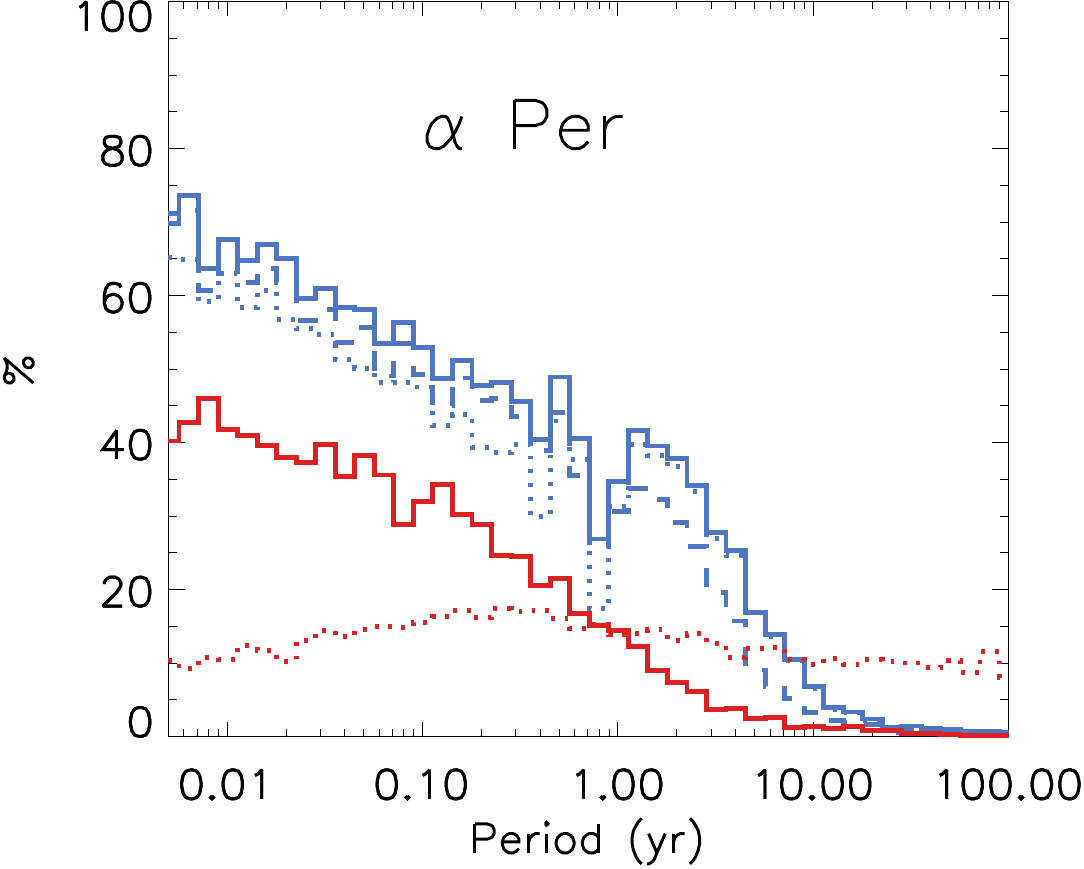}{0.25\textwidth}{}
                  \fig{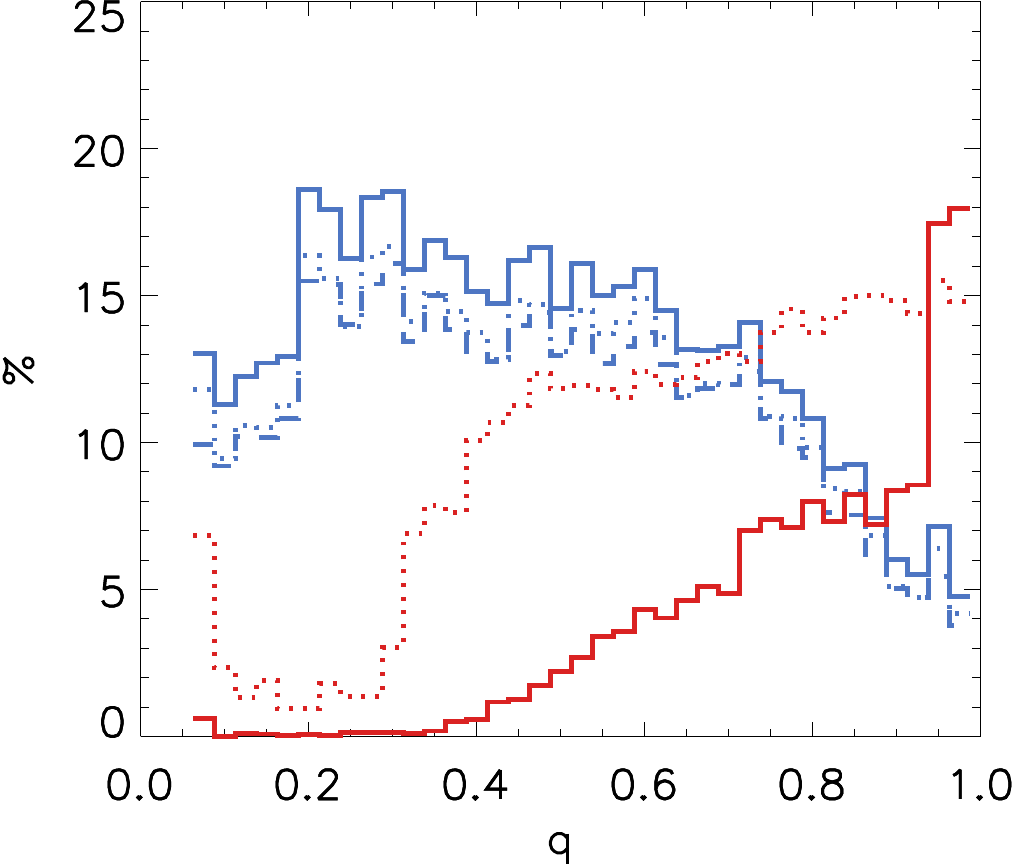}{0.235\textwidth}{}
                  \fig{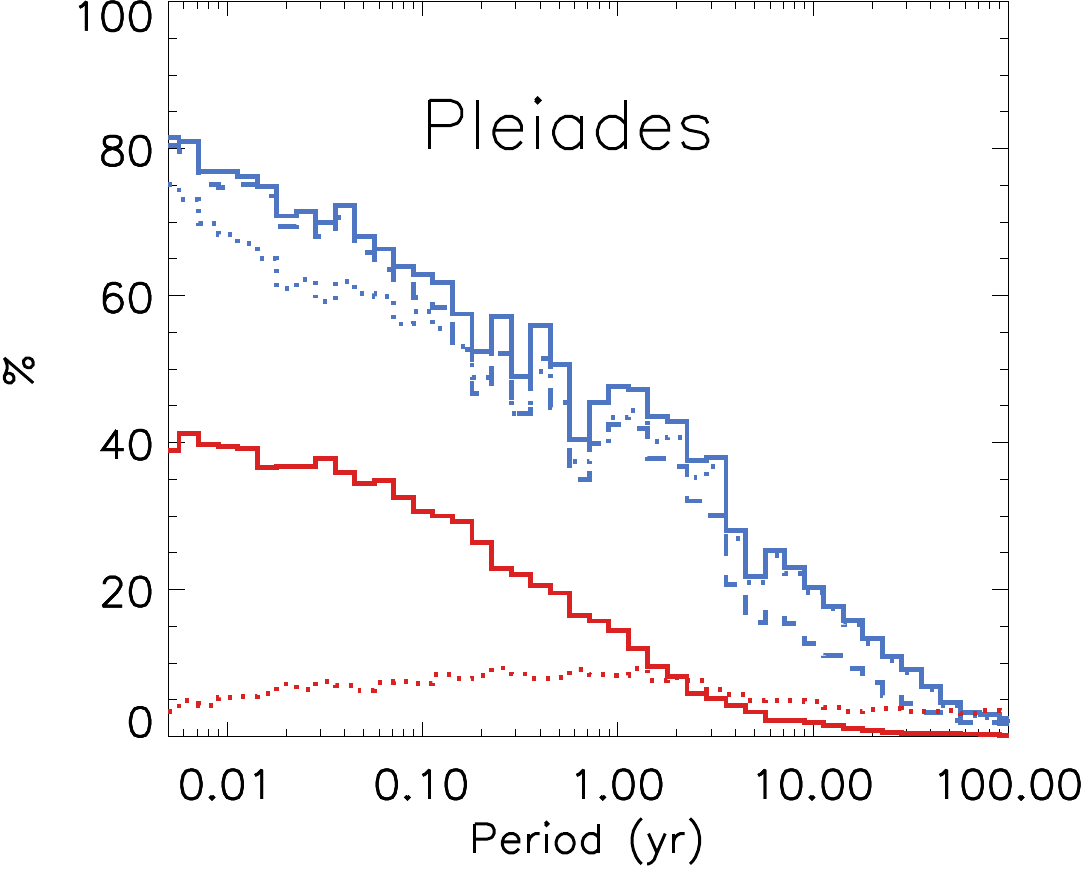}{0.25\textwidth}{}
                  \fig{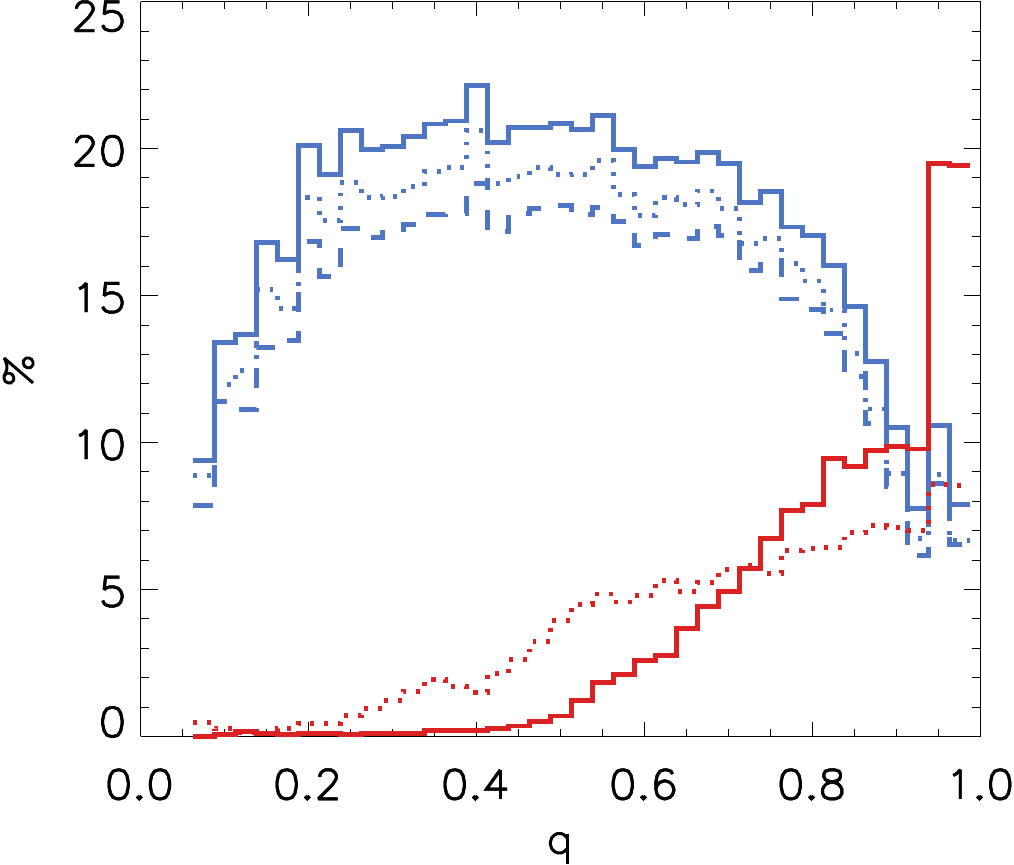}{0.235\textwidth}{}
        }
        \vspace{-0.5cm}
      \caption{Completeness of the recovered binaries (Same as Figure \ref{fig:detection}), but showing in period and mass ratio in all of the individual regions. Blue line shows the fraction of the recovered RV variables, restricted only to the sources with 3 epochs. Red line shows the completeness of SB2s.\label{fig:detection3}}
\end{figure*}

To analyze the dependence of the recovered stars on the configuration of the system, we constructed a set of synthetic binary stars with stellar properties matched to those of the curated sample, and multiplicity properties consistent with the field population as measured by \citet{moe2017}. A total of 950 binary and 50 single systems were generated for each star in the curated sample, for a total of $\sim$5 million systems with $\sim$20 million synthetic spectra. Period distribution in the sample is log-normal, with $\mu_{\log P} = 4.6$ and $\sigma_{\log P} = 2.2$, where P is in days. This distribution is largely consistent with the more commonly used \citet{raghavan2010} for the close binaries, with the primary difference occurring at wider systems due to the difference in accounting for tertiary companions. The minimum period was set at 2 days, which is close to the minimum period observed in YSOs. The main sequence stars do have companions with shorter separations, however, because young stars are still inflated compared to their older counterparts, they cannot yet sustain shorter orbits yet. Some of these companions will migrate closer in through the magnetic breaking and tides. The distribution of mass ratios is uniform above $q=0.08 M_\odot/M_1$, with an excess of twins ($0.95<q<1.00$) found at close separations; the excess is characterized at 20\% probability at the separations $a<0.1$ AU, and 0\% at $a>200$ AU (on top of the underlying uniform distribution), decreasing linearly with respect to $\log a$ (Figure \ref{fig:orbit}a). The eccentricity distribution is uniform from 0 to $e_{max}$, where $e_{max}$ is defined by 

\begin{equation}
e_{max}(P)=1-\left(P/k\right)^{-2/3}
\end{equation}

\noindent and $k$ is the tidal circulization period, which was set to 2 days. For computational efficiency, $e_{max}$ was capped at 0.98 (Figure \ref{fig:orbit}b).

The orbit of each synthetic binary was sampled at epochs consistent with the actual observations of the reference source to obtain the RVs of the simulated primary and companion, computed at each epoch separately. Appropriate PHOENIX synthetic spectra \citep{husser2013} were chosen for each synthetic system: for the primary, both \teff\ and \logg\ were chosen as the best template match to the reference source; the parameters of the secondary were interpolated from the PARSEC-COLIBRI isochrones \citep{marigo2017} for a given estimate of age, primary mass, and mass ratio. Both spectra were doppler shifted to the corresponding RVs, broadened by the kernel corresponding to the IN-SYNC \vsini\ of the reference source (if a reference source was identified to be an SB2, the FWHM of the CCF peak that corresponds to the primary component was used instead). The flux was then scaled from the stellar radius to the appropriate distance\footnote{In the case of Taurus there are two separate populations projected on top of each other \citep{luhman2018}, although the distance and proper motions are the only criterions in which it is possible to reliably separate them, and thus in absence of \textit{Gaia} astrometry for some of the sources for the sake of simplicity both of them were treated as a single region}. Spectra of both components were co-added, interpolated over the relevant wavelenth range, and the pixel noise profile of the reference source was applied. The CCFs of the resulting spectra were computed, and deconvolved into components for SB2 identification using the same routines as applied to the actual APOGEE observations. It proved impractical to process $2\times10^7$ synthetic spectra through the IN-SYNC pipeline to provide RV estimates that could be used, as the IN-SYNC pipeline takes several hours to process a single epoch spectrum. Instead, a single Gaussian was fit to the CCF, and the Gaussian centroid is used as an approximation of the single measured RV in the spectrum. A check was performed to confirm that the recovered properties are consistent with Figure \ref{fig:sb2pr}. Then, the intrinsic properties of the recovered systems, both SB2s and RV variables, are examined.

The recovered $\Delta v$ of SB2s with a quality flag of 4 is well matched by the original input (Figure \ref{fig:synt}), with a false positive identification rate among the single stars of only 0.06\%. The recovered FWHM of flag 4 sources is also typically representative of the injected \vsini. Our recovery procedure overestimates the RV separation of a substantial number of SB2s with quality flag 3, particularly among rapid rotators with the injected \vsini $\sim$60 \kms. This occurs because such binary systems may produce asymmetries in the wings of the CCF even with the intrinsic $\Delta v$ of just a few \kms. While the Gaussians in the best fit do reproduce the shape of the CCF, and the sources involved are real binaries (the false positive rate among single stars is 2.8\%, assuming that none of the systems are spotted), neither the recovered RV separation nor the FWHM are representative of the true RVs of the components or the \vsini\ in Flag=3 systems.

Next we examine the distribution of the orbital parameters of the recovered SB2s, defining completeness as a ratio between the number of systems identified as a binary and the total number of synthetic binary systems, marginalized over a given parameter. In terms of the overall probability of detection of a system as a spectroscopic binary, for SB2s, almost no detections occur at $\Delta v$ between the primary and the secondary (\smax $\equiv \mid v_1-v_2\mid$) below 20 \kms. The completeness of the recovered sample rises rapidly between 20$<$\smax$<$40 \kms, and then remains constant at \smax$>$40 \kms. Among the RV variables, it is possible to detect the amplitude of RV variability between epochs (\vmax$\equiv$ max(RV$_{jd1}$)-min(RV$_{jd2}$)) down to 1 \kms, but the recovered fraction reaches its maximum at \vmax$\sim$10 \kms\, and remains constant at higher \vmax. The highest probability of detection occurs at very short periods, with almost no binaries with separations beyond 10 AU identified (Figure \ref{fig:detection}). Highly eccentric systems increase the likelihood of an identification of SB2s, because only a single epoch is needed to catch such a system at the pericenter to detect the maximum RV separation. On the other hand, RV variables with high eccentricity are more difficult to identify, as multiple epochs with sufficiently high RV offsets are needed. The effect of eccentricity on recovery of both RV variables and SB2s is less pronounced for systems with periods $<$200 days (Figure \ref{fig:detection2}). A high mass ratio increases the probability for detection of SB2s. Among the RV variables, however, detections are maximized at $q\sim0.5$. At higher $q$, the companion makes a stronger contribution to the spectrum, even if the RV separation is too small for the system to be identified as an SB2. In these cases, the pipeline measured RV is influenced by both the primary and the secondary, and tends toward the stable barycentric RV of the system. Binaries whose components have low \vsini\ are more reliably recovered, as their narrow spectral features produce similarly sharp CCF peaks, which allow more accurate RV measurements and a better separation of components even at low $\Delta v$. Multiple epochs do help with identification, but not significantly - sources with 10 epochs are only 1.5 times more likely to be caught in a favorable configuration for detection compared to a single epoch. RV variables with more than 3 epochs are not more easily identified via $\chi^2$, although there is a modest increase of RV variables identified via the slope.

It should be noted that because the synthetic sample was constructed using the real observations as a basis for the distributions of stellar properties, any unresolved SB2s in the data would have a somewhat inflated \vsini, resulting in a lower recovery rate for these sources, thus potentially underestimating the total MF in the model compared to the data. It is not expected, however, that this effect should be significant as the difference in completeness between the true and a slightly inflated \vsini\ is small, and only a few sources will be affected.

As all the sources are individually represented in the model, we also compare the completeness of the overall sample as it is split among the individual regions. In general, the dependence of the completeness on the underlying stellar parameters is consistent among all of the regions, in shape if not in the absolute normalization. The only exception to this is the mass ratio, because it is driven not only by the typical masses of the stars in a given sample, but also by age (Figure \ref{fig:detection3}). This does not strongly affect RV variables, but appears to have a strong effect on SB2s. In the synthetic spectra is easier to recover sources across all mass ratios in younger systems with lower \logg. On the other hand, as the stars shrink to their main sequence sizes, the sensitivity to low $q$ systems decrease significantly. This occurs because in the younger systems, the companion tends to have a lower surface gravity than the primary, thus the luminosity ratio between the secondary and the primary is higher than what it will be in a more evolved system, where the trend of \logg\ is reversed.

\begin{figure}
\epsscale{1.1}
\plotone{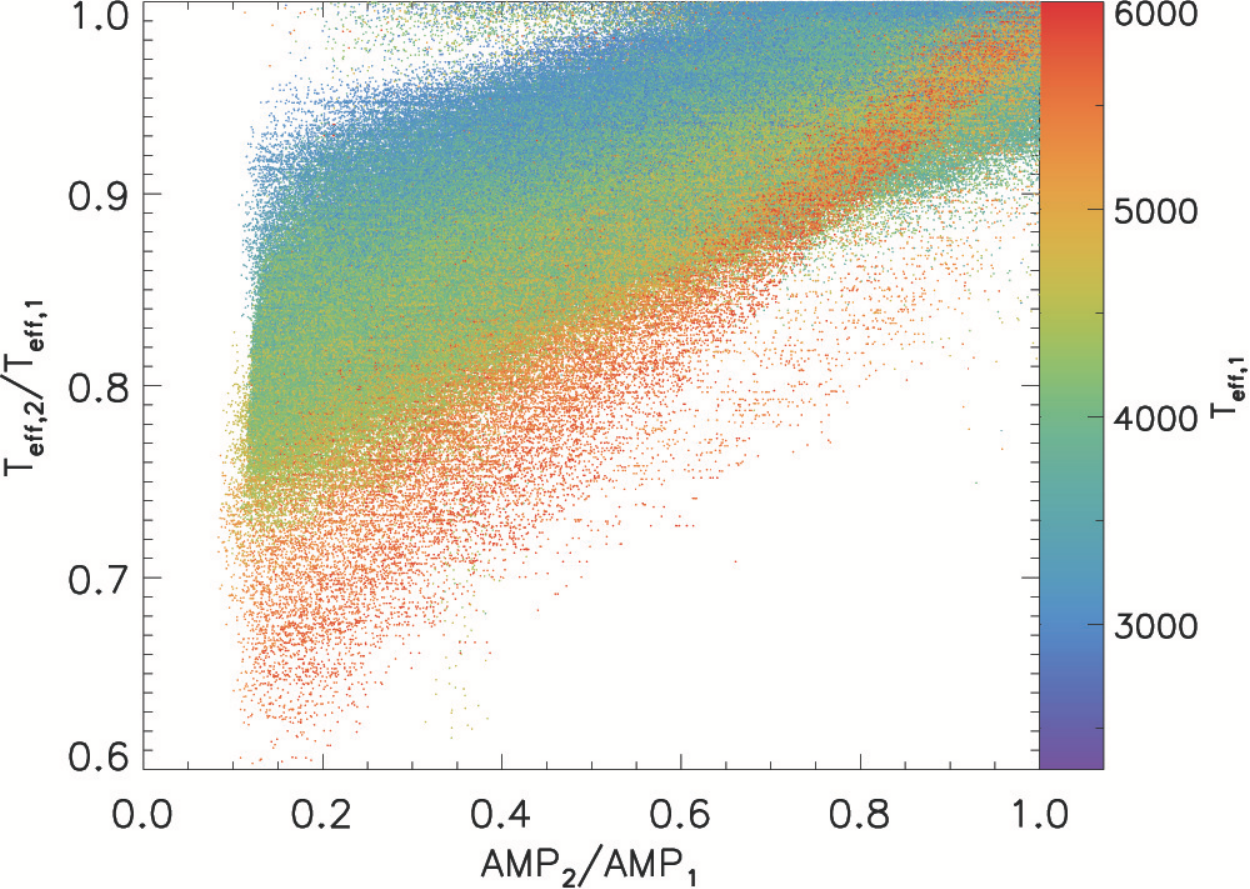}
\caption{Dependence of the amplitude ratio of the components of the SB2s recovered from the CCF on the ratio of the injected \teff\ between the secondary and the primary.
\label{fig:q}}
\end{figure}

Among the SB2s, the amplitude of the components in the CCF typically corresponds to the ratio of the effective temperatures as a function of the \teff\ of the primary (Figure \ref{fig:q}). Instead of the ratio of the amplitudes, it is also possible to construct a similar relation using the integral of the individual Gaussian components, although with a greater degree of scatter. The trends observed in Figure \ref{fig:q} are not as apparent when the \teff\ ratio is replaced by $q$: the direction of dispersion with the relation to \teff\ of the primary reverses half-way through the range.

\section{Results}\label{sec:result}

In this section we analyze the trends in the data in relation to the model of field multiplicity.

\subsection{Mass ratio}

\begin{figure}
\epsscale{1.1}
\plotone{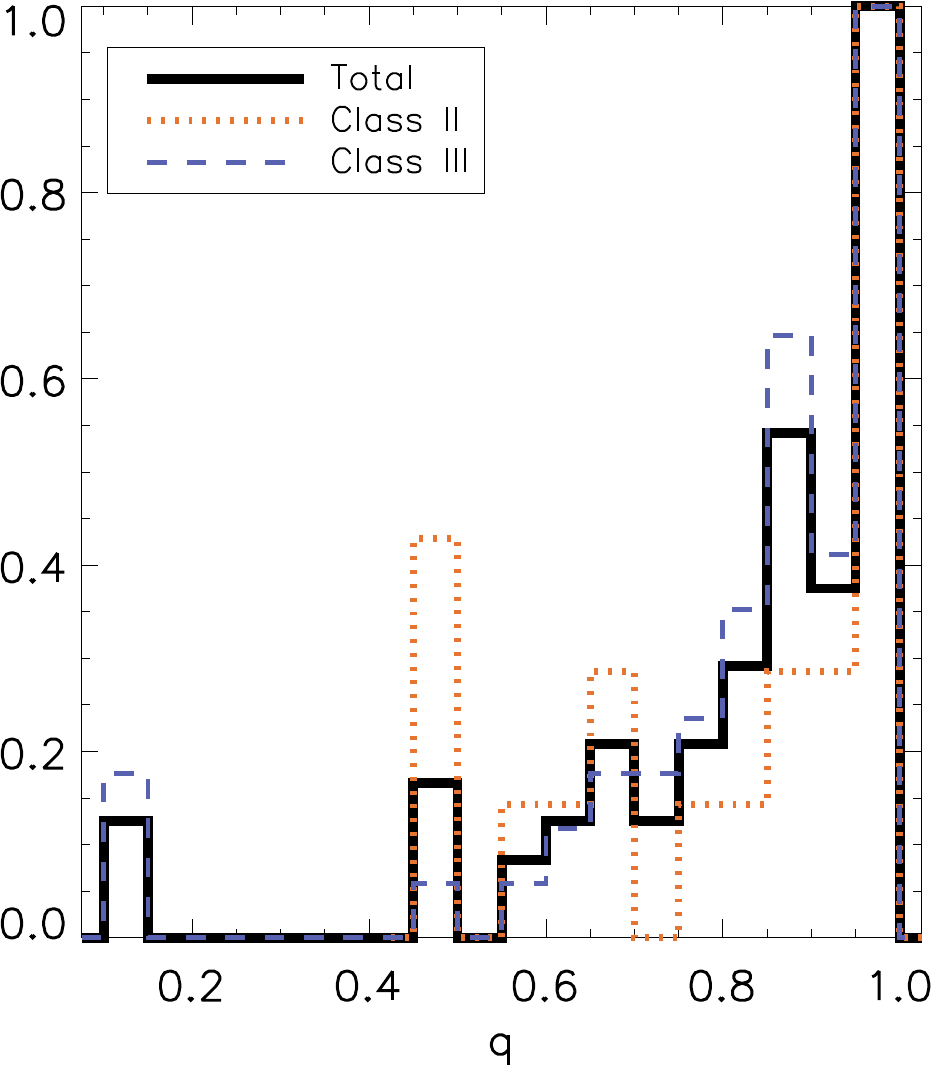}
\caption{Distribution of the mass ratios measured among SB2s, normalized to the peak.
\label{fig:qref}}
\end{figure}

For SB2s that have multiple epochs and substantial $\Delta v$ between components, we measured mass ratios by fitting a slope to RV$_{primary}$ vs RV$_{secondary}$ \citep{wilson1941}. We included in the fit all of the deconvolved components down to Flag=1, assuming that a robust Flag=4 detection was available in at least one epoch. When components were unresolved, both the primary and the secondary were assigned to the same central RV. Any apparent SB3s and SB4s (8 sources, most likely compact higher order multiple systems) were excluded, as the sample does not contain enough epochs to reliably fit their relative motions. If a component was apparent during visual examination but was not recovered by the pipeline, the peak of that component was fitted manually. Fits to determine $q$ could be performed for 78 systems, the resulting mass ratios are included in Table \ref{tab:sample}. In general, there was no confusion in tracking the companions. Primary/secondary misidentification resulted in fits that returned q$>1$ for 16 stars; the identification were consistent across all epochs, however, and we correct the error by reporting the inverse of the derived mass ratio (i.e., $q=1/q_{derived}$). The only confusion occurred in epochs in which companions were unresolved if the CCF had a low SNR component not associated with a real source.

Previously, \citet{fernandez2017} compared the distribution of mass ratios among young SB2s detected by APOGEE with results from the other binary surveys. They found that equal mass binaries dominated their sample, with a tail stretching towards lower $q$. They found that the overall distribution is consistent with what has been observed by \citet{pourbaix2004} for more evolved stars, but they interpreted the excess of twins as a detection bias, where high mass systems have a larger orbital velocity amplitude than lower mass ratio systems. \citet{tokovinin2000} and \citet{moe2017}, however, concluded that at short separations (within 200 AU), the overall distribution of $q$ is not entirely uniform but that there is a significant excess of companions with $0.95<q<1$ on top of the uniform distribution.

Examining the mass ratios measured from the data, we find a similar distribution to that observed by \citet{fernandez2017}, with a large excess of twins (Figure \ref{fig:qref}). Examining the completeness of the recovery of synthetic binaries as SB2s, we find that the probability of detection does rise somewhat for systems with high mass ratio, but it is not the most dominant factor: the completeness only changes from $\sim$5\% of all binaries at $q\sim0.4$ to 7\% at $q\sim0.9$. The reason for the apparent discontinuity at $q=0.95$ in the total distribution in Figure \ref{fig:detection} is due to the fact that this excess of twins at short separations is included in the model population we adopt for our completeness calculations, and it is the small separations, not the near-equal mass ratios, that drives these twins to be twice as likely to be identified as SB2s. Without the intrinsic separation dependent excess of twins, such systems would not dominate the sample, i.e., the observational biases cannot account for their presence. Therefore, the interpretation used by \citet{fernandez2017} regarding their presence does not apply, and the model of mass ratios from \citet{moe2017} does accurately represent the data.

There are no strong differences in $q$ as a function of the evolutionary class, nor as a function of period (Section \ref{sec:period}), nor is there a trend as a function of \teff. However, curiously, there are 3 sources that have a peculiarly low $q\sim0.1$; this is significantly lower than the minimum of 0.2 that is recovered in the synthetic observations. Two of these sources, 2M03281336+4856154 and 2M03301892+4943348, are located in $\alpha$ Per, the third, 2M03432619+2602308, is found in the Pleiades. The primaries of these systems are $\sim$0.2 \msun objects, making their companions to be a part of the brown dwarf desert.

\subsection{Normalization}

\begin{figure*}
\epsscale{1.1}
\plottwo{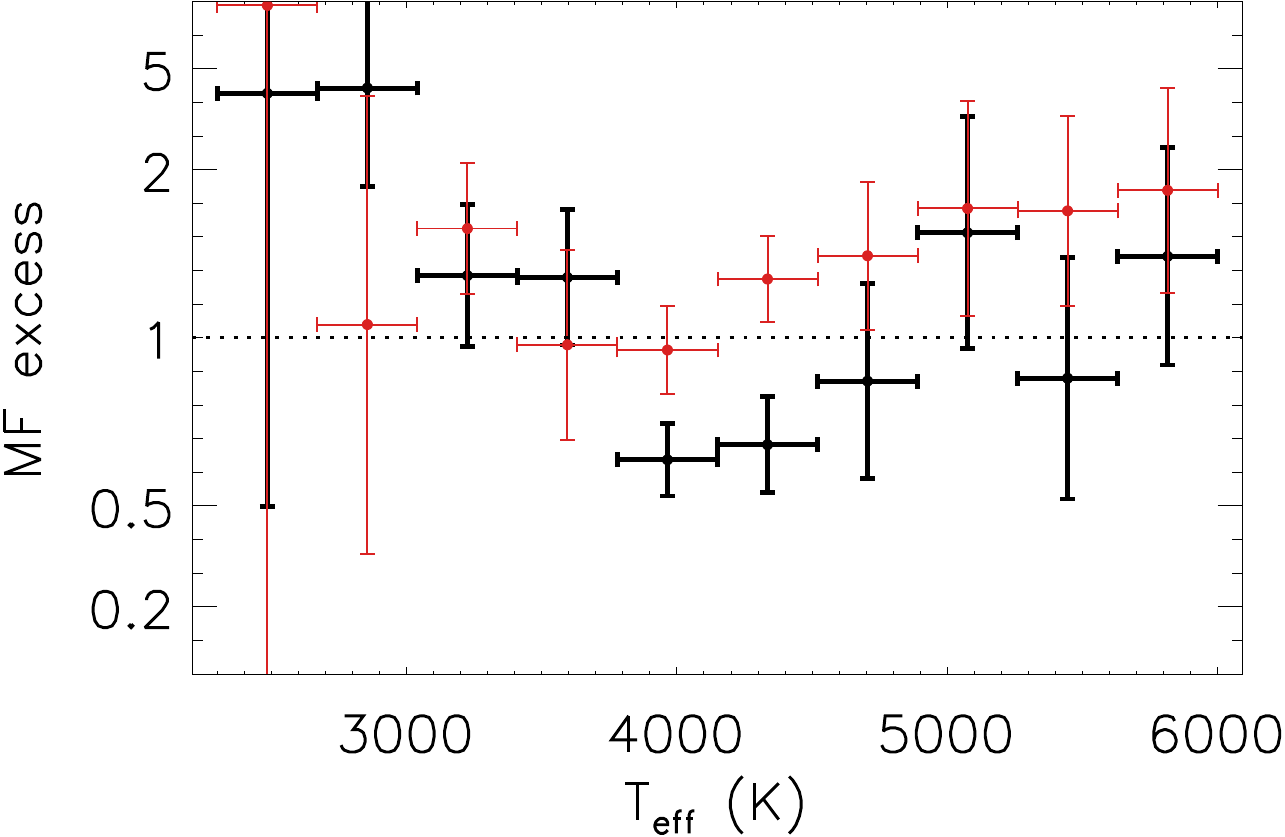}{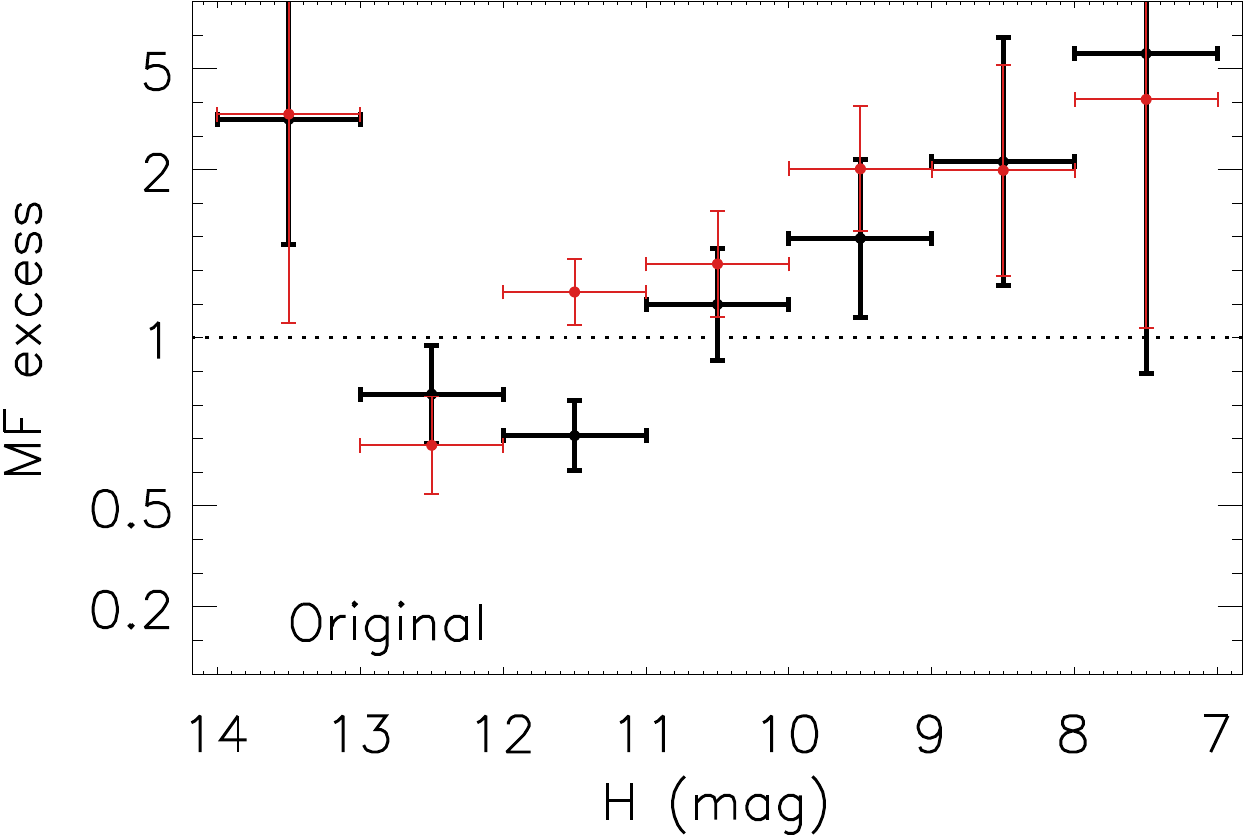}
\plottwo{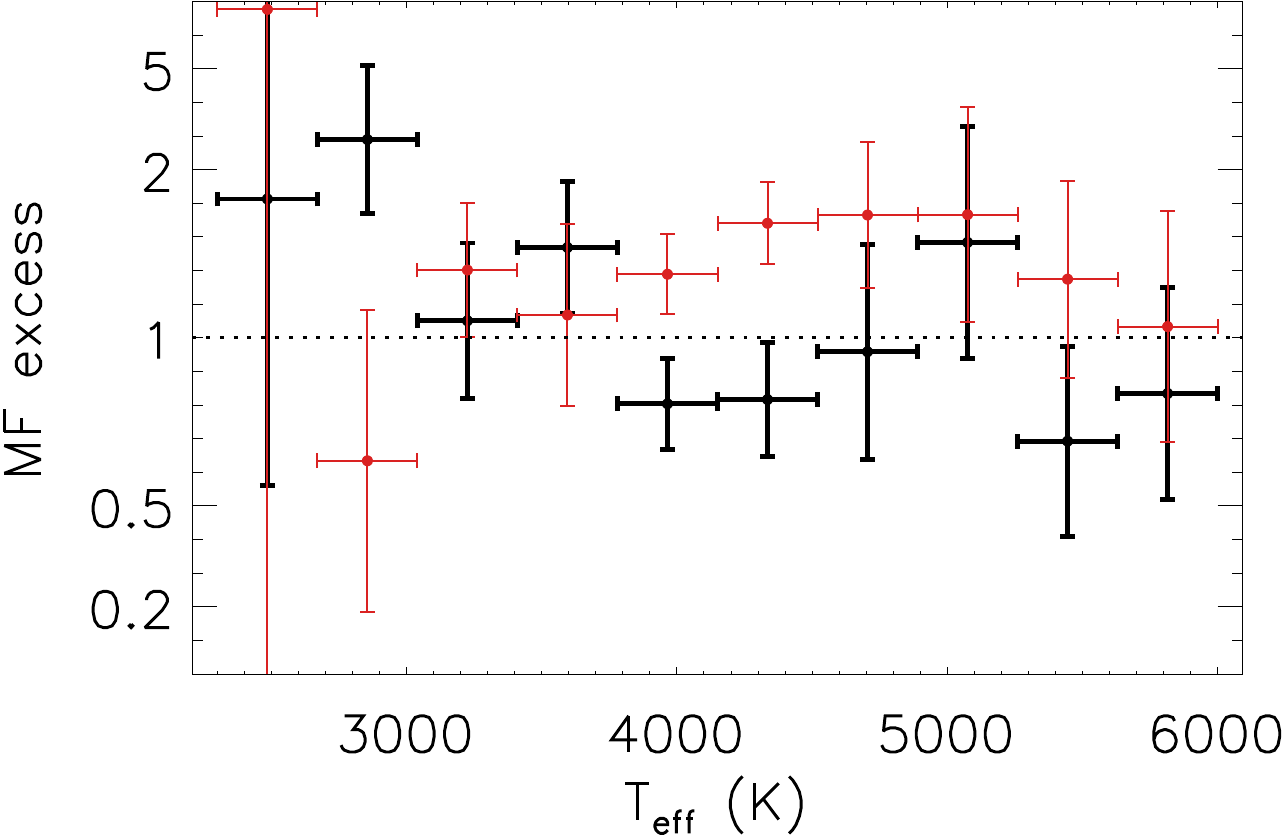}{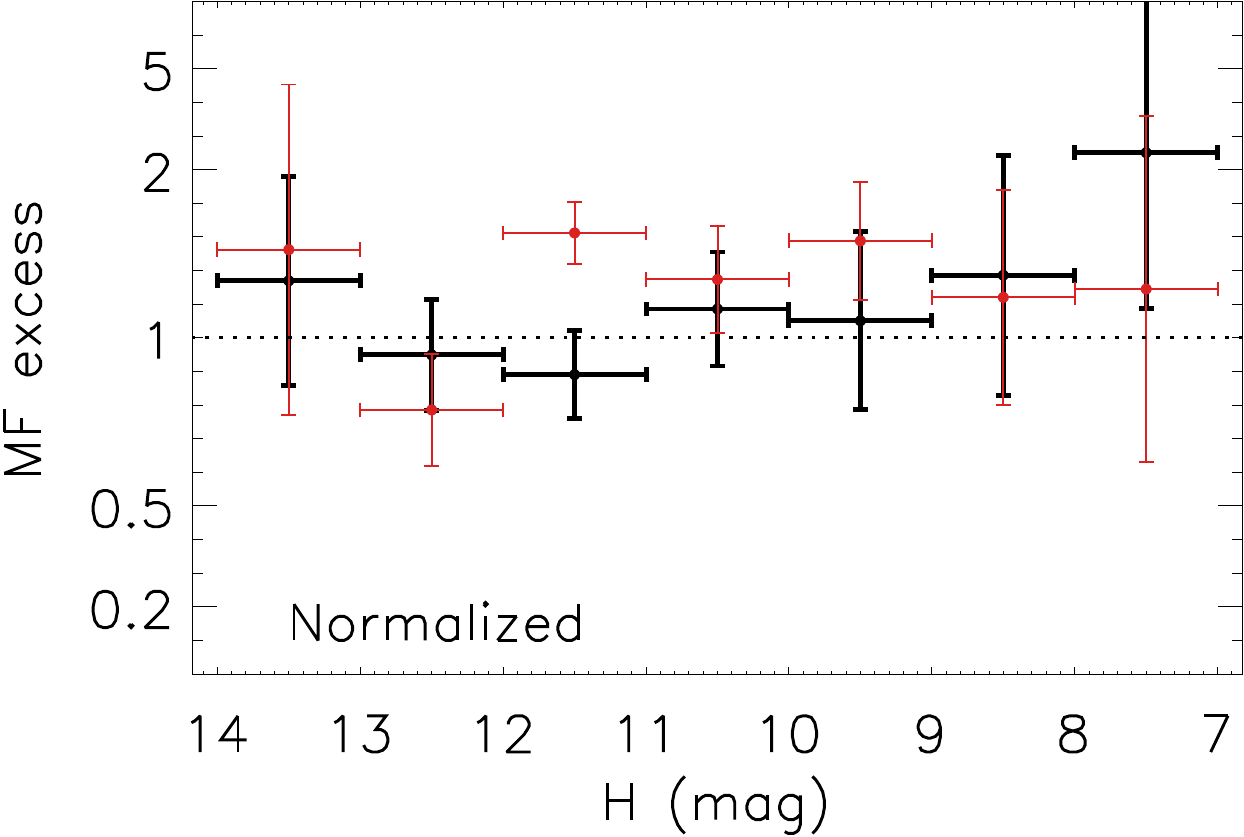}
\caption{Distribution of MFs relative to the model in the sample as a function of \teff\ and $H$ band magnitudes, with the combined data from all clusters (the individual regions do show similar trends). The top row shows the original uniform scaling assuming MF($<$10 AU) of 16.5\%; the excess at low \teff/faint H originates from the Malmquist bias, and at high \teff/bright H it shows the increase of MF with the mass of the primary. The bottom row has the normalization specified in the Equation \ref{eq:normalization}. The distribution measured from SB2s is shown in black, the one that is measured from RV variables is shown in red. The scaling of the plots is such that the MF that is consistent with the model is located at 1, and values above 1 imply that MF in the data is in excess of what is expected by the model. The top and the bottom portion of the plot is symmetric around 1, running from 0 to $\infty$.
\label{fig:mfteff}}
\end{figure*}

The total MF for SB2s across all epochs is 2.8$\pm$0.2\%, this MF rises only slightly to 3.1$\pm$0.3 in SB2s with at least 3 epochs. Among RV variables with at least 3 epochs, the MF is 6.3$\pm$0.5\%. Because there is some overlap between the sources classified as both SB2s and RV variables, the combined multiplicity in the sample is 8.3$\pm$0.5\%.

In order to compare the MF in all of the regions in a uniform manner, we use the recovered fraction of the synthetic binaries as a proxy for what is expected in each region given the underlying model of the distribution of companions. This allows us to take into the account any differences that originate from the different observing strategies between individual regions. We do not report the MFs measured in individual regions but rather compare if the measurement is over or under-abundant in comparison to the model.

For computational expediency, almost no single stars were generated as a part of the synthetic comparison sample. However, in order to calculate the MF in the model that is most representative of the data, it is necessary to incorporate single stars in some way. We account for the presence of the single stars by scaling down the MF from the synthetic sample. We originally considered a normalization that would maintain 16.5\% of systems with separations less than 10 AU (with MF defined as $\frac{B+T}{S+B+T}$, where S is the number of single systems, B is the number of binaries, T is the number of triples with the given separation range, etc). This relation is interpolated from \citet{moe2017}, corrected for the typical masses ($\sim$0.6 \msun) and metallicities (Z=0) of the stars in the primary sample. This scales the overall synthetic MF by a factor of 2.4.

However, the MF does have a strong dependence on the mass of the primary \citep[e.g.,][]{duchene2013,raghavan2010,ward-duong2015}, with the overall MF increasing by a factor of $\sim$1.7 from M to G dwarfs, with the effect most pronounced in wide binaries. Comparatively modest increase is observed in the MF of close systems, with only a factor of $\sim$1.2 increase from 0.6 to 1 \msun, although it jumps by a factor of $\sim$1.9 from 1 to 2 \msun\ primaries \citep{moe2017,murphy2018}. While most of the sample consists of K type objects (56\% of sources have 3900$<$\teff$<$4800 K, correlates to masses of $\sim$0.5--1 \msun), overall the sample spans from close to the substellar boundary (\teff =2300 K) up to early G type stars (\teff = 6000 K, with masses as much as 2 \msun\ -- in YSOs, by the time they reach the main sequence they will become hotter). The exact distribution of primary masses in the APOGEE sample may have an effect on the recovered MF. Furthermore, the APOGEE observations are limited in brightness down to $H\sim13$ mag. Due to the Malmquist bias of the second kind, binaries may be overrepresented among the low mass sources in our sample because they were more likely to meet or exceed our targeting limit due to being brighter than their single counterparts. Due to the complex targeting strategy, it is difficult to reproduce this entirely through forward modeling, but close to the magnitude limit, it may artificially raise the MF by 10--100\% in a given mass or flux range.

In Figure \ref{fig:mfteff} we compare the MF as a function of both \teff\ and the $H$ band. In each of the bins (here, and in the subsequent subsections), we restrict the sample from the model to only those synthetic sources that were produced in reference to the sources that fall into a given bin. Both distributions show a decrease in MF from hotter to cooler stars by a factor of $\sim$2 from 4000 to 6000 K. Through performing linear regression, the rising slope of MF relative to the H band fluxes is inconsistent with being flat at 3.7$\sigma$ level with H$<$13, and 2.8$\sigma$ relative to \teff\ for \teff$>$3800 K. This is largely consistent with what is expected from physical models of the dependence of multiplicity on primary mass. While each individual region has a different conversion from mass to \teff\ and apparent H mag due to different distances and ages, this trend holds across all of the regions. However, it is impossible to correct the increase in the MF in both H band fluxes or just \teff\ by using a normalization that relies on just one of these parameters, necessitating including both of them. This suggests that in addition to the astrophysical reasons for the increase, there might also be an effect from the increased SNR that helps to identify binaries that is not entirely replicated by the model.

The MF also increases for the faintest stars, with the strongest effect occurring at the observational magnitude limit, consistent with the Malmquist bias. This translates to an increase in MF for sources with \teff$<3500$ K in the full sample, and while this is also observed in all of the individual regions covered by the survey, depending on the distance the location of the break would be different.

To correct for both of these effects, we apply a normalization scaling

\begin{multline}
MF (<10 AU)=16.5\times (308.9-119.6H+\\17.46H^2-1.131H^3+0.02735H^4)\times \\ (7.634-2.965T_{eff}\times10^{-3}+3.189T_{eff}^2\times10^{-7})  \label{eq:normalization}
\end{multline}

to the model. This was obtained by fitting the MF dependence on H band across the entire range, and then fitting the deficiencies in \teff. It should be noted, however, that this normalization has very little effect on the results of the subsequent sections as compared to a simpler $MF (<10 AU)=16.5\%$, as both mix high and low mass stars together and average out any strong \teff\ or flux dependence.

\subsection{Overall distribution between the regions}

\begin{figure*}
\epsscale{1.1}
\plotone{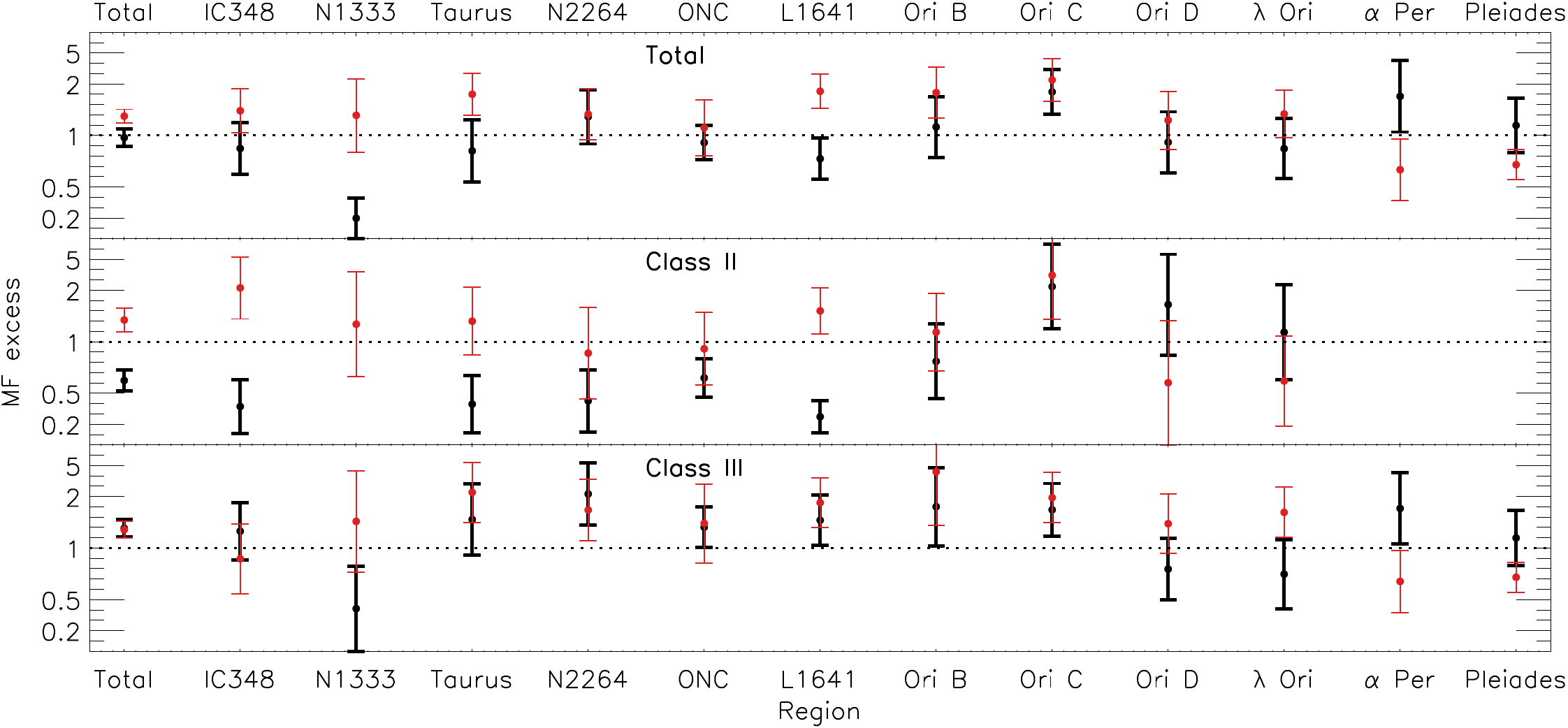}
\caption{Distribution of MF relative to the model separated into the individual regions covered by the survey, averaged across all of the sources in the region. The distribution measured from SB2s is shown in black, the one that is measured from RV variables is shown in red. The top panel includes the full curated sample, second panel is limited only to the sources that have been previously classified as Class II YSOs, third panel is restricted to known Class III sources. Normalization specified by the equation \ref{eq:normalization} is applied. The regions are roughly ordered by an increasing age.
\label{fig:mf}}
\end{figure*}

Overall, the distribution of spectroscopic binaries averaged across all regions appears to be consistent with the model; this holds for both SB2s and RV variables (Figure \ref{fig:mf}). The same applies to all of the individual regions in the survey -- while individual regions may show a slight non-statistically significant excess or deficit in one type of companions, those deviations average out in the combined sample. There are only a few exceptions:

\begin{itemize}
\item SB2s (but not RV variables) are deficient in NGC 1333 by a factor of 5.3 (4.3$\sigma$)
\item RV variables (but not SB2s) are in excess in Taurus and L1641 by a factor of 1.6 (1.5--2$\sigma$)
\item In Orion C, both SB2s and RV variables are in excess by a factor of 1.95 (1.9--2.2$\sigma$)
\item the MF measured from RV variables is somewhat elevated (by as much as $\sim$50\%) compared to what is measured from SB2s, although the trend is reversed in the older regions ($\alpha$ Per and Pleiades).
\end{itemize}

The excess of companions in Orion C may be explained by the membership assignment. Because this region and Orion D are projected on top of each other in the plane of the sky, there is some difficulty in distinguishing them apart. When possible, we used membership from \citet{kounkel2018a}, otherwise a weighted average RV cut of 25 \kms was used. While this cut is sufficient to reliably separate single stars in these populations, in the case of spectroscopic binaries this results in the systems that are currently being blueshifted being assigned to the Orion D, and those that are redshifted being assigned to the Orion C. Because Orion D has more sources in total by a factor of 1.3, the companions will be disproportionally represented in Orion C, resulting in an elevated MF. Averaging these two regions together brings them both into a closer agreement with the model.

\subsection{YSO Class}\label{sec:class}

We examine the effect of the multiplicity on the presence of the protoplanetary disk. Out of 1882 disk-bearing Class II sources, 59 are found to be RV variables, and 38 are SB2s. Out of 3125 diskless Class III sources, 116 are RV variables, and 103 are SB2s. Class IIs are deficient in SB2s compared to the model by the factor of 1.9 (4.2$\sigma$, Figure \ref{fig:mf}); this deficit is recovered in Class III, which show a slight excess over the model of a factor of 1.15 (1.3$\sigma$). On the other hand, RV variables have no effect on the protoplanetary disks: both Class II and Class IIIs show a similar excess of Class III SB2s. The physical implications of this are discussed in Section \ref{sec:concl}.

This does help to shed light on the origin of some of the minor differences in the distribution of companions in the total sample (Figure \ref{fig:mf}). This includes the deficit of SB2s in NGC 1333, as it is one of the youngest clusters in the study. However, this cannot account for the excess of companions in Taurus and L1641. 

It should be noted that while veiling is systematically different between Class II and Class III systems, introducing it as an additional parameter in the construction of the synthetic binaries does not affect the recovery probability of close binaries.

The difference between RV variables and SB2s as a function of the evolutionary type does not appear to be an effect of RVs being artificially scattered by means other than a companion (e.g. star spots). To test this we also performed a comparison restricted only to the systems with low \vsini$<$20 \kms. These slow rotators have a narrow profile of their CCF that does not show as much variability due to spots in the synthetic spectra. While spots may alter RVs, the resulting shift is smaller than the typical RV uncertainties in such systems. The difference between SB2s and RV Variables between Class II and Class III systems remained consistent when the sample was limited only to slow rotators.

Similarly, the difference cannot be explained by the systematic difference in \vsini\ between Class II and Class III systems. Class II systems do tend to have \vsini\ that is higher than that of Class III systems, but only by a few \kms\, not enough to cause significant discrepancy, and the individual \vsini measurements are propagated to the model, with the comparison relying only on those synthetic spectra that are referenced to the sources that are considered in the individual bin.

\subsection{Close systems}\label{sec:period}

\begin{figure}
\epsscale{1.1}
\plotone{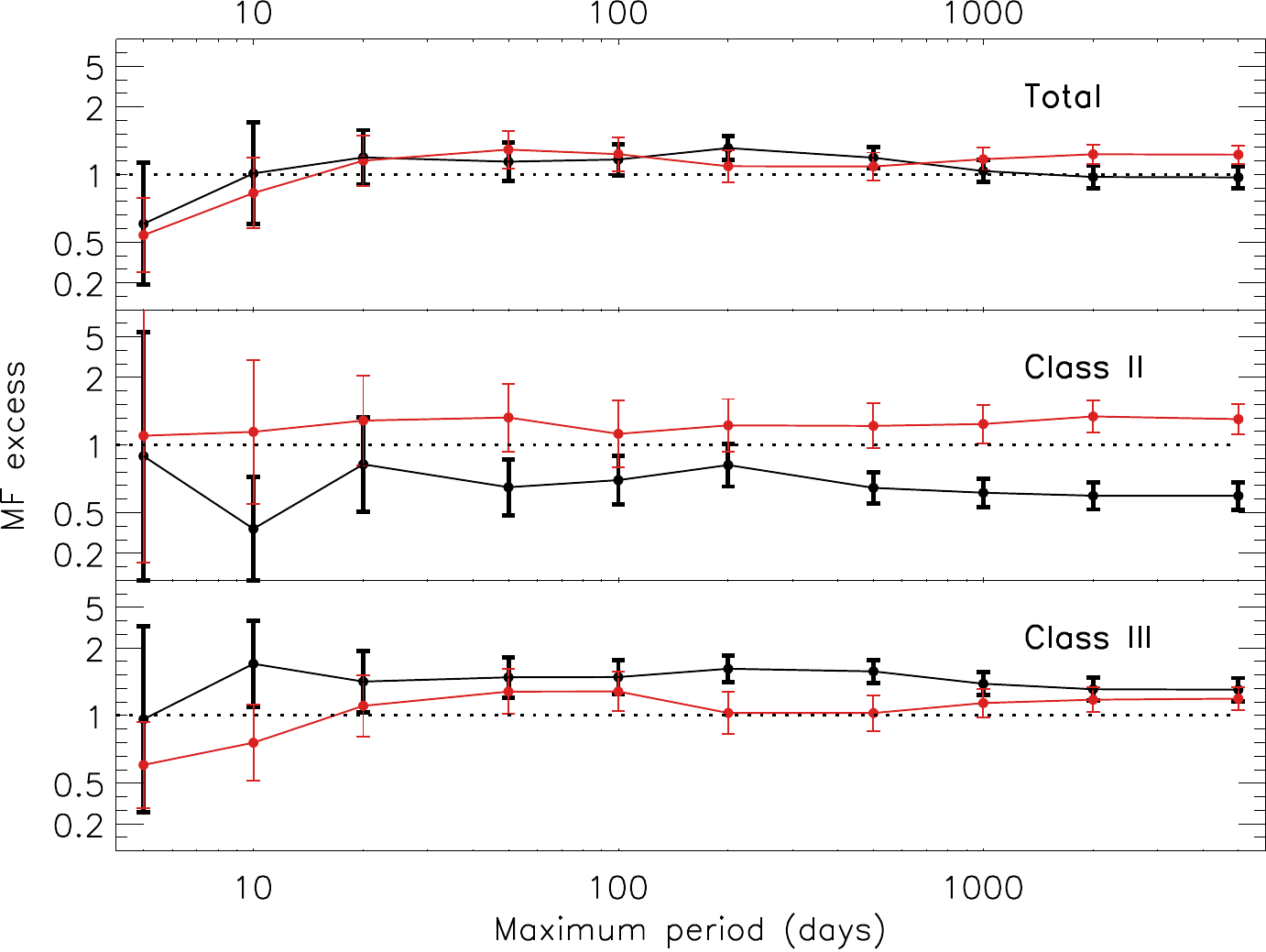}
\caption{Distribution of MF relative to the model split according to the maximum period measured from \vmax\ and \smax\ defined in the Table \ref{tab:close}.
\label{fig:mf1}}
\end{figure}

\begin{deluxetable}{ccc}
\tabletypesize{\scriptsize}
\tablewidth{0pt}
\tablecaption{\vmax\ and \smax\ cuts to identify close binaries\label{tab:close}}
\tablehead{
\colhead{Period} &\colhead{\vmax} & \colhead{\smax}\\
\colhead{(day)} &\colhead{\kms} & \colhead{\kms}
}
\startdata
$<$2000 & 2.6 & 21 \\
$<$1000 & 4.6 & 26 \\
$<$500 & 7.7 & 33 \\
$<$200 & 13 & 46 \\
$<$100 & 20 & 62 \\
$<$50 & 27 & 78 \\
$<$20 & 40 & 110 \\
$<$10 & 60 & 140 \\
$<$5 & 90 & 160 \\
\enddata
\end{deluxetable}

We use the data to test the log-normal period distribution of the model population with $\mu_{\log P} = 4.6$ and $\sigma_{\log P} = 2.2$, and the minimum period of 2 days.

In order to limit the sample to just the closest systems, for each system identified as a binary in both the data and the synthetic sample, we measured \vmax\ for the systems identified as RV variables, and \smax\ for SB2s. These properties are highly dependent on the inclination of the system, the cadence of the observations, and the probability of catching the system near the maximum velocity separation. These factors reduce the observed velocity amplitude, making \vmax\ and \smax\ only lower limits to the intrinsic maximum RV separation of the system. However, sufficiently high amplitudes in either of these parameters will nonetheless identify systems with shorter periods.

Using the synthetic spectra, we measured the \vmax\ and \smax\ cuts that rule out systems with periods longer a particular threshold with 66\% certainty (Table \ref{tab:close}). With this criterion, however, it is impossible to conclusively separate sources into individual period bins, only to give a probability that the period of the system is smaller than a given threshold. For the shortest period systems this introduces an increasingly more stringent cut, making it more likely to discard a system whose actual period is shorter than the desired upper limit. As such, for example, at 200 days it is possible to identify 72\% of all such systems with these cuts, but only 32\% of systems for periods within 10 days can be recovered, increasing a false negative fraction.


Examining the MF as a function of the cuts in \vmax\ and \smax\ shows that the period distribution in the data is consistent with the model at nearly all separations. The exception to this are sources with periods less than 5 days, which show a weak 1$\sigma$ deficit in both RV variables and SB2s (Figure \ref{fig:mf1}). This deficit is not entirely a product of the model's 2 day minimum period, however, as a raising the minimum period would affect other bins with higher maximum period (as they also include systems with short periods), making them less compatible with the model. It is possible that there may be a physical explanation for the difference between the model and the data for these short period systems. For example, because YSOs are inflated compared to main sequence stars, YSOs should originally have wider separations, but some would evolve dynamically and harden to be closer together, lessening the deficit at short periods. It is also possible that there might be some differences in the manner RVs are measured in the synthetic sample could predominantly affect the most widely resolved systems. Additionally, because we probe only the likely maximum period and not the exact period distribution of each system, it is possible to use different statistical weighting for converting from velocity to period, which could push this sub-5 day deficit to either shorter or somewhat longer periods.

Separating the sample according to their evolutionary classification shows a similar distribution; while there might be an excess or a deficit that is seen in the total sample, it remains consistent at all separations with exception of the shortest periods.

\subsection{Stellar Density}

Because the sample spans a large variety of environments, we examine the effect of multiplicity on the local stellar density in the primordial population.

A measurement of stellar density at the position of each source in the sample is a non-trivial task because of the non-uniform membership list in each of the regions in question. Disparate distances and disparate ages result in different completeness limit in each cluster. No single method is capable of identifying all of the members of a population. Moreover, fundamentally, stellar density is a three-dimensional problem. While Gaia DR2 made incredible strides in making stellar distances accessible, how deeply positioned a given source might be in a given cluster still cannot be known as precisely as a relative position in the plane of the sky.

Nonetheless, we attempt to estimate stellar density for the sources in our sample. We begin with identifying the population corresponding to each young cluster and star forming region using the Gaia DR2 data, by using TopCat \citep{topcat} to make a rough selection in the position, parallax, proper motion, and color-magnitude space that includes each cluster's sources in the curated catalog. We also made a cut perpendicular to the main sequence at the magnitude limit of the furthest cluster in the sample (i.e., NGC 2264), and discarded all sources lower than this limit in the closer regions. Similarly, we discard all high mass sources brighter than the sources that correspond to \teff$\sim$6000 K in each cluster. Unfortunately, any sources that are too heavily extinct, or sources that have irregular kinematics despite being cluster members, cannot be counted in this sample.

\begin{figure}
\epsscale{1.1}
\plotone{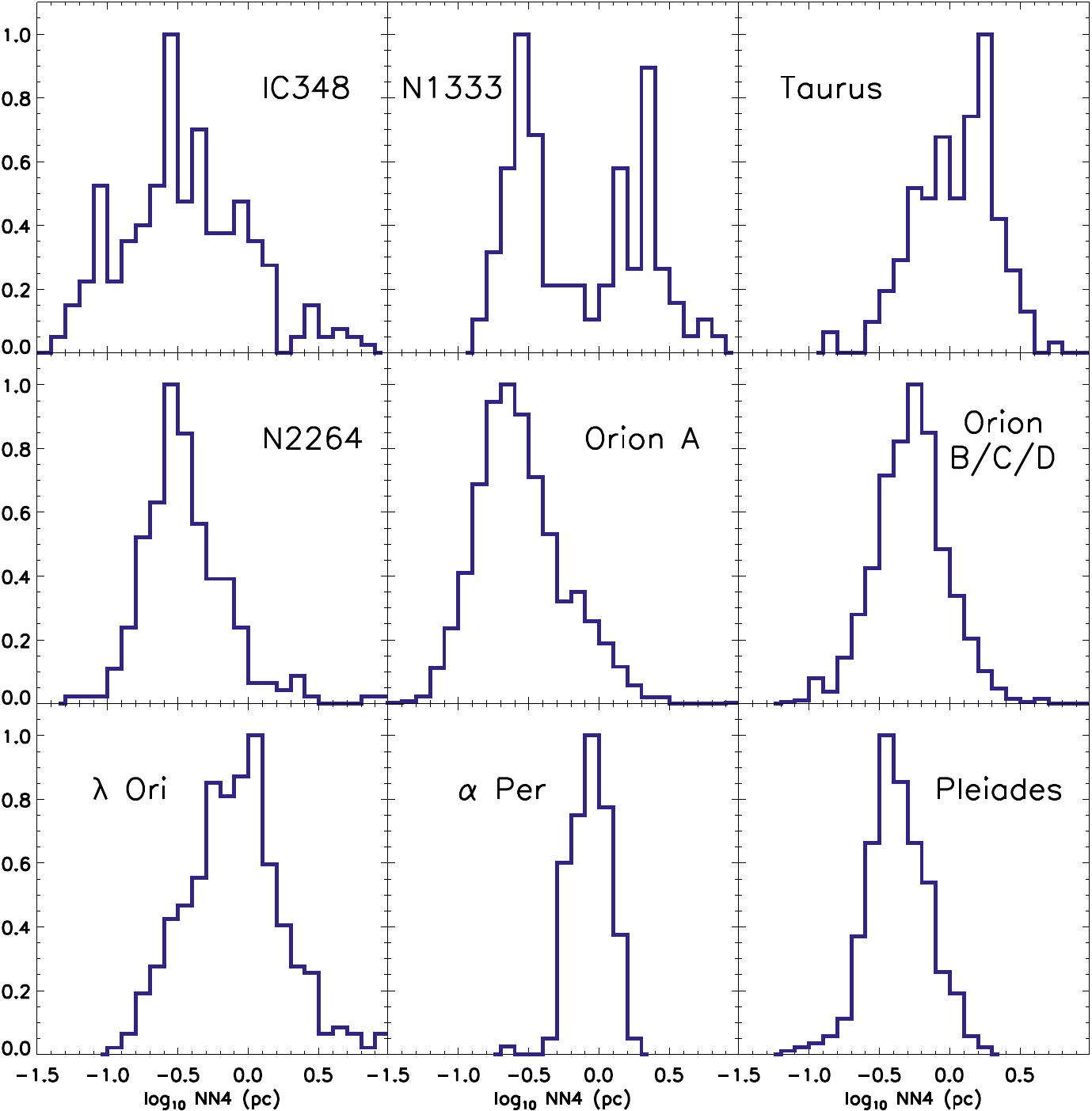}
\caption{Normalized distribution of the fourth nearest neighbor (NN4) separations in all of the regions. NN4 is inversely proportional to the stellar density.
\label{fig:density}}
\end{figure}

We then measure a projected separation between the sources in the curated sample and the resulting catalog. This separation is converted into parsecs using the parallax measurement of the source in question. To estimate the local density, we record the projected separation to the fourth nearest neighbor (NN4), similarly to the approach used by \citet{kounkel2016a}. The relative distribution of densities is shown in figure \ref{fig:density}. Again, however, it should be cautioned that this is just an estimate, made worse by any possible line of sight effects, such as may be the case in Orion C/D and in Taurus.

\begin{figure}
\epsscale{1.1}
\plotone{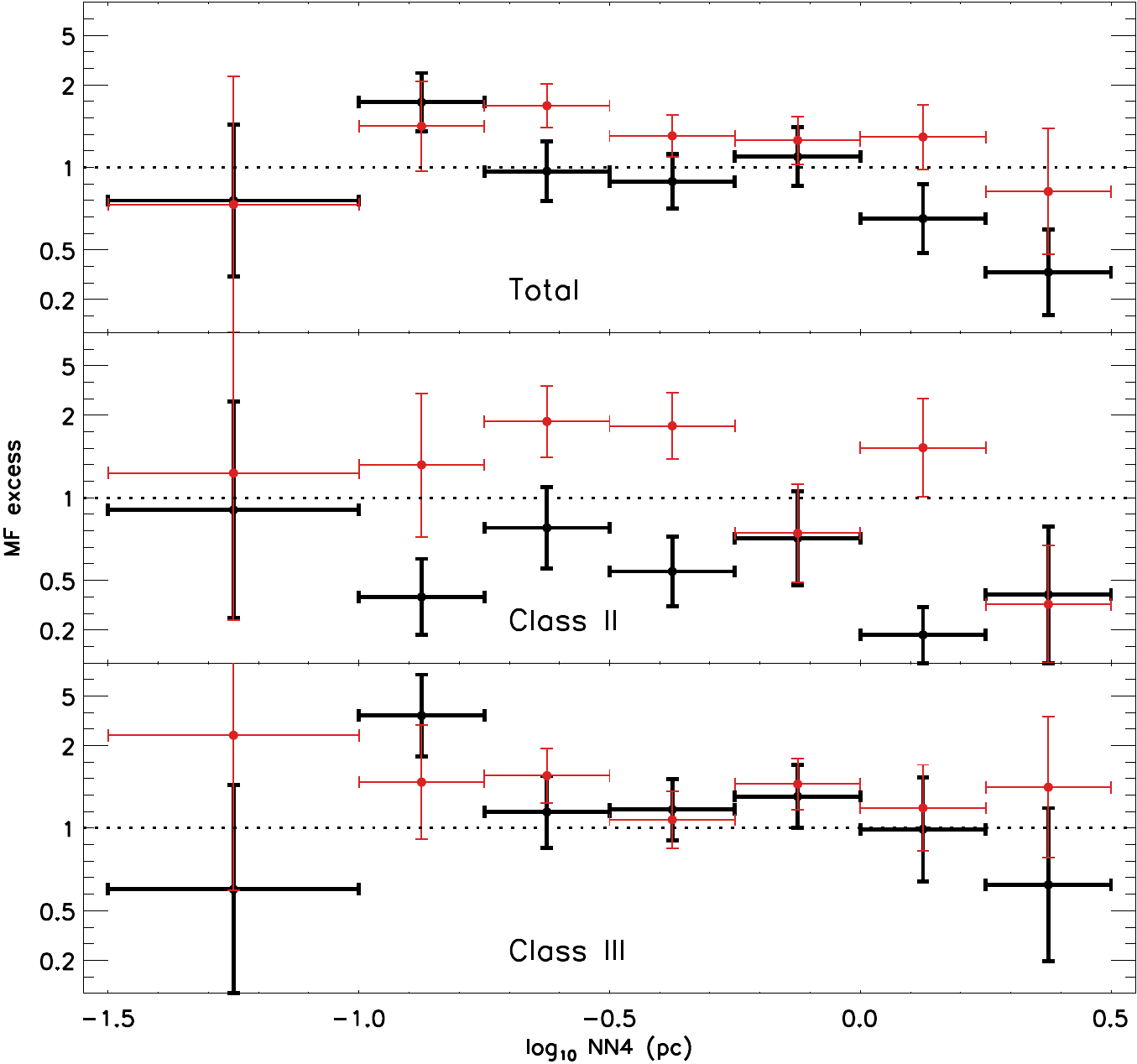}
\caption{Distribution of MF relative to the model separated according to their fourth nearest neighbor (NN4) separations.
\label{fig:mf2}}
\end{figure}

Looking at multiplicity as a function of NN4 (Figure \ref{fig:mf2}), MF appears to have a maximum at NN4$\sim$0.2 pc, or $\Sigma_*\sim30$ pc$^{-2}$ that is primarily driven by Class III SB2s, remaining mostly flat with a possible decline at the lowest densities of $\Sigma_*\sim30$ pc$^{-2}$

\section{Discussion}\label{sec:concl}

The close binary fraction and properties of young stars provide an invaluable insight into the mechanisms behind formation and evolution of the multiples. Using APOGEE spectra of pre-MS stars in various star-forming environments and accounting for selection effects, we have found their close binary fraction is consistent with the field population with separations $<10$ AU. This result is consistent with models in which the majority of close binaries form during the embedded Class 0/I phase within the first $\sim$1--3 Myr \citep{moe2018a,bate2012}. The measured binary fractions inferred from both SB2s and RV variables are consistent with each other, validating our identification pipelines and corrections for incompleteness are robust. We also find the bias-corrected pre-MS spectroscopic binary fraction increases by a factor of $\sim$2 with increasing temperature and brightness up to \teff = 6000 K, which is similar to the observed change in the field close binary fraction with respect to primary mass.

There might be a slight $\sim$30\% deficit at the shortest of periods P$<$5 day. The observed deficit suggests a minority of very close binaries migrated during the main-sequence phase via tides \citep{fabrycky2007,moe2018a}.

The mass ratio distribution of young close binaries is also largely consistent with their field counterparts. Specifically, the mass ratios are uniformly distributed with a small excess of twins with $q > 0.95$. 
Close pre-MS binaries also follow the short-end tail of a log-normal period distribution similar to that observed for low-mass and solar-type MS binaries in the field. There might be a slight deficit at the shortest of periods $P<$5 days.

The overall distribution is largely mirrored by the individual regions. There does not appear to be a dependence in MF on age, as it appears to be consistent between young clusters (e.g., ONC), as well as those that are more evolved (e.g. $\alpha$ Per, Pleiades). This contradicts what has been observed previously by \citet{jaehnig2017}, however, their trend has been primarily driven by a small number of RV variables in Pleiades.

Within each cluster younger than a few Myr, however, there appears to be a strong deficit of SB2s among sources with protoplanetary disks. The most extreme deficit is found in NGC 1333. This deficit of SB2s in disked systems is present across all separations. Close binaries are known to to affect protostellar disks -- disks are half as common in visual binaries with projected separations of $<$40 AU than they are around single stars, and their frequency may further decrease at closer separations \citep{jensen1996, cieza2009, kraus2012,harris2012, rodriguez2015}. Thus, the decreased MF in the Class II SB2s is consistent with the interpretation interpretation that close binaries either disrupt or accelerate the evolution of their natal disks. On the other hand, RV variables do not appear to affect disks, and it is consistent with \citet{kuruwita2018} conclusions regarding the observations of Class II RV variables in Upper Scorpius and Upper Centaurus-Lupus.

It is not clear why there is such a stark difference between RV variables and SB2s regarding their effect on the protoplanetary disks. Observationally, the primary difference between these systems is their mass ratios. A possible interpretation is that more equal mass companions are the result of accreting a substantial fraction of the disk, thereby reducing the disk mass and accelerating the transition to the Class III phase.

Few regions, most notably, Taurus and L1641, do show a slight elevation in MF among RV variables (but not SB2s, thus affecting only the systems with lower $q$). It is not immediately apparent why this might be the case.

There may be some weak trends with stellar density, with multiples being most common at the projected $\Sigma_*\sim$30 pc$^{-2}$, and declining at the lower end of the density distribution. While this trend may not be statistically significant, such observations pose important constraints in analyzing the degree to which the primordial binaries are disrupted in dense environments.

\software{IN-SYNC pipeline \citep{cottaar2014}, GaussPy \citep{lindner2015}, PHOENIX spectral library \citep{husser2013}, PARSEC-COLIBRI isochrones \citep{marigo2017}, TOPCAT \citep{topcat}}

\acknowledgments
We acknowledge Adrian Price-Whelan for the discussion, and Claire Murray for the clarification of the functionality of GaussPy. 
M.K. and K.C. acknowledge support provided by the NSF through grant AST-1449476, and from the Research Corporation via a Time Domain Astrophysics Scialog award (\#24217). M.M. and K.M.K. acknowledge financial support from NASA under grant ATP-170070. C.R-Z acknowledges support from program
UNAM-DGAPA-PAPIIT IN 108117, Mexico. A.R-L acknowledges financial support provided in Chile by Comisi\'{o}n Nacional de Investigaci\'{o}n Cient\'{i}fica y Tecnol\'{o}gica (CONICYT) through the FONDECYT project 1170476 and by the QUIMAL project 130001. KPR acknowledges CONICYT PAI Concurso Nacional de Insercio ́n en la Academia, Convocatoria 2016 Folio PAI79160052. Support for JB is provided by the Ministry for the Economy, Development and Tourism, Programa Iniciativa Cientica Milenio grant IC120009, awarded to the Millennium Institute of Astrophysics (MAS).

Funding for the Sloan Digital Sky Survey IV has been provided by the Alfred P. Sloan Foundation, the U.S. Department of Energy Office of Science, and the Participating Institutions. SDSS-IV acknowledges
support and resources from the Center for High-Performance Computing at
the University of Utah. The SDSS web site is www.sdss.org.
SDSS-IV is managed by the Astrophysical Research Consortium for the 
Participating Institutions of the SDSS Collaboration including the 
Brazilian Participation Group, the Carnegie Institution for Science, 
Carnegie Mellon University, the Chilean Participation Group, the French Participation Group, Harvard-Smithsonian Center for Astrophysics, 
Instituto de Astrof\'isica de Canarias, The Johns Hopkins University, 
Kavli Institute for the Physics and Mathematics of the Universe (IPMU) / 
University of Tokyo, Lawrence Berkeley National Laboratory, 
Leibniz Institut f\"ur Astrophysik Potsdam (AIP),  
Max-Planck-Institut f\"ur Astronomie (MPIA Heidelberg), 
Max-Planck-Institut f\"ur Astrophysik (MPA Garching), 
Max-Planck-Institut f\"ur Extraterrestrische Physik (MPE), 
National Astronomical Observatories of China, New Mexico State University, 
New York University, University of Notre Dame, 
Observat\'ario Nacional / MCTI, The Ohio State University, 
Pennsylvania State University, Shanghai Astronomical Observatory, 
United Kingdom Participation Group,
Universidad Nacional Aut\'onoma de M\'exico, University of Arizona, 
University of Colorado Boulder, University of Oxford, University of Portsmouth, 
University of Utah, University of Virginia, University of Washington, University of Wisconsin, 
Vanderbilt University, and Yale University.
This work has made use of data from the European Space Agency (ESA)
mission {\it Gaia} (\url{https://www.cosmos.esa.int/gaia}), processed by
the {\it Gaia} Data Processing and Analysis Consortium (DPAC,
\url{https://www.cosmos.esa.int/web/gaia/dpac/consortium}). Funding
for the DPAC has been provided by national institutions, in particular
the institutions participating in the {\it Gaia} Multilateral Agreement.

\bibliographystyle{aasjournal.bst}
\bibliography{apbin.bbl}

\end{document}